\documentclass[aps,prxquantum,reprint,superscriptaddress,amsmath,amssymb,pra,showkeys,longbibliography]{revtex4-2}

\usepackage{graphicx}
\usepackage{dcolumn}
\usepackage{bm}
\usepackage[colorlinks=true,linkcolor=blue, citecolor=blue, urlcolor=blue]{hyperref}
\usepackage{braket}
\usepackage{appendix}
\usepackage{natbib}
\usepackage{physics}
\usepackage{float}
\usepackage{booktabs}
\usepackage[T1]{fontenc}
\usepackage{booktabs}
\usepackage{adjustbox}
\usepackage{float}
\usepackage{multirow}
\usepackage{gensymb}
\usepackage{array}
\usepackage{siunitx}
\usepackage{xr}
\usepackage[sort&compress]{cleveref}
\usepackage[caption=false]{subfig}
\usepackage{cleveref}

\newcolumntype{L}{>{$}l<{$}}
\newcolumntype{C}{>{$}c<{$}}
\newcolumntype{R}{>{$}r<{$}}

\usepackage[hang,flushmargin,symbol*]{footmisc}
\DefineFNsymbolsTM{otherfnsymbols}{
	\textdagger    \dagger
}

\begin{document}
	\title{Quantum geometrical description of hole spin qubits far away from the $\Gamma$-point}
	\author{Zoltán György}
    \email{zoltan.gyoergy@unibas.ch}
\affiliation{Department of Physics, University of Basel, Klingelbergstrasse 82, CH-4056 Basel, Switzerland} 
	\author{Dmitry Miserev}
\affiliation{Department of Physics, University of Basel, Klingelbergstrasse 82, CH-4056 Basel, Switzerland}
    \author{Jelena Klinovaja}
\affiliation{Department of Physics, University of Basel, Klingelbergstrasse 82, CH-4056 Basel, Switzerland}
    \author{Daniel Loss} 
\affiliation{Department of Physics, University of Basel, Klingelbergstrasse 82, CH-4056 Basel, Switzerland}
\affiliation{Physics Department, King Fahd University of Petroleum and Minerals, 31261, Dhahran, Saudi Arabia}
\affiliation{Quantum Center, KFUPM, Dhahran, Saudi Arabia}
\affiliation{RDIA Chair in Quantum Computing}

	\date{\today}            
	
\begin{abstract}
		Hole spin qubits provide one of the leading platforms for spin-based quantum computing due to their large intrinsic spin-orbit interaction (SOI), which enables fast electrical manipulation. The SOI of planar quantum dots has mostly been investigated in theoretical studies by examining the SOI already present in the two-dimensional hole gas (2DHG). Here, we study the SOI created by the in-plane confinement by deriving non-perturbative effective Hamiltonians numerically for hole spin qubits. We find that the quantum geometry of the 2DHG naturally emerges, leading to a meaningful non-perturbative definition of pseudospin valid far away from the $\Gamma$-point. The SOI of the 2DHG and of the in-plane confinement have different forms; therefore, they cannot be turned off simultaneously, ruining the perfect spin-orbit switch functionality of spin qubits. We construct effective Hamiltonians using the symmetry approach for various low-dimensional hole systems: (i) a heavy-hole confined in a SiGe/Ge/SiGe heterostructure, (ii) a light-hole confined in SnGe/Ge, (iii) a gate-defined nanowire in SiGe/Ge/SiGe, and (iv) a hole confined in a Ge/Si core/shell nanowire. The non-perturbative effective Hamiltonians provide results with excellent agreement with the full Hamiltonians.
\end{abstract}
	
	\maketitle

	\section{\label{sec:intro}Introduction}
Spin qubits confined in semiconductor quantum dots \cite{PhysRevA.57.120,Fang_2023,RevModPhys.95.025003} provide one of the most promising platforms for quantum computing due to their small footprint \cite{philips2022universal, hendrickx2021four,borsoi2024shared,zhang2025universal,wang2024operating, john2024two,george202412, lim20242x2}, high-fidelity single- and two-qubit operations \cite{veldhorst2014addressable,yoneda2018quantum,lawrie2023simultaneous,mills2022high,huang2024high, wu2025simultaneous,xue2022quantum,mills2022two,noiri2022fast,huang2024high}, their compatibility with semiconductor manufacturing processes \cite{zwerver2022qubits,steinacker2025industry,george202412}, and the prospect of creating entanglement between distant qubits via the integration of superconducting elements \cite{zwerver2022qubits,eggli2025coupling,bottcher2022parametric,noirot2025coherence,de2024strong,mi2018coherent} or long-distance shuttling \cite{bosco2024high,kunne2024spinbus,van2024coherent,de2025high,yoneda2021coherent,ademi2025distributing}. Today, the state-of-the-art devices are based on electrons confined in Si or holes in Ge heterostructures. In particular, holes confined in SiGe/Ge/SiGe heterostructures \cite{scappucci2021germanium} have shown rapid progress in recent years, from the demonstration of single- and two-qubit control \cite{hendrickx2020single,watzinger2018germanium,jirovec2021singlet,hendrickx2020fast} to multi-qubit processors \cite{ivlev2025operating,hendrickx2021four,john2024two,zhang2025universal} and quantum dot arrays \cite{borsoi2024shared}. 

Using holes instead of electrons has multiple advantages, including the reduced hyperfine interaction with spinful nuclei due to their $p$-like character, the strong intrinsic spin-orbit interaction (SOI) that allows fast electric control \cite{bulaev2007electric,wang2022ultrafast,froning2021ultrafast,bosco2021hole,abadillo2023hole,martinez2022hole,terrazos2021theory}, the possibility of $g$-tensor tuning via gate voltages \cite{liles2021electrical,bassi2024optimal,mauro2025hole,seidler2025spatial, sommer2026disentangling}, and the implementation of single-qubit gates using hopping spins \cite{wang2024operating}. Moreover, holes have relaxation times comparable to those of electrons \cite{bulaev2005spin}. The holes confined in SiGe/Ge/SiGe planar quantum dots have mainly a heavy-hole character, which leads to a cubic Rashba SOI \cite{winkler2001spin,marcellina2017spin}. Although linear-in-momentum SOI can arise from inhomogeneous shear strain or interface effects \cite{abadillo2023hole, rodriguez2023linear,sarkar2025effect}, one might look for other implementations where linear Rashba SOI naturally appears. In particular, light holes confined to structures $\mathrm{Sn_xGe_{1-x}/Ge}$ possess a strong linear-in-k Rashba SOI, with the SOI and the out-of-plane $g$ factor tunable by gate voltages, and with larger in-plane $g$ factor, overall resulting in a less anisotropic $g$-tensor \cite{assali2022light,del2023light,del2024light,de2024strong}.

\begin{figure}[h!]
\centering
\includegraphics[width=\columnwidth]{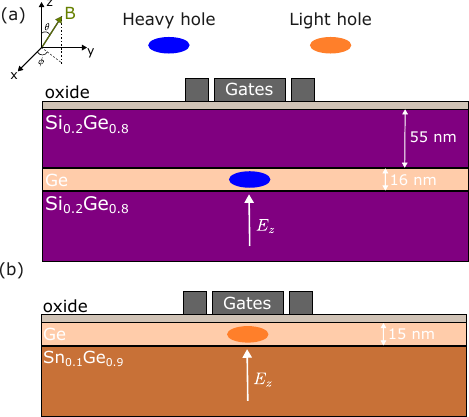}

\caption {\label{fig:devices}Planar heterostructures considered in this work. A homogeneous electric field $E_z$ is applied perpendicularly, while a homogeneous magnetic field $\bm{B}$ is applied in an arbitrary direction. In both cases, the metallic gates on top confine a single hole in the quantum dot. \small{(a) Schematic representation of a planar heterostructure hosting a heavy hole. The compressive biaxial strain pushes a heavy hole state as the ground state. (b) Schematic representation of a planar heterostructure hosting a light hole. The tensile strain pushes the light-hole states into the ground state.}}
\refstepcounter{subfigure}\label{fig:devicesa}
    \refstepcounter{subfigure}\label{fig:devicesb}
\end{figure}

For Si and Ge devices, SOI can only originate from a structural inversion asymmetry. The holes confined in nanowire quantum dots exhibit a very large direct Rashba SOI \cite{kloeffel2011strong,kloeffel2018direct,carballido2025compromise}, due to the strong confinement in two dimensions, while along the nanowire there is a weaker confinement. For lateral quantum dots, SOI emerges due to the strong electric field applied perpendicularly to the plane of the heterostructure \cite{terrazos2021theory, sarkar2023electrical,wang2024modeling}. Then, to create the quantum dot, a weaker in-plane confinement is applied. 
In both cases, SOI mostly originates from strong confinement. However, an additional SOI is expected to emerge due to weak quantum dot confinement. 

Most theoretical studies of holes in planar quantum dots and nanowires focused on effective 2D or nanowire Hamiltonians describing the 2DHG or the nanowire itself without considering the additional confinement that creates the quantum dot \cite{del2024light,kloeffel2018direct,miles2025effective}. Other works derived perturbative quantum dot effective Hamiltonians valid close to the $\Gamma$-point \cite{del2023light,terrazos2021theory,adelsberger2022hole}. In this paper, we present a method to derive non-perturbative effective quantum dot Hamiltonians. Including the quantum dot confinement in the model sheds light on its interplay with the SOI already present in the 2DHG (nanowire), and points to the proper definition of pseudospin far away from the $\Gamma$-point (even though pseudospin is a property of the 2DHG/nanowire, not of the confinement). 

The remainder of the paper is structured as follows: in Sec.~\ref{sec:model} we present our model and assumptions. In Sec. \ref{sec:Heffband} we derive the effective Hamiltonian in the so-called band basis. In Sec.~\ref{sec:Heffpseudo} we define pseudospin and map the effective Hamiltonian from the band basis to the pseudospin basis. In Sec.~\ref{sec:planar} we apply our method to planar quantum dots. In Sec.~\ref{sec:nanowires} we apply the method to nanowires. Finally, in Sec.~\ref{sec:Conclusion} we discuss our results and conclude.  Additional technical details are provided in the Appendices.

\section{Model and assumptions}\label{sec:model}

We aim to describe hole spin qubits confined in different nanostructures. Our main assumption is that the hole spin is confined strongly in one (two) dimension, while in the remaining two (one) dimensions it is more weakly confined. This assumption holds for lateral quantum dots (nanowires). We start with the 3D Hamiltonian and first consider the strong confinement, leading to subbands. We derive a non-perturbative effective Hamiltonian that describes the physics of the lowest two spin-split subbands and neglects all other subbands. As the remaining quantum dot confinement is much weaker, it does not mix the high-energy subbands with the two subbands of the effective Hamiltonian. Therefore, we can describe hole spin qubits using the effective Hamiltonian of the lowest two subbands and the weak confinement potential. A detailed derivation of the effective Hamiltonian can be found in Appendices~\ref{App:HeffBand} and~\ref{App:Apseudo}.

Throughout the paper, we consider germanium hole spin qubits, which we model using the 6-band Luttinger-Kohn-Bir-Pikus (LKBP) Hamiltonian. However, our method should apply to hole spin qubits based on any cubic semiconductor, e.g., Si hole spin qubits.

To be more concrete, here we present the method assuming planar heterostructures. The application of the method to nanowires can be found in Sec.~\ref{sec:nanowires}. The 3D Hamiltonian is the following: 
\begin{equation}\label{eq:H3D}
    H_\mathrm{3D}=H_\mathrm{LKBP}+V_\mathrm{qw}(z)+V(x,y),
\end{equation}
where $H_\mathrm{LKBP}$ is the 6-band LKBP Hamiltonian, $V_\mathrm{qw}$ contains the strong quantum well confinement and an electric field $E_z$ along $z$, and $V(x,y)$ is the weak in-plane confinement. We further assume a homogeneous magnetic field $B$ with an arbitrary direction. For more details on the full Hamiltonian, see Appendix \ref{app:Ham}.

The two planar heterostructures that we study in detail in Sec.~\ref{sec:planar} can be seen in Fig.~\ref{fig:devices}. Figure~\ref{fig:devicesa} shows a heavy hole confined in SiGe/Ge/SiGe, with a 16 nm quantum well and 55 nm barrier, while Fig.~\ref{fig:devicesb} shows a light hole confined in a SnGe/Ge structure, with a 15 nm Ge layer. The metallic gates on top provide the in-plane confinement $V(x,y)$. 

\begin{figure}[h!]
\centering
\includegraphics[width=\columnwidth]{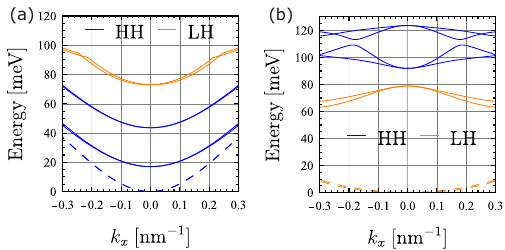}

\caption {\label{fig:dispersion2DHG} \small{Subbands created by a strong $z$-confinement in planar heterostructures shown as a function of $k_x$, while $k_y=0$. (a) Subbands of a $\mathrm{Si_{0.2}Ge_{0.8}/Ge/Si_{0.2}Ge_{0.8}}$ heterostructure shown in Fig.~\ref{fig:devicesa}. The electric field is $E_z=1$ mV/nm, and the magnetic field is zero. (b) Subbands of a $\mathrm{Sn_{0.1}Ge_{0.9}}\mathrm{/Ge}$ heterostructure from Fig.~\ref{fig:devicesb}. The electric field is $E_z=6$ mV/nm, and the magnetic field is zero. In both cases, the blue (orange) curves show subbands with dominantly heavy-hole (light-hole) character. The effective Hamiltonian from Eq.~\eqref{eq:Heff} is constructed based on the lowest two spin-split subbands (dashed lines).}}
\refstepcounter{subfigure}\label{fig:dispersion2DHGa}
    \refstepcounter{subfigure}\label{fig:dispersion2DHGb}
\end{figure}

We assume a magnetic field $\bm{B}$ with a small out-of-plane component. This means that the Landau quantization can be neglected, and the out-of-plane Zeeman effect can be taken into account via the direct Zeeman coupling. Under this approximation, the Hamiltonian along $z$ is decoupled from $x$ and $y$ directions, therefore, we can solve the confinement problem along $z$ numerically and obtain subbands as a function of in-plane momenta $k_x$ and $k_y$. These subbands can be seen in Fig.~\ref{fig:dispersion2DHG}. The subbands are split by Rashba SOI, originating from the electric field $E_z$. The full 2D Hamiltonian with the $z$-dependence integrated out becomes: 
\begin{equation}\label{eq:Hfull}
    H_\mathrm{full}=H_0(k_x,k_y)+V(x,y),
\end{equation}
where $H_0(k_x,k_y)$ is the Hamiltonian that describes the subbands of the 2DHG. We note that the Hamiltonian from Eq.~\eqref{eq:Hfull} is still the full Hamiltonian of the system. It contains the same information as Eq.~\eqref{eq:H3D}, but with the $z$-dependence integrated out. More details on the solution of the quantum well confinement can be found in the Appendix~\ref{app:conf}.

\section{Effective Hamiltonian in the band basis}\label{sec:Heffband}

Our goal is to derive an effective Hamiltonian of the hole spin qubit using only the lowest two spin-split subbands, which are plotted in Fig.~\ref{fig:dispersion2DHG} with dashed lines. This is a good approximation if the separation of these two subbands from the higher ones is much larger than the energy scale of the in-plane confinement $V(x,y)$. This assumption holds for real devices, where the orbital level spacing is on the order of 1 meV \cite{hendrickx2020single, hendrickx2024sweet,brauns2016anisotropic,froning2018single, sommer2026disentangling}. 

We numerically diagonalize the 2DHG Hamiltonian $H_0$ in Eq.~\eqref{eq:Hfull} to find the energies and eigenstates of the two lowest bands. 
\begin{equation}
    H_0(\bm{k})\ket{\psi_\sigma(\bm{k})}=E_\sigma(\bm{k})\ket{\psi_\sigma(\bm{k})}, 
\end{equation}
where $\sigma \in \{1,2\}$ is the subband index and $\bm{k}=(k_x,k_y)$ is the in-plane momentum. 
In Appendix~\ref{App:HeffBand} we show that the effective Hamiltonian can be written as follows: 
\begin{equation}\label{eq:Heff}
    H_\mathrm{eff}=H^{(d)}(\bm{k})+V\left(x+A^{(x)}(\bm{k}),y+A^{(y)}(\bm{k})\right),
\end{equation}
where $H^{(d)}(\bm{k})$ is the dispersion Hamiltonian that describes the spin-split pair of bands of the 2DHG, while $A^{(\mu)}(\bm{k})$ ($\mu=\{x,y\}$) is the Berry connection operator restricted to the spin-split pair of bands.
These two operators can be represented as matrices on the so-called band basis, on the 2D subspace $S(\bm{k})=\mathrm{span}\{\psi_1(\bm{k}),\psi_2(\bm{k})\}$. On the band basis, the dispersion Hamiltonian is a diagonal matrix: 
\begin{equation}\label{eq:Hdband}
    H_\mathrm{band}^{(d)}(\bm{k})=\begin{pmatrix}
      E_1(\bm{k}) & 0 \\ 
      0 & E_2(\bm{k})
    \end{pmatrix},
\end{equation}
while the Berry connection matrix can be written as: 
\begin{equation}\label{eq:Aband}
    A^{(\mu)}_\mathrm{band}(\bm{k})=i\begin{pmatrix}
        \bra{\psi_1(\bm{k})}\partial_{k_\mu}\ket{\psi_1(\bm{k})} & \bra{\psi_1(\bm{k})}\partial_{k_\mu}\ket{\psi_2(\bm{k})} \\ 
        \bra{\psi_2(\bm{k})}\partial_{k_\mu}\ket{\psi_1(\bm{k})} & \bra{\psi_2(\bm{k})}\partial_{k_\mu}\ket{\psi_2(\bm{k})}
    \end{pmatrix}.
\end{equation}

In the in-plane confinement potential of the effective Hamiltonian from Eq.~\eqref{eq:Heff}, the position operators $x$ and $y$ are shifted by the Berry connections, which are $2\times 2$ $\bm{k}$-dependent Hermitian operators. The position and momentum operators in Eq.~\eqref{eq:Heff} need to be treated as operators that do not commute. The emergence of Berry connections upon projecting operators to a subset of bands has already been studied in Ref.~\cite{xu2025quantum}. However, in Ref.~\cite{xu2025quantum}, the authors do not focus on constructing effective Hamiltonians.

We can define Pauli operators on the band subspace $S(\bm{k})$: 
\begin{subequations}\label{eq:Pauliband}
\begin{align}
\sigma_x(\bm{k})&=|\psi_1(\bm{k})\rangle\!\langle \psi_2(\bm{k})|+|\psi_2(\bm{k})\rangle\!\langle \psi_1(\bm{k})|, \\
\sigma_y(\bm{k})&=-i|\psi_1(\bm{k})\rangle\!\langle \psi_2(\bm{k})|+i|\psi_2(\bm{k})\rangle\!\langle \psi_1(\bm{k})|, \\
\sigma_z(\bm{k})&=|\psi_1(\bm{k})\rangle\!\langle \psi_1(\bm{k})|-|\psi_2(\bm{k})\rangle\!\langle \psi_2(\bm{k})|.
\end{align}
\end{subequations}
The Berry connections in the band basis from Eq.~\eqref{eq:Aband} can be expressed using the Pauli operators in Eq.~\eqref{eq:Pauliband}:
\begin{equation}\label{eq:AbandPauli}
\begin{aligned}
    A^{(\mu)}_\mathrm{band}(\bm{k})&=A^{(\mu)}_\mathrm{band,x}(\bm{k})\sigma_x(\bm{k})+A^{(\mu)}_\mathrm{band,y}(\bm{k})\sigma_y(\bm{k})+\\&+A^{(\mu)}_\mathrm{band,z}(\bm{k})\sigma_z(\bm{k}). 
\end{aligned}
\end{equation}
Note that we can Taylor-expand the effective potential from the effective Hamiltonian in Eq.~\eqref{eq:Heff} around some point $x_0$ and $y_0$ to obtain the following: 
\begin{equation}\label{eq:HeffExpand}
\begin{aligned}
    H_\mathrm{eff}&=H^{(d)}_\mathrm{band}(\bm{k})+V(x_0,y_0)+\frac{1}{2}\{\partial_xV,A^{(x)}_\mathrm{band}(\bm{k})\}+\\& +\frac{1}{2}\{\partial_yV,A^{(y)}_\mathrm{band}(\bm{k})\}+ \hdots ,
\end{aligned}
\end{equation}
where the derivatives are taken at $x_0$ and $y_0$ and $\{A,B\}=AB+BA$. The last two terms of Eq.~\eqref{eq:HeffExpand} are new spin-orbit terms originating from the in-plane confinement. We remark that keeping only the linear terms from the expansion does not lead to a gauge-covariant effective Hamiltonian, as the Berry connection is not a gauge-covariant quantity. However, the full form of $H_\mathrm{eff}$ in Eq.~\eqref{eq:Heff} is gauge covariant, because $i\partial_{k_\mu}+A^{(\mu)}_\mathrm{band}(\bm{k})$ transforms covariantly, where $i\partial_{k_\mu}$ is the position operator in momentum representation. 

\section{Effective Hamiltonian in the pseudospin basis}\label{sec:Heffpseudo}
So far, we represented the effective Hamiltonian $H_\mathrm{eff}$ in Eq.~\eqref{eq:Heff} on the adiabatically changing band basis, on the subspace $S(\bm{k})=\mathrm{span}\{\psi_1(\bm{k}),\psi_2(\bm{k})\}$. We can also represent $H_\mathrm{eff}$ on a constant basis, using the two eigenstates at the $\Gamma$ point that spans the subspace $S_0=\mathrm{span}\{\psi_1(\bm{0}),\psi_2(\bm{0})\}$. We call this basis the pseudospin basis. 
To map the effective Hamiltonian from the adiabatic to the pseudospin basis, we construct an isomorphism between the subspaces $S(\bm{k})$ and $S_0$, using the overlap matrix $M(\bm{k})$:
\begin{equation}\label{eq:M}
    M(\bm{k})=\begin{pmatrix}
        \bra{\psi_1(\bm{0})}\ket{\psi_1(\bm{k})} & \bra{\psi_1(\bm{0})}\ket{\psi_2(\bm{k})} \\
        \bra{\psi_2(\bm{0})}\ket{\psi_1(\bm{k})} & \bra{\psi_2(\bm{0})}\ket{\psi_2(\bm{k})} 
    \end{pmatrix}.
\end{equation}
 Then, we construct the unitary matrix $X(\bm{k})$
 \begin{equation}\label{eq:X}
     X(\bm{k})=M(\bm{k})[M^\dag(\bm{k}) M(\bm{k})]^{-1/2}, 
 \end{equation}
 which can be used to transform the operators and Berry connections from the band basis. Under this new unitary $X(\bm{k})$ calculated using polar decomposition, regular operators such as $H^{(d)}_\mathrm{band}$ from Eq.~\eqref{eq:Hdband} transform the following way: 
\begin{equation}\label{eq:dispersion_band}
    H^{(d)}_\mathrm{pseudo}(\bm{k})=X(\bm{k})H^{(d)}_\mathrm{band}(\bm{k})X^\dag(\bm{k}), 
\end{equation}
while the Berry connections from Eq.~\eqref{eq:Aband} transform as
\begin{equation}\label{eq:AbandTransform}
    A^{(\mu)}_\mathrm{pseudo}(\bm{k})=X(\bm{k})A^{(\mu)}_\mathrm{band}(\bm{k})X^\dag(\bm{k})+iX(\bm{k})\partial_{k_\mu}X^\dag(\bm{k}).
\end{equation}
We note that the $X(\bm{k})$ isomorphism in Eq.~\eqref{eq:X} between subspaces $S_0$ and $S(\bm{k})$ is not unique; there are infinitely many ways to construct isomorphisms between two-dimensional subspaces. However, the $X(\bm{k})$ transformation is a natural choice, due to its simplicity and because it approximately cancels the Berry connections. In the pseudospin basis, only the physical Berry connections remain, which describe the additional SOI created by the in-plane confinement. The Berry connections originating from the adiabatically changing basis are eliminated.

The effective Hamiltonian always has the form as in Eq.~\eqref{eq:Heff}, but the dispersion Hamiltonian $H^{(d)}(\bm{k})$ and the Berry connections $A^{(x)}(\bm{k})$ and $A^{(y)}(\bm{k})$
have different matrix representations in the band basis and in the pseudospin basis. In the band basis, the dispersion Hamiltonian is diagonal, and the Berry connections are large, whereas in the pseudospin basis, the dispersion Hamiltonian becomes off-diagonal and the Berry connections are small. For more details on the mapping from the band basis to the pseudospin basis, see Appendix~\ref{App:Apseudo}.

We note that the transformation of Hamiltonians from an adiabatic basis to a diabatic one is commonly employed in chemical physics \cite{lin2018explicit, fatehi2013derivative, subotnik2008constructing, littlejohn2022parallel}, where the eliminated adiabatic parameter is the nuclear position, instead of the wavevector. Similar transformations are also used in condensed matter physics, where a basis of $\bm{k}$-dependent Bloch functions is transformed onto a diabatic basis \cite{souza2001maximally, mostofi2008wannier90, yates2007spectral, ozaki2024closest}. 

The Berry connections and the dispersion Hamiltonian can be expressed in the pseudospin basis. Analogously to the band basis, we can construct the Pauli operators $\Sigma_x$, $\Sigma_y$ and $\Sigma_z$ using the states $\ket{\psi_1(\bm{0})}$ and $\ket{\psi_2(\bm{0})}$.
These Pauli operators, contrary to the band basis, are $\bm{k}$-independent. The Berry connection in the pseudospin basis: 
\begin{equation}\label{eq:Apseudo}
\begin{aligned}
    A_\mathrm{pseudo}^{(\mu)}(\bm{k})&=A_\mathrm{pseudo,x}^{(\mu)}(\bm{k})\Sigma_x+A_\mathrm{pseudo,y}^{(\mu)}(\bm{k})\Sigma_y+ \\ &+A_\mathrm{pseudo,z}^{(\mu)}(\bm{k})\Sigma_z.
\end{aligned}
\end{equation}
The dispersion Hamiltonian can also be expressed as
\begin{equation}\label{eq:Hdpseudo}
    H^{(d)}_\mathrm{pseudo}(\bm{k})=E_0(\bm{k})+a_x(\bm{k})\Sigma_x+a_y(\bm{k})\Sigma_y+a_z(\bm{k})\Sigma_z.
\end{equation}
The effective Hamiltonian in the pseudospin basis written in Eq.~\eqref{eq:Heff} can be expressed using the operators from Eqs.~\eqref{eq:Apseudo} and \eqref{eq:Hdpseudo}. 

The main advantage of using the pseudospin basis is that the Pauli operators $\Sigma_x$, $\Sigma_y$, and $\Sigma_z$ possess spin transformation properties. This means that the dispersion Hamiltonian in Eq.~\eqref{eq:Hdpseudo} satisfies the symmetry constraints of the host material, even at large momentum $\bm{k}$, when the overlap between the states $\ket{\psi_{1(2)}(\bm{k})}$ and $\ket{\psi_{1(2)}(\bm{0})}$ is not close to 1. 

We note that the dimensional reduction of Hamiltonians from 3D to 2D has been studied in~\cite{miserev2017dimensional} for planar GaAs and InAs heterostructures. The calculations presented in that paper are based on an isotropic approximation. Our method is more general and rigorous, and it works without isotropic approximation. Therefore, throughout the paper, we include the anisotropy arising from the $\gamma_2\neq \gamma_3$ Luttinger parameters. 

\begin{figure}[t]
\centering
\includegraphics[width=\columnwidth]{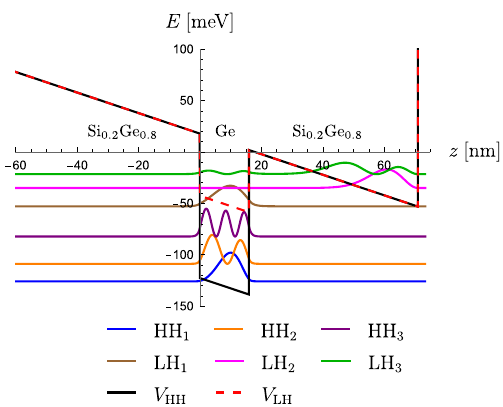}
\caption {\label{fig:QWHH} \small{Quantum well and lowest six wavefunctions (absolute value squares) in a SiGe/Ge/SiGe planar heterostructure with $x=0.2$ Si composition, sketched in Fig. \ref{fig:devicesa}. The applied electric field is $E_z=1$ mV/nm. The heavy-hole (light-hole) states experience the potential drawn with black (red).}}
\end{figure}

\section{Application to planar quantum dots}\label{sec:planar}
In this section, we apply our method to derive effective Hamiltonians for hole spin qubits confined in planar heterostructures. We model the heterostructure using the 6-band LKBP Hamiltonian, assuming biaxial strain only in the Ge quantum well, which arises from the lattice mismatch between Ge and SiGe or SnGe. We further assume that the heterostructure is grown along the [001] crystal direction.

\begin{figure*}[!t]
  \centering
   \includegraphics[width=2\columnwidth]{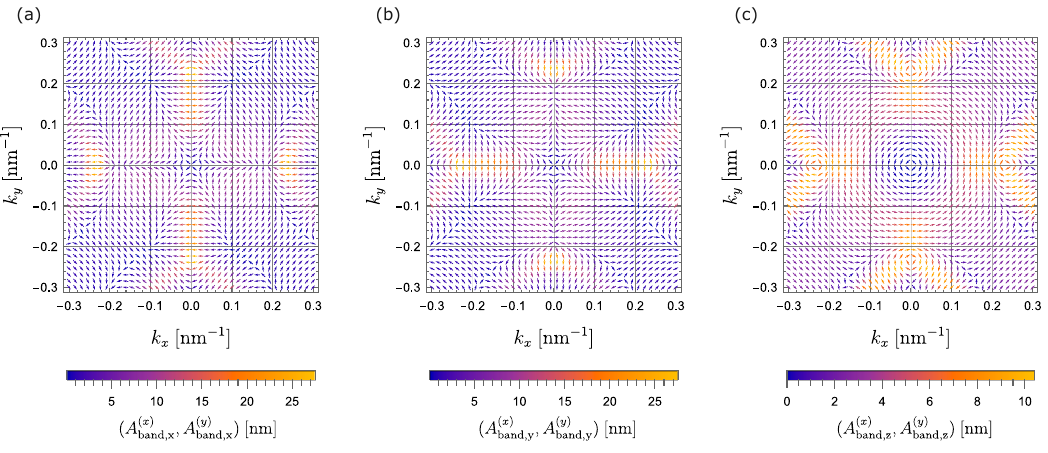}
    \caption {\label{fig:AHHband}\small{Berry connections of a heavy hole 2DHG, see Fig.~\ref{fig:devicesa}. The Berry connections are expanded in the band basis, according to Eq.~\eqref{eq:AbandPauli}. The electric field is $E_z=1$ mV/nm, and the magnetic field is out-of-plane $B=20$ mT. \small{(a) The $\Sigma_x$, (b) the $\Sigma_y$ and (c) the $\Sigma_z$ components of the Berry connections. The Berry connections in the band basis have large values, they cannot be neglected.}}}
    \refstepcounter{subfigure}\label{fig:AHHbanda}
    \refstepcounter{subfigure}\label{fig:AHHbandb}
    \refstepcounter{subfigure}\label{fig:AHHbandc}
\end{figure*}

First, we solve the problem of $z$-confinement numerically, assuming a finite quantum well set by the valence band offset and strain. In the calculations, we take into account the different effective masses of Ge and SiGe or SnGe, since we treat the Luttinger and other material parameters as $z$-dependent operators \cite{del2024light,del2025fully}. Then, we expand the full Hamiltonian from Eq.~\eqref{eq:H3D} on the solutions of the quantum well, to obtain a Hamiltonian that depends only on the in-plane momentum $\bm{k}$. This Hamiltonian $H_0(\bm{k})$ appears in Eq.~\eqref{eq:Hfull} and describes the 2DHG in the absence of in-plane confinement, resulting in the subbands in Fig.~\ref{fig:dispersion2DHG}. More details on the numerical solution of the $z$-confinement problem can be found in Appendix~\ref{app:conf}, while the material parameters used in the calculations are presented in Appendix~\ref{App:parameters}.

\subsection{Heavy-hole spin qubit}\label{section:HH}

First, we consider the state-of-the-art platform for hole spin qubits, a heavy hole confined in a $\mathrm{Si_{0.2} Ge_{0.8}/Ge/Si_{0.2} Ge_{0.8}}$ planar heterostructure, see Fig.~\ref{fig:devicesa}. The quantum well profile is shown in Fig.~\ref{fig:QWHH}. The heavy-hole and light-hole states experience different confinement potentials in the Ge quantum well due to the strain; the former is shown with a black line, while the latter has a dashed red line. The electric field $E_z$ pushes the states closer to the upper $\mathrm{Si_{0.2}Ge_{0.8}}$ layer, close to the oxide, which is modeled as an infinite potential barrier. We note that the $z$-confinement mixes the light hole states with the split-off states. The eigenstates obtained from the confinement problem are either pure heavy-hole states or superpositions of light-hole and split-off states. However, the lower-lying mixed subbands have a predominantly light-hole character; therefore, we simply refer to them as light-hole states.

 In the absence of magnetic and electric fields, the time-reversal and structural inversion symmetries yield two-fold degenerate subbands as functions of $\bm{k}$. A transverse electric field $E_z$ leads to Rashba splitting of the bands. This splitting is always zero in the $\Gamma$-point. A degeneracy in the $\Gamma$-point would lead to singular Berry connections. Therefore, when we calculate Berry connections, we always assume a small Zeeman-field to lift the degeneracy. The 2DHG spectrum obtained can be seen in Fig.~\ref{fig:dispersion2DHGa}.

We now turn to calculating the effective Hamiltonian in the band basis. To calculate the Berry connections in the band basis in Eq.~\eqref{eq:Aband}, we have to calculate the momentum derivatives of the eigenstates of the lowest two bands $\psi_1(\bm{k})$ and $\psi_2(\bm{k})$. As the eigenstates can only be determined numerically, we need a consistent numerical gauge-fixing technique to approximate the derivatives using finite differences. We expand the two eigenstates at finite $\bm{k}$ in terms of the eigenstates at the $\Gamma$-point
\begin{equation}\label{eq:gaugeFixing}
    \psi_\sigma(\bm{k})=\sum\limits_{i=1}^Nc_{\sigma,i}(\bm{k})\psi_i(\bm{0}), 
\end{equation}
where $\sigma\in\{1,2\}$, and $N$ is the total number of subbands included in the calculations. At $\bm{k}=0$, $c_{\sigma,i}(\bm{0})=\delta_{\sigma i}$. Because the global phase of $\psi_\sigma(\bm{k})$ is arbitrary, we can redefine the phase such that $c_{\sigma,\sigma}(\bm{k})$ becomes positive real. In this way, the eigenstates $\psi_\sigma(\bm{k})$ will have a smooth gauge, and the Berry connections can be calculated using finite differences. 

We expand the Berry connections in the band basis according to Eq.~\eqref{eq:AbandPauli}. The coefficients $A_\mathrm{band,x}^{(\mu)}(\bm{k})$, $A_\mathrm{band,y}^{(\mu)}(\bm{k})$ and $A_\mathrm{band,z}^{(\mu)}(\bm{k})$ with $\mu \in \{x,y\}$ can be represented as three different two-dimensional vector fields in $k$-space. These vector fields are plotted in Fig.~(\ref{fig:AHHband}). The Berry connection contributions in Fig.~\ref{fig:AHHband} are large. Therefore, in the band basis, the Berry connections cannot be neglected in the effective Hamiltonian written in Eq.~\eqref{eq:Heff}. In fact, without considering the Berry connection contributions to Eq.~\eqref{eq:Heff}, the time-reversal symmetry (in the absence of a magnetic field) is violated. The Berry connections show vortices around the $\Gamma$-point, because the SOI at $\bm{k}=0$ is zero, and the two bands are close. 

\begin{figure}[t]
\centering
\includegraphics[width=\columnwidth]{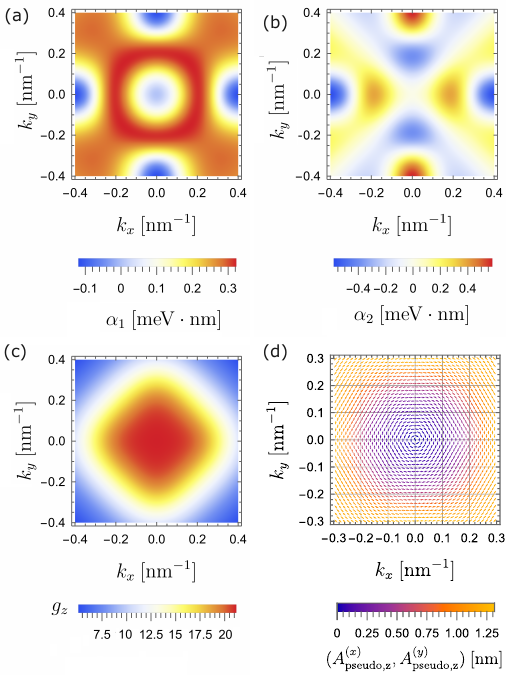}
\caption {\label{fig:alphaHHpseudo} \small{Calculated parameters of the effective Hamiltonian in the pseudospin basis for a heavy hole 2DHG, see Eq.~\eqref{eq:dispersionPseudo}. The applied electric field is $E_z=1$ mV/nm. (a) $\alpha_1(\bm{k})$ spin-orbit function, (b) $\alpha_2(\bm{k})$ spin-orbit function, (c) $g_z(\bm{k})$ $g$-function in the $z$ direction and (d) the $\Sigma_z$ components of the Berry connection in the pseudospin basis. The spin-orbit functions $\alpha_1(\bm{k})$ and $\alpha_2(\bm{k})$ are small in the vicinity of the $\Gamma$-point, due to the absence of linear SOI. In addition, $g_z(\bm{k})$ has a large value around 22, which slowly decreases with larger momentum. }}
\refstepcounter{subfigure}\label{fig:alphaHHpseudoa}
    \refstepcounter{subfigure}\label{fig:alphaHHpseudob}
    \refstepcounter{subfigure}\label{fig:alphaHHpseudoc}
    \refstepcounter{subfigure}\label{fig:alphaHHpseudod}
\end{figure}

Next, we transform the effective Hamiltonian from the band basis to the pseudospin basis by transforming the dispersion Hamiltonian, Eq.~\eqref{eq:dispersion_band}, and the Berry connections, Eq.~\eqref{eq:AbandTransform}. The transformed Berry connections are almost zero, as expected. The $\Sigma_x$ and $\Sigma_y$ components are practically zero, the $\Sigma_z$ contributions can be seen in Fig.~\ref{fig:alphaHHpseudod}. The Berry connections in the pseudospin basis are small even at large $k$, far from the $\Gamma$-point. Therefore, the Berry connections in the pseudospin basis can usually be neglected in Eq. (\ref{eq:Heff}). However, for light holes, these contributions can be non-negligible, as shown in Sec.~\ref{sec:LH}.

In the pseudospin basis, the rich spin-orbit physics is inherited by the pseudospin Hamiltonian, which becomes off-diagonal. Based on the symmetry properties of cubic semiconductors and the transformation properties of a heavy-hole ground state, the dispersion Hamiltonian in the pseudospin basis has the following form: 

\begin{figure}[t]
\centering
\includegraphics[width=\columnwidth]{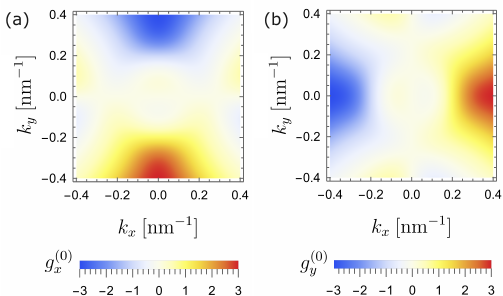}
\caption {\label{fig:g0xyHH} \small{Spin-independent $g$-functions of the dispersion Hamiltonian in the pseudospin basis for a heavy hole 2DHG, see Eq.~\eqref{eq:dispersionPseudo}. The electric field is $E_z=1$ mV/nm. (a) $g_x^{(0)}(\bm{k})$ function, (b) $g_y^{(0)}(\bm{k})$ function. For finite momentum, the magnetic fields can have significant spin-independent contributions.}}
\refstepcounter{subfigure}\label{fig:g0xyHHa}
\refstepcounter{subfigure}\label{fig:g0xyHHb}
\end{figure}

\begin{equation}\label{eq:dispersionPseudo}
\begin{aligned}
H^{(d, \mathrm{2DHG})}_\mathrm{pseudo}(\bm{k})&=E_0(\bm{k})+g_x^{(0)}(\bm{k})\mu_B B_x+g_y^{(0)}(\bm{k})\mu_B B_y+\\&+\alpha_1(\bm{k})(k_x\Sigma_y+k_y\Sigma_x)+\frac{g_z(\bm{k})}{2}\mu_B B_z\Sigma_z+\\&+\alpha_2(\bm{k})(k_x\Sigma_y-k_y\Sigma_x)+\frac{g_x^{(1)}(\bm{k})}{2}\mu_BB_x\Sigma_x+\\&+\frac{g_y^{(1)}(\bm{k})}{2}\mu_B B_y\Sigma_y+\frac{g_x^{(2)}(\bm{k})}{2}\mu_B B_x\Sigma_y+\\&+\frac{g_y^{(2)}(\bm{k})}{2}\mu_BB_y\Sigma_x, 
\end{aligned}
\end{equation}
where the functions have the reflection properties; 
\allowdisplaybreaks
\begin{subequations}\label{eq:pseudoHHsymmetries1}
\begin{align}
E_0(-k_x,k_y)&=E_0(k_x,k_y), \\E_0(k_x,-k_y)&=E_0(k_x,k_y), \\
g_x^{(0)}(-k_x,k_y)&=g_x^{(0)}(k_x,k_y),\\
g_x^{(0)}(k_x,-k_y)&=-g_x^{(0)}(k_x,k_y),\\
g_y^{(0)}(-k_x,k_y)&=-g_y^{(0)}(k_x,k_y),\\
g_y^{(0)}(k_x,-k_y)&=g_y^{(0)}(k_x,k_y),\\
\alpha_{1(2)}(-k_x,k_y)&=\alpha_{1(2)}(k_x,k_y), \\\alpha_{1(2)}(k_x,-k_y)&=\alpha_{1(2)}(k_x,k_y), \\ 
g_z(-k_x,k_y)&=g_z(k_x,k_y), \\
g_z(k_x,-k_y)&=g_z(k_x,k_y),\\
g_{x(y)}^{(1)}(-k_x,k_y)&=g_{x(y)}^{(1)}(k_x,k_y), \\
g_{x(y)}^{(1)}(k_x,-k_y)&=g_{x(y)}^{(1)}(k_x,k_y), \\
g_{x(y)}^{(2)}(-k_x,k_y)&=-g_{x(y)}^{(2)}(k_x,k_y), \\
g_{x(y)}^{(2)}(k_x,-k_y)&=-g_{x(y)}^{(2)}(k_x,k_y),
\end{align}
\end{subequations}

and the rotational properties; 
{
\allowdisplaybreaks
\begin{subequations}\label{eq:pseudoHHsymmetries2}
\begin{align}
E_0(k_y,k_x)&=E_0(k_x,k_y), \\
g_x^{(0)}(k_y,k_x)&=-g_y^{(0)}(k_x,k_y),\\
\alpha_1(k_y,k_x)&=\alpha_1(k_x,k_y), \\
\alpha_2(k_y,k_x)&=-\alpha_2(k_x,k_y), \\
g_x^{(1)}(k_y,k_x)&=-g_y^{(1)}(k_x,k_y), \\
g_x^{(2)}(k_y,k_x)&=-g_y^{(2)}(k_x,k_y), \\
g_z(k_y,k_x)&=g_z(k_x,k_y).
\end{align}
\end{subequations}
}

\begin{figure}[b]
\centering
\includegraphics[width=\columnwidth]{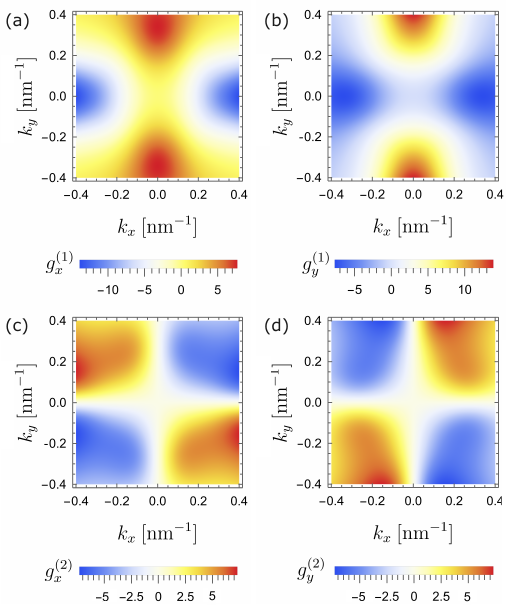}
\caption {\label{fig:gxyHH} Calculated \small{$g$-functions of the dispersion Hamiltonian in the pseudospin basis for a heavy hole 2DHG, see Eq~\eqref{eq:dispersionPseudo}. The applied electric field is $E_z=1$ mV/nm. (a) $g_x^{(1)}(\bm{k})$, (b) $g_y^{(1)}(\bm{k})$, (c) $g_x^{(2)}(\bm{k})$ and (d) $g_y^{(2)}(\bm{k})$. The in-plane $g$-functions show a high degree of anisotropy, as expected for a heavy-hole ground state.}}
\refstepcounter{subfigure}\label{fig:gxyHHa}
\refstepcounter{subfigure}\label{fig:gxyHHb}
\refstepcounter{subfigure}\label{fig:gxyHHc}
\refstepcounter{subfigure}\label{fig:gxyHHd}
\end{figure}

The non-perturbative dispersion Hamiltonian in Eq.~\eqref{eq:dispersionPseudo} contains all orders in $k$; however, we kept only the terms linear in the magnetic field. Aside from the neglected higher-order magnetic contributions, it is an exact Hamiltonian of the lowest two bands of the 2DHG. 

We construct the effective Hamiltonian in Eq.~\eqref{eq:dispersionPseudo} using Eq.~\eqref{eq:dispersion_band}. To identify the different functions in Eq.~\eqref{eq:dispersionPseudo}, we first calculate the effective Hamiltonian with a small magnetic field along $z$, which makes it possible to identify $E_0(\bm{k})$, $\alpha_{1(2)}(\bm{k})$, and $g_z(\bm{k})$. Then, we apply a magnetic field along $x$, which allows us to extract $g_x^{(0)}(\bm{k})$, $g_x^{(1)}(\bm{k})$ and $g_x^{(2)}(\bm{k})$, using the already determined $E_0(\bm{k})$ and $\alpha_{1(2)}(\bm{k})$ functions.
Finally, we apply a small magnetic field along $y$ to obtain $g_y^{(0)}(\bm{k})$, $g_y^{(1)}(\bm{k})$, and $g_y^{(2)}(\bm{k})$.

The spin-orbit coefficients and $g_z(\bm{k})$ are plotted in Fig.~\ref{fig:alphaHHpseudo}. We can see that the functions satisfy the required symmetry relations from Eqs.~\eqref{eq:pseudoHHsymmetries1} and \eqref{eq:pseudoHHsymmetries2}. The spin-orbit functions $\alpha_1(\bm{k})$ and $\alpha_2(\bm{k})$ are small in the vicinity of the $\Gamma$-point, this is a consequence of the absence of linear-in-k SOI. However, at large k, they show a non-monotonic behaviour, as a result of heavy-hole/light-hole mixing. The $g_z(\bm{k})$ function from Fig.~\ref{fig:alphaHHpseudoc} around the $\Gamma$-point has an approximate value of 22, which means that a not too strongly confined quantum dot should have a $g$-factor around 22, compatible with experiments. 

Note that in the dispersion Hamiltonian in Eq.~\eqref{eq:dispersionPseudo}, two additional terms $g_x^{(0)}(\bm{k})$ and $g_y^{(0)}(\bm{k})$ appear beside the standard Zeeman terms. These are also proportional to the magnetic field but are not coupled to the spin; they originate from the orbital magnetic contributions. The functions $g_x^{(0)}(\bm{k})$ and $g_y^{(0)}(\bm{k})$ are plotted in Fig.~\ref{fig:g0xyHH}. We can see that at larger $k$ these $g$-functions are significant, they cannot be neglected in the effective Hamiltonian, Eq.~\eqref{eq:dispersionPseudo}. 
For small $k$, the spin-independent magnetic terms in Eq.~\eqref{eq:dispersionPseudo} have the following kinematic structure
\begin{equation}\label{eq:spinIndOrigin}
\begin{aligned}
    &g_x^{(0)}(\bm{k})\mu_BB_x+g_y^{(0)}(\bm{k})\mu_BB_y \propto (\bm{B}\times \bm{k})\cdot \bm{E}=\\&=(B_xk_y-B_yk_x)E_z, 
\end{aligned}
\end{equation}
which is a signature of magnetochiral anisotropy leading to nonreciprocal currents. 

Figure~\ref{fig:gxyHH} shows the in-plane $g$-functions in Eq.~\eqref{eq:dispersionPseudo}. These functions show a high degree of anisotropy, which, together with an additional anisotropic quantum dot confinement, leads to an anisotropic in-plane $g$-factor, as observed in experiments \cite{hendrickx2024sweet,seidler2025spatial}.

To validate our effective model and to understand its limitations, we confine a single hole from the 2DHG to a quantum dot. We assume an anisotropic harmonic in-plane confinement potential
\begin{equation}\label{eq:ConfHarm}
    V(x,y)=\frac{m^*}{2}(\omega_x^2 x^2+\omega_y^2 y^2), 
\end{equation}
where $m^*$ is the effective mass at $k=0$, and is extracted numerically from the spectrum of the 2DHG, see Fig.~\ref{fig:dispersion2DHGa}.
The effective Hamiltonian of the heavy-hole qubit (lateral quantum dot) can be written as 
\begin{equation}\label{eq:Heff_lateral}
H_\mathrm{eff}^\mathrm{lat}=H_\mathrm{pseudo}^{(d,\mathrm{2DHG})}(\bm{k})+V(x+A_\mathrm{pseudo}^{(x)}(\bm{k}),y+A_\mathrm{pseudo}^{(y)}(\bm{k})),
\end{equation}
where $H_\mathrm{pseudo}^{(d,\mathrm{2DHG})}(\bm{k})$ is from Eq.~\eqref{eq:dispersionPseudo}, $V(x,y)$ from Eq.~\eqref{eq:ConfHarm}, and $A_\mathrm{pseudo}^{(x)}(\bm{k})$ and $A_\mathrm{pseudo}^{(y)}(\bm{k})$ are Berry connections in the pseudospin basis.

\begin{figure}[t]
\centering
\includegraphics[width=\columnwidth]{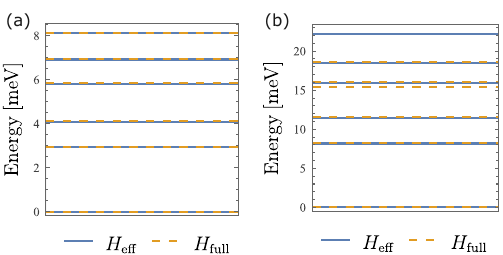}
\caption {\label{fig:orbitalsHH} \small{Calculated orbital levels of a heavy-hole spin qubit. The electric field is $E_z=1$ mV/nm, and the magnetic field is zero. (a) The confinement energies are $\hbar\omega_x=3$ meV and $\hbar\omega_y=4.2$ meV, we see excellent agreement between the effective (Eq.~\eqref{eq:Heff_lateral}) and the full Hamiltonian (Eq.~\eqref{eq:Hfull}). (b) The confinement energies are $\hbar\omega_x=9$ meV and $\hbar\omega_y=12.6$ meV. We see that even though the higher orbitals obtained using the effective Hamiltonian deviate from the full Hamiltonian, the lowest three orbitals match well.  }}
\refstepcounter{subfigure}\label{fig:orbitalsHHa}
\refstepcounter{subfigure}\label{fig:orbitalsHHb}
\end{figure}

In the case of the heavy-hole qubit, we can safely neglect the Berry connections in Eq.~\eqref{eq:Heff_lateral}, see Fig.~\ref{fig:alphaHHpseudod}. This makes the calculations with the effective Hamiltonian for different magnetic fields convenient. The functions in Eq.~\eqref{eq:dispersionPseudo} need to be calculated only once because they are independent of the magnetic field. In contrast, the Berry connections depend on the magnitude and direction of the magnetic field. Keeping the Berry connections in the effective Hamiltonian would require their recalculation for different magnetic fields, which is time-consuming.

We numerically solve the effective Hamiltonian from Eq.~\eqref{eq:Heff_lateral} in the pseudospin basis, using finite differences and momentum representation. In momentum representation, the momentum operators in the dispersion Hamiltonian in Eq.~\eqref{eq:dispersionPseudo} become scalars, while the position operators $x$ and $y$ can be written as $x=i\partial_{k_x}$ and $y=i\partial_{k_y}$. 

\begin{figure}[t]
\centering
\includegraphics[width=\columnwidth]{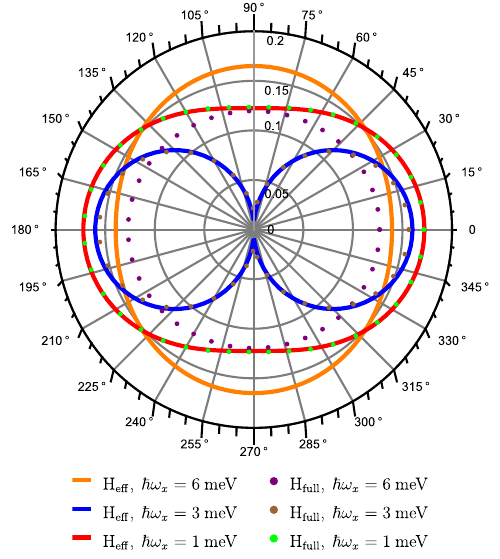}
\caption {\label{fig:gplaneHH} \small{Calculated in-plane $g$-factors as a function of $\phi$, in-plane angle of the magnetic field. The full lines show the result of the effective Hamiltonian, Eq.~\eqref{eq:Heff_lateral}, in the pseudospin basis, neglecting the Berry connection contributions, while the dots show the results from the full Hamiltonian in Eq.~\eqref{eq:Hfull}. The electric field is $E_z=1$ mV/nm and $\omega_y=1.4$~$\omega_x$ for all cases. For strong in-plane confinement, when $\hbar\omega_x=6$ meV, the difference between the effective Hamiltonian and the exact approach is notable.}}
\end{figure}

In Fig.~\ref{fig:orbitalsHH}, we benchmark our effective Hamiltonian in Eq.~\eqref{eq:Heff_lateral} against the full Hamiltonian in Eq.~\eqref{eq:Hfull} by calculating the orbital levels of the quantum dot in the absence of a magnetic field. Figure~\ref{fig:orbitalsHHa} shows a quantum dot confined with $\hbar\omega_x=3$ meV and $\hbar\omega_y=4.2$ meV, we see excellent agreement between the effective and full Hamiltonians. For stronger confinement, when $\hbar\omega_x=9$ meV and $\hbar\omega_y=12.6$ meV, we see a good agreement for the first three orbitals. High-energy orbitals deviate significantly from the exact Hamiltonian. The deviation starts around 15 meV, which is close to the energy difference between the lowest two subbands of the 2DHG and the higher subbands; see Fig.~\ref{fig:dispersion2DHGa}.

To further demonstrate that the effective Hamiltonian retains the spin-orbit physics and anisotropy of heavy holes, we calculate the dependence of the in-plane $g$-factor of the quantum dot on the in-plane angle $\phi$ of the magnetic field. Figure~\ref{fig:gplaneHH} shows the $g$-factors for different confinement strengths, with the ratio $\omega_y/\omega_x=1.4$ fixed. When $\hbar\omega_x=1$ meV, we observe excellent agreement between the effective [see Eq.~\eqref{eq:Heff_lateral}] and full Hamiltonians [see Eq.~\eqref{eq:Hfull}]. The $g$-factors obtained in this case are between 0.1 and 0.2, in good agreement with the experiments \cite{hendrickx2021four, seidler2025spatial}.

When the quantum dot has a stronger confinement, $\hbar\omega_x=3$ meV, we still see a good agreement between the results of the effective and full Hamiltonians. In this case, the $g$-factors are smaller due to the larger orbital contributions, which compete with the direct Zeeman contributions. The $g$-factor changes sign as the confinement increases. At even stronger confinement, when $\hbar\omega_x=6$ meV, the $g$-factors are larger than for $\hbar\omega_x=3$ meV. At this point, the orbital contributions are stronger than the direct Zeeman contributions. We note that in Fig.~\ref{fig:gplaneHH} we show the absolute value of the $g$-factor. 

\subsection{Light-hole spin qubit}
\label{sec:LH}

\begin{figure}[t]
\centering
\includegraphics[width=\columnwidth]{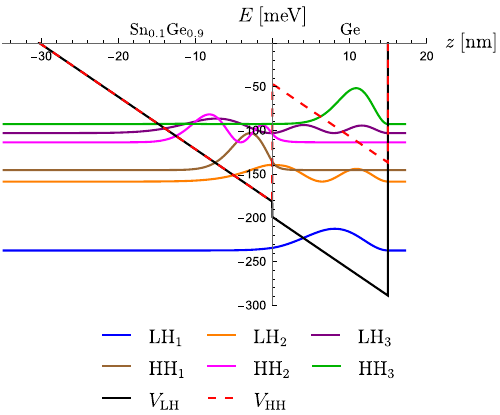}

\caption {\label{fig:QWLH} \small{Quantum well and first six wavefunctions (absolute value squares) in a $\mathrm{Sn_{0.1}Ge_{0.9}/Ge}$ planar heterostructure with an applied electric field of $E_z=6$ mV/nm. For the light-hole states, the strained Ge layer acts as a quantum well, see the black line for the potential. For the heavy-hole states, however, the strained Ge layer acts as a potential barrier, see the red dashed line. Thus, the first two heavy-hole wavefunctions (brown and magenta) reside in the $\mathrm{Sn_{0.1}Ge_{0.9}}$ layer, while the third one (green) is pushed into the Ge layer by the electric field.}}
\end{figure}

Now, we apply our method presented in Sec.~\ref{sec:Heffband} and Sec.~\ref{sec:Heffpseudo} to a light hole confined in a $\mathrm{Sn_{0.1}Ge_{0.9}/Ge}$ heterostructure, as shown in Fig.~\ref{fig:devicesb}. The description is the same as for heavy holes discussed in Sec.~\ref{section:HH}. We construct the Hamiltonian of the 2DHG $H_0$ from the 3D Hamiltonian in Eq.~\eqref{eq:H3D} by solving the problem of $z$-confinement and expanding the 3D Hamiltonian on the solutions. The quantum well and the first few solutions of the $z$-confinement problem can be seen in Fig.~\ref{fig:QWLH}. 

\begin{figure}[t]
\centering
\includegraphics[width=\columnwidth]{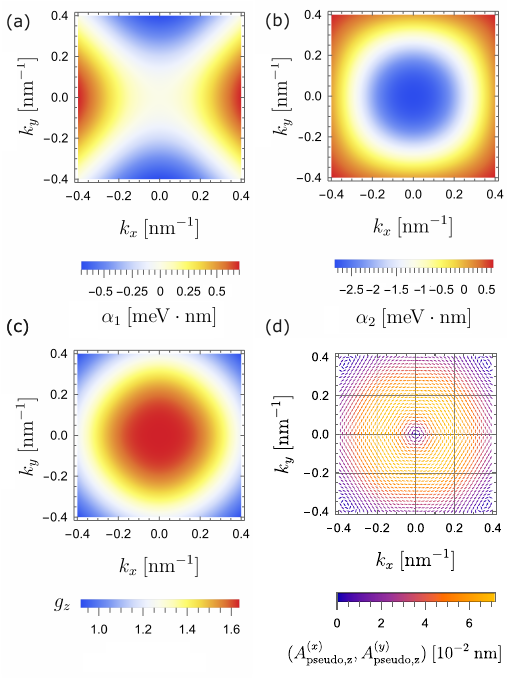}
\caption {\label{fig:alphaLHpseudo} \small{Calculated parameters of the effective Hamiltonian in the pseudospin basis for a light hole 2DHG, see Eq.~\eqref{eq:dispersionPseudo}. The electric field is $E_z=6$ mV/nm. (a) $\alpha_1(\bm{k})$ function, (b) $\alpha_2(\bm{k})$, (c) $g_z(\bm{k})$ and (d) the $\Sigma_z$ components of the Berry connection in the pseudospin basis. We can see strong linear SOI in (a) and (b), a small out-of-plane $g$-factor in (c), and small Berry connections in pseudospin basis in (d).}}
\refstepcounter{subfigure}\label{fig:alphaLHpseudoa}
\refstepcounter{subfigure}\label{fig:alphaLHpseudob}
\refstepcounter{subfigure}\label{fig:alphaLHpseudoc}
\refstepcounter{subfigure}\label{fig:alphaLHpseudod}
\end{figure}

\begin{figure*}[tb]
\centering
\includegraphics[width=2\columnwidth]{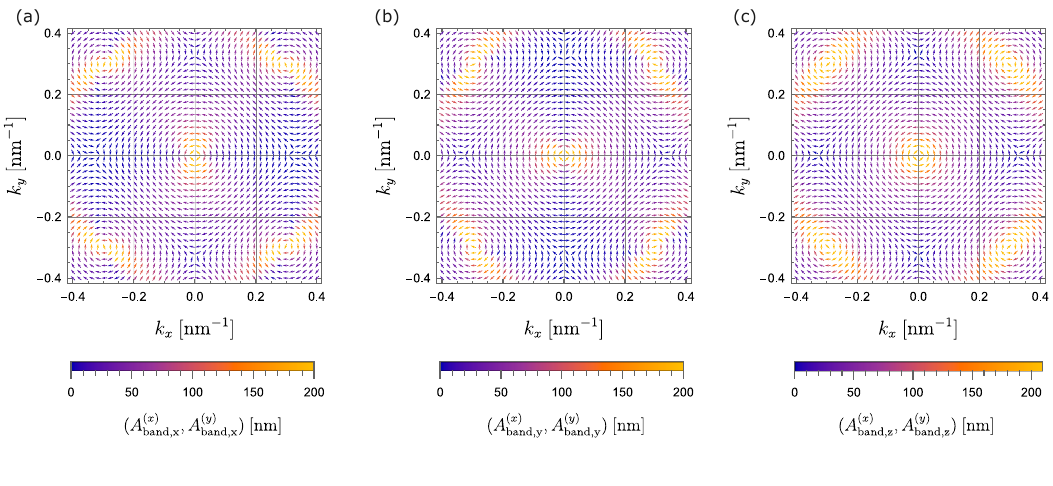}
\caption {\label{fig:ALHband}\small{Berry connections of a light hole 2DHG, see Fig.~\ref{fig:devicesb}. The Berry connections are expanded in the band basis, according to Eq.~\eqref{eq:AbandPauli}. The electric field is $E_z=6$ mV/nm, the magnetic field is $B=20$ mT out-of plane. \small{(a)  The $\Sigma_x$, (b) the $\Sigma_y$ and (c) the $\Sigma_z$ components of the Berry connections. In the band basis, the Berry connections are large, comparable to the dot size, and cannot be neglected. The Berry connections have vortices at the $\Gamma$-point and also at four points at finite $k$, due to the presence of  both linear and cubic SOI.}}}
\refstepcounter{subfigure}\label{fig:ALHbanda}
\refstepcounter{subfigure}\label{fig:ALHbandb}
\refstepcounter{subfigure}\label{fig:ALHbandc}
\end{figure*}

We solve the 2DHG Hamiltonian $H_0$ in Eq.~\eqref{eq:Hfull} to obtain the subbands, which are plotted in Fig.~\ref{fig:dispersion2DHGb}. Note that this time the lowest two subbands have a predominantly light-hole character.
Using the eigenstates of the 2DHG, we calculate the Berry connections in the band basis according to Eq.~\eqref{eq:Aband}. We use the same gauge-fixing technique as outlined in Sec.~\ref{section:HH}, the resulting Berry connections are plotted in Fig.~\ref{fig:ALHband}. There are two main differences between the Berry connections for light holes (see Figs.~\ref{fig:alphaLHpseudo} and ~\ref{fig:ALHband}) and for heavy holes (see Fig.~\ref{fig:AHHband}). First, the $\Sigma_x$ and $\Sigma_y$ components for the light holes have a winding number of one, whereas for heavy holes the winding number is three. Second, the Berry connections for light holes have a large value close to the $\Gamma$-point, whereas for heavy holes, the Berry connections can be large only far from the $\Gamma$-point. These arise from the fact that light holes possess strong linear-in-k SOI \cite{del2023light, del2025fully}, while heavy holes do not (in the absence of shear strains or interface effects.).

The vortices appearing far from the $\Gamma$-point in Fig.~\ref{fig:ALHband} are the result of the interplay between linear and cubic Rashba SOI. In the absence of magnetic field, the two bands could be degenerate at finite $\bm{k}$ values, where the linear and cubic spin-orbit interactions cancel out. A finite magnetic field creates a splitting, but at these points, the two bands are still close to each other, leading to the vortices. Similar vortices appear in Fig.~\ref{fig:AHHband}, as a result of HH-LH mixing at larger k.

Now, we transform the dispersion Hamiltonian [see Eq.~\eqref{eq:Hdband}] and the Berry connections [see Eq.~\eqref{eq:Aband}] from the band basis to the pseudospin basis, according to Eqs.~\eqref{eq:dispersion_band} and ~\eqref{eq:AbandTransform}. The new pseudospin Pauli operators $\Sigma_x$, $\Sigma_y$, and $\Sigma_z$ represent an effective spin 1/2, contrary to the heavy hole case with effective spin 3/2 and -3/2. Both spins have the same reflection properties with respect to the $x\rightarrow -x$ and $y\rightarrow -y$ operations, which means that the dispersion Hamiltonian has the same form as for heavy holes; see Eq.~\eqref{eq:dispersionPseudo}. In addition to this, the parameters of the dispersion Hamiltonian in Eq.~\eqref{eq:dispersionPseudo} have the same reflection properties, see Eq.~\eqref{eq:pseudoHHsymmetries1}. The differences between heavy and light holes lie in their transformation properties under rotations around the $z$-axis. Under a rotation with angle $\varphi$ around $z$, the light hole pseudospin operators transform as 
\begin{equation}\label{eq:rotationLH}
    \Sigma_\pm \rightarrow e^{\pm i\varphi}\Sigma_\pm, 
\end{equation}
while for heavy holes
\begin{equation}
    \Sigma_\pm \rightarrow e^{\pm3 i\varphi}\Sigma_\pm, 
\end{equation}
where $\Sigma_\pm=\Sigma_x\pm i\Sigma_y$.

The transformation rule in Eq.~\eqref{eq:rotationLH} results in the following rotational properties of the parameters of the effective Hamiltonian [see Eq.~\eqref{eq:dispersionPseudo}] for light holes:

\begin{subequations}\label{eq:pseudoLHsymmetries2}
\begin{align}
E_0(k_y,k_x)&=E_0(k_x,k_y), \\
g_x^{(0)}(k_y,k_x)&=-g_y^{(0)}(k_x,k_y),\\
\alpha_1(k_y,k_x)&=-\alpha_1(k_x,k_y), \\
\alpha_2(k_y,k_x)&=\alpha_2(k_x,k_y), \\
g_x^{(1)}(k_y,k_x)&=g_y^{(1)}(k_x,k_y), \\
g_x^{(2)}(k_y,k_x)&=g_y^{(2)}(k_x,k_y), \\
g_z(k_y,k_x)&=g_z(k_x,k_y).
\end{align}
\end{subequations}

Figure~\ref{fig:alphaLHpseudo} shows the calculated spin-orbit functions $\alpha_1(\bm{k})$ and $\alpha_2(\bm{k})$, $g_z(\bm{k})$ and the $\Sigma_z$ component of the Berry connection in the pseudospin basis. Here we see a linear Rashba SOI, which is large already in the vicinity of the $\Gamma$-point, contrary to heavy holes. The $g_z(\bm{k})$ function in Fig.~\ref{fig:alphaLHpseudoc} around the $\Gamma$-point is approximately 1.6, which is an order of magnitude smaller than for heavy holes, see Fig.~\ref{fig:alphaHHpseudoc}. The $\Sigma_x$ and $\Sigma_y$ components of the Berry connections in the pseudospin basis are negligible compared to the $\Sigma_z$ components and therefore not plotted.

\begin{figure}[b]
\centering
\includegraphics[width=\columnwidth]{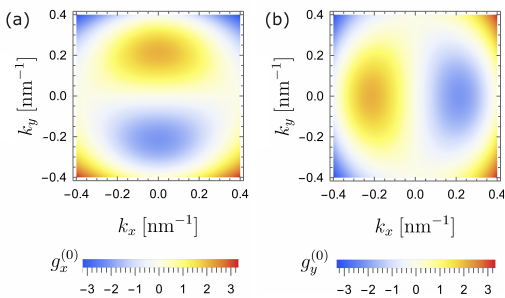}
\caption {\label{fig:g0xyLH} \small{Numerically found $g$-functions of the dispersion Hamiltonian in the pseudospin basis for a light hole 2DHG, see Eq.~\eqref{eq:dispersionPseudo}. The electric field is $E_z=6$ mV/nm. (a) $g_x^{(0)}(\bm{k})$ function, (b) $g_y^{(0)}(\bm{k})$ function.}}
\refstepcounter{subfigure}\label{fig:g0xyLHa}
\refstepcounter{subfigure}\label{fig:g0xyLHb}
\end{figure}

Figure~\ref{fig:g0xyLH} shows the $g_x^{(0)}(\bm{k})$ and $g_y^{(0)}(\bm{k})$ functions of the dispersion Hamiltonian, Eq.~\eqref{eq:dispersionPseudo}. These functions, similar to those for heavy holes, are significant.  

\begin{figure}[t]
\centering
\includegraphics[width=\columnwidth]{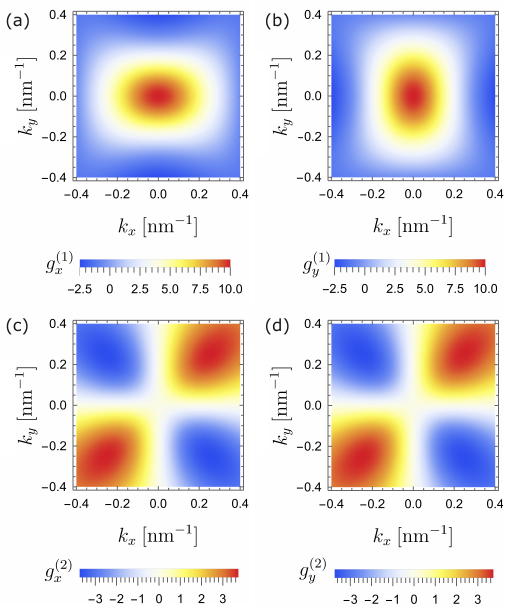}
\caption {\label{fig:gxyLH} Numerically found  \small{$g$-functions of the dispersion Hamiltonian in the pseudospin basis for a light hole 2DHG, see Eq.~\eqref{eq:dispersionPseudo}. The electric field is $E_z=6$ mV/nm. (a) $g_x^{(1)}(\bm{k})$, (b) $g_y^{(1)}(\bm{k})$ function, (c) $g_x^{(2)}(\bm{k})$ function, and d) $g_y^{(2)}(\bm{k})$ function. The in-plane $g$-functions in (a) and (b) are large, around -12 at the $\Gamma$-point and also much more isotropic than for heavy holes, see Fig.~\ref{fig:gxyHH}.}}
\refstepcounter{subfigure}\label{fig:gxyLHa}
\refstepcounter{subfigure}\label{fig:gxyLHb}
\refstepcounter{subfigure}\label{fig:gxyLHc}
\refstepcounter{subfigure}\label{fig:gxyLHd}
\end{figure}

The in-plane $g$-functions are plotted in Fig.~\ref{fig:gxyLH}. We can see that the functions $g_x^{(1)}(\bm{k})$ and $g_y^{(1)}(\bm{k})$ have an approximate value 10 around the $\Gamma$-point, much larger than for heavy holes. We can also observe that these $g$-functions are much less anisotropic compared to the $g$-functions of heavy holes; see Fig.~\ref{fig:gxyHH}. 

\begin{figure}[t]
\centering
\includegraphics[width=\columnwidth]{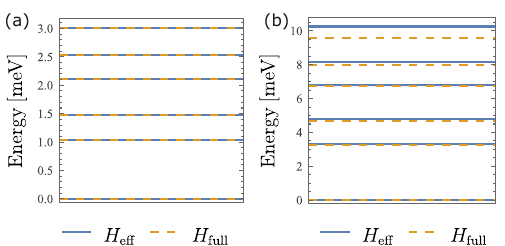}
\caption {\label{fig:orbitalsLH} \small{Calculated orbital levels of a light-hole spin qubit, with an electric field $E_z=6$ mV/nm and zero magnetic field. The blue lines show the result of the effective Hamiltonian in Eq.~\eqref{eq:Heff_lateral} in the pseudospin basis, neglecting the Berry connection contributions, while the dashed lines show the results from the full Hamiltonian in Eq.~\eqref{eq:Hfull}. (a) The confinement energies are $\hbar\omega_x=1$ meV and $\hbar\omega_y=1.4$ meV. (b) The confinement energies are $\hbar\omega_x=3$ meV and $\hbar\omega_y=4.2$ meV.}}
\refstepcounter{subfigure}\label{fig:orbitalsLHa}
\refstepcounter{subfigure}\label{fig:orbitalsLHb}
\end{figure}

To test the effective Hamiltonian in Eq.~\eqref{eq:Heff_lateral} in the case of light holes, we confine a single light hole in a quantum dot using the harmonic confinement in Eq.~\eqref{eq:ConfHarm}, with the corresponding effective mass of the light hole dispersion, see Fig.~\ref{fig:dispersion2DHGb}. First, we calculate the quantum dot orbital levels and compare it to the solution of the full Hamiltonian, see Fig. \ref{fig:orbitalsLH}. In Fig.~\ref{fig:orbitalsLHa}, the confinement is moderate, $\hbar\omega_x=1$ meV and $\hbar\omega_y=1.4$ meV, and we see excellent agreement between the effective and full Hamiltonians (see Eqs.~\eqref{eq:Heff_lateral} and~\eqref{eq:Hfull}). In Fig.~\ref{fig:orbitalsLHb}, the confinement is strong, $\hbar\omega_x=3$ meV and $\hbar\omega_y=4.2$ meV, but the agreement between the effective and full Hamiltonians for the lowest orbitals is still great. For the higher orbitals, the discrepancy is getting larger and larger, as expected.

In Fig.~\ref{fig:orbitalsLH}, we can see that the quantum dot spectrum predicted by the effective Hamiltonian in Eq.~\eqref{eq:Heff_lateral} is actually worse than the predicted spectrum for the heavy hole qubit, see Fig.~\ref{fig:orbitalsHH}. This is a consequence of the split-off hole contributions to the light-hole bands at finite $k$. In the case of a heavy-hole ground state, the LKBP Hamiltonian couples it only to the light-hole bands. However, a light-hole ground state in addition to the heavy-hole bands is also coupled to the light-hole bands via the split-off holes. If the split-off holes were not considered, the light-hole quantum dot spectrum predicted by the effective Hamiltonian in Eq.~\eqref{eq:Heff_lateral} would show excellent agreement with the full Hamiltonian in Eq.~\eqref{eq:Hfull} at $\hbar\omega_x=3$ meV and $\hbar\omega_y=4.2$ meV.

\begin{figure}[b]
\centering
\includegraphics[width=\columnwidth]{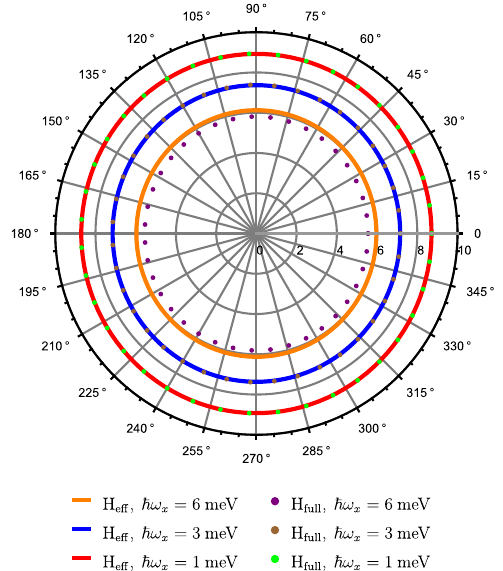}
\caption {\label{fig:gplaneLH} \small{Calculated in-plane $g$-factors as a function of $\phi$ in-plane angle of the magnetic field for a light-hole spin qubit. The full lines show the result of the effective Hamiltonian in Eq.~\eqref{eq:Heff_lateral} in the pseudospin basis, neglecting the Berry connection contributions, while the dots show the results from the full Hamiltonian in Eq.~\eqref{eq:Hfull}. The parameters are $E_z=6$ mV/nm and $\omega_y=1.4$~$\omega_x$.}}
\end{figure}

To demonstrate that the effective Hamiltonian contains not only the orbital physics but also the important spin properties, we calculate the in-plane $g$-factor of a light-hole spin qubit as a function of in-plane magnetic field angle; see Fig.~\ref{fig:gplaneLH}. For moderate confinement strengths, when $\hbar\omega_x=1$ meV or $\hbar\omega_x=3$ meV, we see excellent agreement between the results of the effective and full Hamiltonians. The discrepancy becomes noticeable for strong confinement, when $\hbar\omega_x=6$ meV and $\hbar\omega_y=8.4$ meV. The $g$-factors are much more isotropic for the light-hole qubit compared to the heavy-hole qubit, see Fig.~\ref{fig:gplaneHH}, and decrease with the in-plane confinement strength.

We note that in Figs. \ref{fig:orbitalsLH} and \ref{fig:gplaneLH} we neglected the contributions of the Berry connection $A_\mathrm{pseudo}^{(\mu)}$ to the confinement potential in Eq.~\eqref{eq:ConfHarm}. This makes the calculations with the effective Hamiltonian much faster, and in these cases, the Berry connection contributions are small. By neglecting them, we make less than 1\% error, even when the confinement is strong. 

\begin{figure}[t]
\centering
\includegraphics[width=\columnwidth]{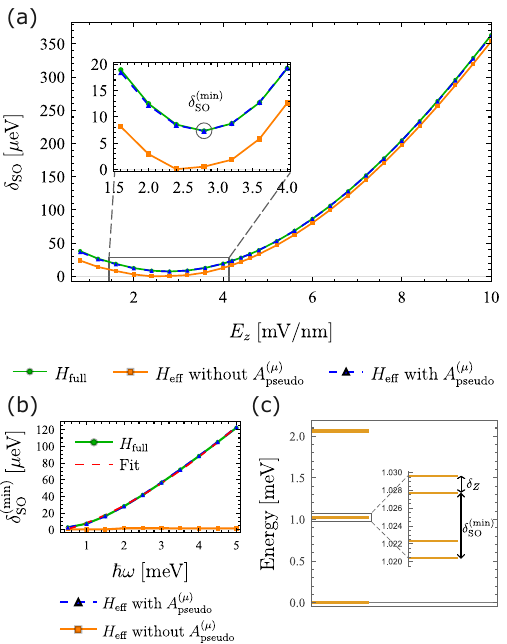}
\caption {\label{fig:LHSO} \small{Spin-orbit splitting of quantum dot orbitals of a light-hole spin qubit, with an applied out-of-plane magnetic field of $B_z=20$ mT and $\omega_x=\omega_y=\omega$ confinement frequency. (a) The spin-orbit splitting as a function of the electric field $E_z$, with a confinement frequency of $\hbar\omega=1$ meV. $\delta_\mathrm{SO}$ has a minimum value at $E_z=2.8$ mV/nm. The green line shows the solution of the full Hamiltonian in Eq.~\eqref{eq:Hfull}, the blue one is the solution of Eq.~\eqref{eq:Heff_lateral} including the Berry connections, while the orange one neglects the Berry connections. 
In this case, the 2DHG contributes to a practically zero (0.5 $\mu$eV) splitting (orange line). If we include the Berry connections, we obtain a splitting of approximately 7.3 $\mu$eV (blue and green lines). (b) The minimal spin-orbit splitting with electric field $E_z=2.8$ mV/nm, as a function of in-plane confinement energy. The solution of the full Hamiltonian is fitted with Eq.~\eqref{eq:LHSOfit}. (c) Quantum dot orbitals with $\hbar\omega=1$ meV and $E_z=2.8$ mV/nm. The inset shows the spin-orbit and Zeeman split $p_x$ and $p_y$ orbitals, where $\delta_\mathrm{SO}^\mathrm{(min)}$ is the minimum spin-orbit splitting, while $\delta_z$ is the Zeeman-splitting.}}
\refstepcounter{subfigure}\label{fig:LHSOa}
\refstepcounter{subfigure}\label{fig:LHSOb}
\refstepcounter{subfigure}\label{fig:LHSOc}
\end{figure}
To demonstrate a case where the Berry connection contributions in the in-plane confinement potential  (Eq.~\eqref{eq:Heff_lateral}) are not negligible in the pseudospin basis, we consider a light hole confined in an isotropic quantum dot, $\omega_x=\omega_y=\omega$. In the absence of SOI, the quantum dot orbitals $p_x$ and $p_y$ orbitals are degenerate. This degeneracy is lifted by SOI, the fourfold-degenerate levels are split into two Kramers doublets. With an applied magnetic field, the Kramers doublets further split, and we obtain four distinct levels; see Fig.~\ref{fig:LHSOc}. We denote the spin-orbit splitting of $p_x$ and $p_y$ orbitals with $\delta_\mathrm{SO}$. This splitting $\delta_\mathrm{SO}$ for light holes is tunable with the perpendicular electric field $E_z$, as shown in Fig.~\ref{fig:LHSOa}. We denote the minimum value $\delta_\mathrm{SO}^\mathrm{(min)}$. 

We calculate $\delta_\mathrm{SO}$ with or without the Berry connection contributions in Eq.~\eqref{eq:Heff_lateral}. These two cases are compared in Fig.~\ref{fig:LHSOa}. In general, the Berry connections have negligible contributions compared to the SOI intrinsic to the 2DHG. However, the Berry connections are dominant in the region where this splitting is close to its minimum, around $E_z=2.8$ mV/nm. At this point, the spin-orbit splitting originating from the 2DHG can be turned off almost exactly; there is a negligible remaining separation of $0.5$ $\mu$eV. However, if we treat the effect of the confinement potential correctly, either by solving the full Hamiltonian (blue line) or by including the Berry connections in the effective Hamiltonian (green line), we see that the SOI is significantly larger. The tunability of the SOI and the possibility to turn it off are great advantages of light holes \cite{del2025fully} and of Si FinFETs \cite{bosco2021hole}. Figure~\ref{fig:LHSOa} shows that it is not enough to study only the tunability of the SOI of the 2DHG or the nanowire. The interplay between the Berry connections and the confinement potential must also be considered. We expect that if the Berry connections are also considered, completely switching off the SOI for a hole spin qubit becomes much more challenging than previously predicted. 

In Fig.~\ref{fig:LHSOa}, we can see that for the electric field value $E_z=2.8$ mV/nm, the spin-orbit splitting $\delta_\mathrm{SO}$ is dominated by the in-plane confinement. Figure~\ref{fig:LHSOb} shows $\delta_\mathrm{SO}^\mathrm{(min)}$ as a function of the in-plane confinement frequency $\omega_x=\omega_y=\omega$. The effective Hamiltonian without considering the Berry connections in the pseudospin basis (orange line) fails to capture the splitting created by the in-plane confinement. The full Hamiltonian (green line) and the effective Hamiltonian with the Berry connections (blue line) show a strong frequency dependence of the splitting. We fit the  $\delta_\mathrm{SO}^\mathrm{(min)}(\omega)$ function obtained using the full Hamiltonian of Eq.~\eqref{eq:Hfull} with the following: 
\begin{equation}\label{eq:LHSOfit}
    \delta_\mathrm{SO}^\mathrm{(min)}(\omega)=\delta_\mathrm{SO}^{(0)}+a \left(\frac{\omega}{\omega_0}\right)^b, 
\end{equation}
where $\delta_\mathrm{SO}^{(0)}$, $a$ and $b$ are fitting parameters and $\hbar\omega_0=1$ meV. We note that we could fix $\delta_\mathrm{SO}^{(0)}$ as the residual 0.5 $\mu$eV splitting; however, we obtain a better fit if we treat it as a fitting parameter. The parameters obtained: 
\begin{subequations}
\begin{align}
    \delta_\mathrm{SO}^{(0)}&=-3.039 \hspace{1mm} \mu\mathrm{eV}, \\ 
    a&=11.375 \hspace{1mm} \mu\mathrm{eV}, \\
    b&=1.497.
\end{align}
\end{subequations}
The offset value $\delta_\mathrm{SO}^{(0)}$, strictly speaking, cannot be negative. However, because it is close to zero, we can practically consider it zero. The important result is that the splitting approximately scales as $\delta_\mathrm{SO}^\mathrm{(min)}\propto \omega^{3/2}$ and can become large for a stronger confinement. This is expected, as the isotropic harmonic confinement potential in Eq.~\eqref{eq:Heff_lateral} yields the following direct-Rashba term leading to the splitting: 
\begin{equation}\label{eq:HDRgeneral}
\begin{aligned}
    H_\mathrm{DR}=\frac{m^*\omega^2}{2}&\left[\left\{x,A_\mathrm{pseudo}^{(x)}(\bm{k})\right\}+\left\{y,A_\mathrm{pseudo}^{(y)}(\bm{k})\right\}\right],
\end{aligned}
\end{equation}
We can arrive at the $\delta_\mathrm{SO}^\mathrm{(min)}\propto \omega^{3/2}$ relation using dimensional analysis. $A^{(x)}_\mathrm{pseudo}(\bm{k})$ and $A^{(y)}_\mathrm{pseudo}(\bm{k})$ are independent of $\omega$, as they are properties of the 2DHG. The relevant scales $x$ and $y$ are proportional to $\omega^{-1/2}$, therefore $\delta_\mathrm{SO}^\mathrm{(min)}\propto \omega^{3/2}$. 

\begin{figure}[t]
\centering
\includegraphics[width=\columnwidth]{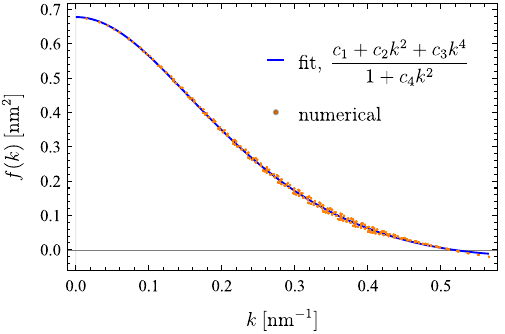}
\caption {\label{fig:fk} \small{$f(k)$ function of the Berry connection in Eq.~\eqref{eq:BerryAnsatz}. The orange markers show the values calculated according to Eq.~\eqref{eq:fk} for points uniformly sampled in $k$-space. The blue line is the fit of the analytic formula in the legend, the parameters are $c_1=0.678$ $\mathrm{nm^2}$, $c_2=-3.689$ $\mathrm{nm^4}$, $c_3=4.310$ $\mathrm{nm^6}$, and $c_4=13.563$ $\mathrm{nm^2}$.}}

\end{figure}

The Berry connections in the pseudospin basis only have $\Sigma_z$ components; the $\Sigma_x$ and $\Sigma_y$ components are practically zero. Taking this into account, Eq.~\eqref{eq:HDRgeneral} becomes 
\begin{equation}\label{eq:HDRZ}
\begin{aligned}
    H_\mathrm{DR}=\frac{m^*\omega^2}{2}&\left[\left\{x,A_\mathrm{pseudo,z}^{(x)}(\bm{k})\right\}+\right.\\+&\left.\left\{y,A_\mathrm{pseudo,z}^{(y)}(\bm{k})\right\}\right]\Sigma_z.
\end{aligned}
\end{equation}
To shed light on the origin of this spin-orbit term, we approximate the Berry connections in the following way 
\begin{equation}\label{eq:BerryAnsatz}
    \left(A_\mathrm{pseudo,z}^{(x)}(\bm{k}),A_\mathrm{pseudo,z}^{(y)}(\bm{k})\right)=f(k)(-k_y,k_x),
\end{equation}
where $f(k)$ is a function of the magnitude of $\bm{k}$. This approximation is suggested by the clear vortex structure of the Berry connections, see Fig.~\ref{fig:alphaLHpseudod}. To test the validity of this approximation, we multiply Eq.~\eqref{eq:BerryAnsatz} by $(-k_y,k_x)$ and express $f(k)$ as

\begin{equation}\label{eq:fk}
    f(k)=\frac{A_y(\bm{k})k_x-A_x(\bm{k})k_y}{k_x^2+k_y^2}. 
\end{equation}
We uniformly sample the $k$-space in $40\times40$ points, in the $k_x \in [-0.4,0.4]$ $\mathrm{nm^{-1}}$ and $k_y \in [-0.4,0.4]$ $\mathrm{nm^{-1}}$ regions. Then we calculate the corresponding pairs $k$ and $f(k)$, according to Eq.~\eqref{eq:fk}. The results are plotted in Fig.~\ref{fig:fk}, where the orange markers show the numerically calculated values, and the blue line is the fit of an analytical expression; see Fig.~\ref{fig:fk}. The numerically obtained values have a small spread, which means that the approximation from Eq.~\eqref{eq:BerryAnsatz} is valid. The function $f(k)$ is positive for most $k$ values, but only becomes slightly negative above $0.5$ $\mathrm{nm^{-1}}$, where the simple approximation from Eq.~\eqref{eq:BerryAnsatz} becomes invalid, see Fig.~\ref{fig:alphaLHpseudod}.

Using Eq.~\eqref{eq:BerryAnsatz}, the direct-Rashba term in Eq.~\eqref{eq:HDRZ} can be rewritten as 
\begin{equation}\label{eq:HDRLS}
    H_\mathrm{DR}=-m^*\omega^2 f(k) L_z \Sigma_z,
\end{equation}
where $L_z=xk_y-yk_x$ is the angular momentum operator. The term in Eq.~\eqref{eq:HDRLS} is analogous to the well-known $\bm{L}\cdot \bm{S}$ interaction from atomic physics.  

The SOI induced by the in-plane confinement (Eq.~\eqref{eq:HDRLS}) can hardly be turned off, as this would require a zero expectation value for $f(k)$, but $f(k)$ is positive for most values $k$. The effect of the in-plane confinement-induced SOI cannot be canceled by the SOI of the 2DHG, because the former couples to $\Sigma_z$, and the latter couples to $\Sigma_x$ and $\Sigma_y$, see Eq.~\eqref{eq:dispersionPseudo}. Therefore, completely turning off the total SOI of planar dots appears unfeasible. 

\section{Application to nanowires}\label{sec:nanowires}
\begin{figure}[h!]
\includegraphics[width=\columnwidth]{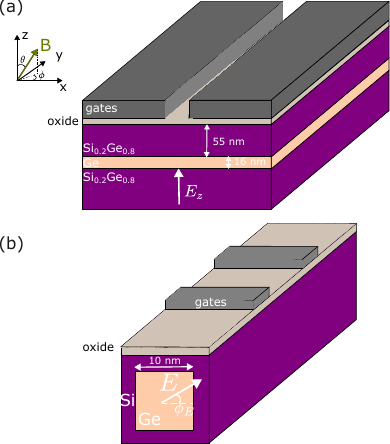}
\caption{\label{fig:nanowires} \small{Schematic representation of the nanowires considered in this work. (a) A gate-defined nanowire in a Ge 2DHG, hosting heavy holes. The gates on top of the heterostructure create a narrow channel along the $y$ direction. (b) A Ge/Si square core/shell nanowire. The gates on top confine a single hole in the nanowire.}}
\refstepcounter{subfigure}\label{fig:nanowiresa}
\refstepcounter{subfigure}\label{fig:nanowiresb}
\end{figure}

So far, we have applied our method to derive effective Hamiltonians of hole spin qubits confined in planar heterostructures. To demonstrate the versatility of our method, we also apply it to two different nanowires: a gate-defined nanowire in a SiGe/Ge/SiGe planar heterostructure and a Ge/Si core/shell nanowire. The schematic representation of these devices can be seen in Fig.~\ref{fig:nanowires}. 

\subsection{Gate-defined nanowire}
Here we consider a gate-defined nanowire created in a SiGe/Ge/SiGe planar heterostructure grown along [001], as shown in Fig.~\ref{fig:nanowiresa}. A 2DHG is created in the planar heterostructure, and two top-gates confine the holes to a small channel, effectively creating a nanowire. For simplicity, we assume that the nanowire is oriented along the [010] crystal direction. We further assume that the gates create a harmonic confinement potential
\begin{equation}\label{eq:Vx}
    V_\mathrm{nw}(x)=\frac{m^*}{2}\omega_x^2x^2. 
\end{equation}
Our goal is to derive an effective Hamiltonian for the nanowire using the method presented in Sections \ref{sec:Heffband} and \ref{sec:Heffpseudo}. We derive the nanowire effective Hamiltonian in two different ways. 

\begin{figure}[t]
\centering
\includegraphics[width=\columnwidth]{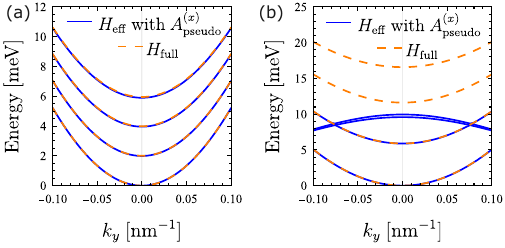}
\caption {\label{fig:orbitalsGate} \small{Calculated subbands of a gate-defined nanowire in SiGe/Ge/SiGe 2DHG. The electric field is $E_z=1$ mV/nm, and the magnetic field is zero. The orange curves are obtained using the Hamiltonian in Eq.~\eqref{eq:Hnano3D}, while the blue curves are obtained using Eq.~\eqref{eq:Hnano2D}. (a) The confinement energy is $\hbar\omega_x=2$ meV. (b) The confinement energy is $\hbar\omega_x=6$ meV.}}
\refstepcounter{subfigure}\label{fig:orbitalsGatea}
\refstepcounter{subfigure}\label{fig:orbitalsGateb}
\end{figure}

First, we can start with the full LKBP Hamiltonian in Eq.~\eqref{eq:H3D}, but instead of taking a confinement potential $V(x,y)$ that would create a quantum dot, we consider $V_\mathrm{nw}(x)$ from Eq.~\eqref{eq:Vx}
\begin{equation}\label{eq:Hnano3D}
    H^{\mathrm{(nw)}}_\mathrm{full}(k_y)=H_\mathrm{LKBP}(k_x,k_y,k_z)+V_\mathrm{qw}(z)+V_\mathrm{nw}(x). 
\end{equation}
We can solve this Hamiltonian numerically to obtain the nanowire subbands, which can be seen in Fig.~\ref{fig:orbitalsGate} with orange. Then, using our method, we can construct an effective Hamiltonian of the nanowire based on the lowest two spin-split subbands of the nanowire.  

The second way to derive the nanowire Hamiltonian is by starting with the effective Hamiltonian of the 2DHG in Eq.~\eqref{eq:dispersionPseudo}. The gate-defined nanowire can be considered as a 2DHG with additional confinement in the $x$ direction. In this case, the nanowire Hamiltonian can be written as 
\begin{equation}\label{eq:Hnano2D}
    H^{\mathrm{(nw)}}_\mathrm{eff}(k_y)=H_\mathrm{pseudo}^{(d,\mathrm{2DHG)}}(k_x,k_y)+V_\mathrm{nw}\left(x+A_\mathrm{pseudo}^{(x)}\right), 
\end{equation}
where $H_\mathrm{pseudo}^{(d,\mathrm{2DHG)}}$ is from Eq.~\eqref{eq:dispersionPseudo}, and $A_\mathrm{pseudo}^{(x)}$ is the Berry connection of the 2DHG in the pseudospin basis. The nanowire Hamiltonian in Eq.~\eqref{eq:Hnano2D} is defined using the pseudospin operators of the 2DHG and yields the whole nanowire spectrum. 

The nanowire subbands can be calculated in both ways, using the Hamiltonian in Eq.~\eqref{eq:Hnano3D} or the Hamiltonian in Eq.~\eqref{eq:Hnano2D}, the results are compared in Fig.~\ref{fig:orbitalsGate}. If the nanowire confinement is not too strong, we expect that the higher-energy subbands of the 2DHG do not contribute, and the two results agree well. In Fig.~\ref{fig:orbitalsGatea} for $\hbar\omega_x$=2 meV, we see excellent agreement, while in Fig.~\ref{fig:orbitalsGateb}, when $\hbar\omega_x=6$ meV, only the lowest four spin-split bands are obtained correctly using the Hamiltonian in Eq.~\eqref{eq:Hnano2D}.

\begin{figure}[t]
\centering
\includegraphics[width=\columnwidth]{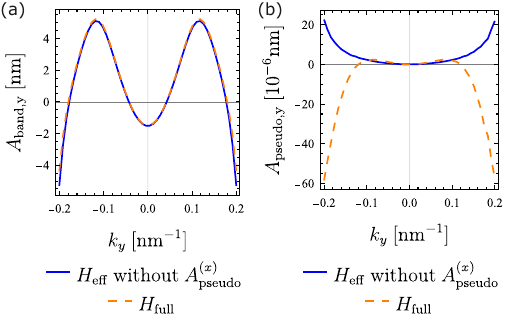}
\caption {\label{fig:BerryGate} \small{Calculated Berry connection of a gate-defined nanowire in SiGe/Ge/SiGe 2DHG. The electric field is $E_z=1$~mV/nm, $\hbar\omega_x=2$ meV, and the magnetic field is out-of-plane, $B_z=20$~mT. The blue curves are obtained using the Hamiltonian in Eq.~\eqref{eq:Hnano3D}, while the orange curves are calculated using Eq.~\eqref{eq:Hnano2D}. (a) The $y$-component of the Berry connection in the band basis. (b) The $y$-component of the Berry connection in the pseudospin basis. The $x$ and $z$ components of the Berry connection are found to be negligible compared to the $y$ component shown here. }}
\refstepcounter{subfigure}\label{fig:BerryGatea}
\refstepcounter{subfigure}\label{fig:BerryGateb}
\end{figure}

The Hamiltonians in Eqs.~\eqref{eq:Hnano3D} and \eqref{eq:Hnano2D} both describe the full nanowire spectrum with multiple subbands. However, we aim to construct an effective Hamiltonian that captures the physics of the lowest two spin-split subbands. We apply the same procedure as before. We solve the Hamiltonians from Eqs.~\eqref{eq:Hnano3D} and \eqref{eq:Hnano2D} to obtain the eigenstates and eigenfunctions $\psi_i(k_y)$, with $i\in \{1,\hdots ,N\}$. The dispersion Hamiltonian that describes the lowest two subbands in the band basis becomes: 
\begin{equation}\label{eq:HbandGate}
    H_\mathrm{band}^{(d,\mathrm{nw})}(k_y)=\begin{pmatrix}
        E_1(k_y) & 0 \\ 
        0 & E_2(k_y)
    \end{pmatrix}, 
\end{equation}
where $E_1(k_y)$ and $E_2(k_y)$ are the energies of the two lowest bands. If we considered nanowire quantum dots, we would have to add a new confinement potential along the $y$ axis, with the $y$ coordinate replaced by $y+A^{(y)}_\mathrm{band}(k_y)$, where $A_{\mathrm{band},\sigma\sigma'}=i\bra{\psi_\sigma(k_y)}\partial_{k_y}\ket{\psi_{\sigma'}(k_y)}$ is the Berry connection in the band basis. The $\Sigma_y$ component of the calculated Berry connection in the band basis is plotted in Fig.~\ref{fig:BerryGatea}. The blue curve shows the result using the Hamiltonian in Eq.~\eqref{eq:Hnano3D}, while the orange curve shows the result of the Hamiltonian in Eq.~\eqref{eq:Hnano2D}, we see excellent agreement. In the calculation of the orange curve, we neglected the Berry connection contribution of the 2DHG, $A_\mathrm{pseudo}^{(x)}$. 

In the calculation of the Berry connection plotted in Fig~\ref{fig:BerryGate}, we used a gauge fixing similar to that used for planar heterostructures. Here, the complication is that we solve the Hamiltonians from Eqs.~\eqref{eq:Hnano3D} and \eqref{eq:Hnano2D} numerically using a finite difference method in momentum representation. The resulting eigenstates that produce the bands in Fig.~\ref{fig:orbitalsGate} are wave functions in momentum representation. Our gauge fixing technique requires us to express the eigenstates at finite $k_y$ in terms of the eigenstates at $k_y=0$. Because in our case all eigenstates are in momentum representation, we apply a unitary basis transformation to the eigenstates at $k_y=0$, to transform them to their eigenbasis. Similarly, we transform the eigenstates at finite $k_y$ onto this basis. In this way, the eigenstates at finite $k_y$ are expressed in terms of the eigenstates at $k_y=0$, and we can apply the gauge fixing technique used previously, see Eq.~\eqref{eq:gaugeFixing}.

Now, we transform the effective Hamiltonian from the band basis, see Eq.~\eqref{eq:HbandGate}, to the pseudospin basis by calculating the overlap matrix $M(k_y)$ from Eq.~\eqref{eq:M} and then the matrix $X(k_y)$. The Hamiltonian of Eq.~\eqref{eq:HbandGate} is transformed according to Eq.~\eqref{eq:dispersion_band}. Here we note that simply calculating the overlap matrix according to Eq.~\eqref{eq:M} might yield pseudospin operators misaligned with the physical coordinate system. This problem can be overcome by redefining the states at $\bm{k}=0$ using a unitary transformation $U_\mathrm{rot}$: 
\begin{equation}
    \begin{pmatrix}
        \Tilde{\psi}_1(\bm{0}) \\ 
        \Tilde{\psi}_2(\bm{0})
    \end{pmatrix}=\begin{pmatrix}
        U_\mathrm{rot,11} & U_\mathrm{rot,12} \\ 
        U_\mathrm{rot,21} & U_{\mathrm{rot},22}
    \end{pmatrix} \begin{pmatrix}
        \psi_1(\bm{0}) \\ 
        \psi_2(\bm{0})
    \end{pmatrix},  
\end{equation}
which leads to a transformed matrix $X$, $\Tilde{X}=U_\mathrm{rot}^*X$, where $U^*_\mathrm{rot}$ is the complex conjugate of $U_\mathrm{rot}$. Then the pseudospin operators transform as follows: 
\begin{equation}
    \Tilde{\Sigma}_i=U_\mathrm{rot}^*\Sigma_iU_\mathrm{rot}^\intercal, 
\end{equation}
where $U_\mathrm{rot}^\intercal$ is the transpose of $U_\mathrm{rot}$. In this way, the pseudospin axes can be aligned to the physical axes. 

For a nanowire along the $y$ direction in a cubic semiconductor, the effective Hamiltonian has the following form allowed by symmetries
\begin{equation}\label{eq:gatedispersion}
\begin{aligned}
    H^{(d,\mathrm{nw})}_\mathrm{pseudo}(k_y)&=E_0(k_y)+\alpha_x(k_y)k_y\Sigma_x-\alpha_z(k_y)k_y\Sigma_z+\\&+g_x^{(0)}(k_y)\mu_B B_x+g_z^{(0)}(k_y)\mu_B B_z+\\&+\frac{g_x^{(1)}(k_y)}{2}\mu_B B_x\Sigma_x+\frac{g_y^{(1)}(k_y)}{2}\mu_B B_y\Sigma_y+\\&+\frac{g_z^{(1)}(k_y)}{2}\mu_B B_z \Sigma_z+\frac{g_x^{(2)}(k_y)}{2}\mu_B B_x\Sigma_z+\\&+\frac{g_z^{(2)}(k_y)}{2}\mu_B B_z \Sigma_x,
\end{aligned}
\end{equation}
where the functions have the following reflection properties:
\begin{subequations}\label{eq:Gatesymmetries}
\begin{align}
E_0(-k_y)&=E_0(k_y), \\
\alpha_{x(z)}(-k_y)&=\alpha_{x(z)}(k_y), \\
g_z^{(1)}(-k_y)&=g_z^{(1)}(k_y), \\
g_{x(z)}^{(0)}(-k_y)&=-g_{x(z)}^{(0)}(k_y), \\
g_{x(y)}^{(1)}(-k_y)&=g_{x(y)}^{(1)}(k_y), \\
g_{x(z)}^{(2)}(-k_y)&=g_{x(z)}^{(2)}(k_y).
\end{align}
\end{subequations}
Equation~\eqref{eq:gatedispersion} describes the general Hamiltonian of a nanowire in a cubic semiconductor, keeping only linear magnetic terms. For the gate-defined nanowire considered in this work, $g_x^{(2)}(k_y)=g_z^{(2)}(k_y)=0$, because the nanowire confinement potential in Eq.~\eqref{eq:Vx} is symmetric under $x\rightarrow -x$ operation. Furthermore, in our case $g_z^{(0)}(k_y)=0$, as such a term would only originate from an electric field along $x$, analogous to Eq.~\eqref{eq:spinIndOrigin}.

In Eq.~\eqref{eq:gatedispersion} there are two spin-orbit terms allowed by symmetry, $\alpha_x(k_y)$ and $\alpha_z(k_y)$. For a gate-defined nanowire along the [010] axis, we can redefine the pseudospin operators using a global transformation such that $\alpha_z(k_y)=0$ for every $k_y$, and the only non-zero spin-orbit function becomes $\alpha_x(k_y)$. 

In Eq.~\eqref{eq:gatedispersion}, there are four different non-zero $g$-functions for the gate-defined nanowire considered here. Three of them are the conventional Zeeman couplings $g_x^{(1)}(k_y)$, $g_y^{(1)}(k_y)$, and $g_z^{(1)}(k_y)$. We also have a function $g_x^{(0)}(k_y)$ that is not coupled to the spin. 

\begin{figure}[t]
\centering
\includegraphics[width=\columnwidth]{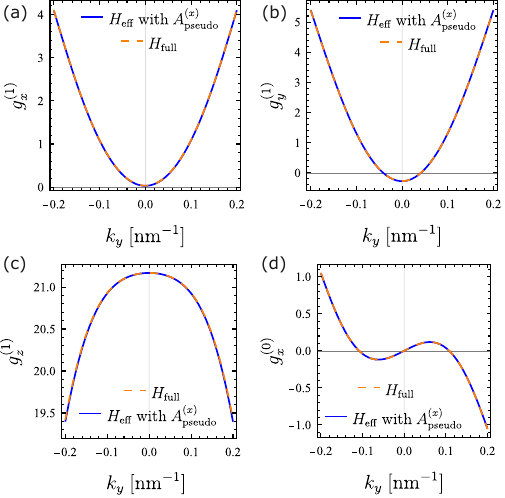}
\caption {\label{fig:Gatexy} \small{Calculated $g$-functions of a gate-defined nanowire in SiGe/Ge/SiGe 2DHG, see Eq.~\eqref{eq:gatedispersion}. The blue curves are obtained using the Hamiltonian in Eq.~\eqref{eq:Hnano3D}, while the orange curves are calculated using Eq.~\eqref{eq:Hnano2D}. The electric field is $E_z=1$ mV/nm, and the confinement strength is $\hbar\omega_x=2$ meV.}}
\refstepcounter{subfigure}\label{fig:Gatexya}
\refstepcounter{subfigure}\label{fig:Gatexyb}
\refstepcounter{subfigure}\label{fig:Gatexyc}
\refstepcounter{subfigure}\label{fig:Gatexyd}
\end{figure}

\begin{figure}[t]
\centering
\includegraphics[width=\columnwidth]{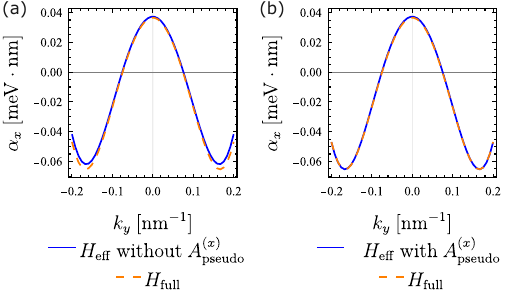}
\caption {\label{fig:alphaGate} \small{Calculated $\alpha_x(k_y)$ spin-orbit function of a gate-defined nanowire in SiGe/Ge/SiGe 2DHG, see Eq.~\eqref{eq:gatedispersion}. The electric field is $E_z=1$ mV/nm, and the confinement is $\hbar\omega_x=2$ meV. (a) The blue curve is calculated using the effective Hamiltonian, neglecting the Berry connection contributions from the confinement along $x$, see Eq.~\eqref{eq:Hnano2D}. The orange curve is obtained using the full Hamiltonian in Eq.~\eqref{eq:Hnano3D}.  (b) In the effective Hamiltonian, the Berry connections are included, leading to an excellent agreement with the full Hamiltonian, even at $\lvert k_y\rvert=0.2$ $\mathrm{nm^{-1}}$.}}
\refstepcounter{subfigure}\label{fig:alphaGatea}
\refstepcounter{subfigure}\label{fig:alphaGateb}
\end{figure} 

The $g$-functions are plotted in Fig.~\ref{fig:Gatexy}, calculated from the Hamiltonian in Eq.~\eqref{eq:Hnano3D} (orange lines) or from the Hamiltonian in Eq.~\eqref{eq:Hnano2D} (blue lines). We see excellent agreement between the results of the two Hamiltonians. Figure~\ref{fig:Gatexya} shows the $g_x^{(1)}(k_y)$ function, which is small for small $k_y$, expected from the moderate confinement along $x$, $\hbar\omega_x=2$ meV. Figure~\ref{fig:Gatexyb} shows the $g$-function along the nanowire direction, which is small for small $k_y$. Figure~\ref{fig:Gatexyc} shows the $g$-function along $z$, which is similar to the $g$-function of the 2DHG along the $z$ axis, see Fig.~\ref{fig:alphaHHpseudoc}. Figure~\ref{fig:Gatexyd} shows the spin-independent $g$-function $g_x^{(0)}(k_y)$, which is also similar to the result of the 2DHG, see Fig.~\ref{fig:g0xyHHa}.

The spin-orbit function $\alpha_x(k_y)$ of the effective Hamiltonian in Eq.~\eqref{eq:gatedispersion} is shown in Fig.~\ref{fig:alphaGate}. The blue lines are obtained by using the Hamiltonian in Eq.~\eqref{eq:Hnano2D}. In that Hamiltonian $x$ is shifted by the Berry connection of the 2DHG, $A_\mathrm{pseudo}^{(x)}(\bm{k})$. Figure~\ref{fig:alphaGate} shows that this contribution to the Berry connection has only a minor effect: in Fig.~\ref{fig:alphaGatea}, the Berry connection is neglected, while in Fig.~\ref{fig:alphaGateb}, it is taken into account. We can see that for larger values $k_y$, there is a discrepancy between the blue and orange lines when the Berry connection is neglected (see Fig.~\ref{fig:alphaGatea}). When the Berry connection is included in the calculation (see Fig.~\ref{fig:alphaGateb}), the blue and orange curves match even at larger $k_y$. This means that the Berry connection corrections to the effective Hamiltonian in the pseudospin basis are usually negligible. However, one should be aware of these contributions, as they might become important in certain cases, as seen in Fig.~\ref{fig:LHSO} for light holes.

\subsection{Ge/Si core/shell nanowire}
So far we have looked at devices based on planar heterostructures. To demonstrate the versatility of our method, we apply it to a Ge/Si core/shell nanowire with a square cross-section. The general requirement for the applicability of our method is that the hole spin qubit is confined strongly in one or two directions, while in the remaining two or one direction it is weakly confined. This condition should result in a large splitting of subbands, which makes it possible to define the effective Hamiltonian using the two lowest subbands well separated from the rest. 

Core/shell nanowires with a Ge core and a Si shell satisfy this requirement if the strain is not too small \cite{kloeffel2011strong,kloeffel2018direct}. We assume such a strained core/shell nanowire with a square cross section with 10 nm core length, schematically represented in Fig.~\ref{fig:nanowiresb}. For simplicity, we assume that the nanowire is grown along the [010] crystal direction, while the side facets are parallel to the planes (100) and (001). An electric field $E$ is applied perpendicularly to the nanowire, parameterized by the angle $\phi_E$ in the $x$-$z$ plane; see Fig.~\ref{fig:nanowiresb}. The two gates on top of the nanowire confine a single hole. 

The strain created in the nanowire by the lattice mismatch between Ge and Si is predicted to be constant for cylindrical geometries \cite{kloeffel2014acoustic}, and it is believed that even for a square geometry, it is a reasonable approximation to assume constant strain in the nanowire \cite{kloeffel2018direct}. Here, similarly to~\cite{kloeffel2018direct}, we assume the non-zero strain tensor elements $\epsilon_{yy}=-21.8\times 10^{-3}$ and $\epsilon_{xx}=\epsilon_{zz}=-5.8 \times 10^{-3}$. 

For a Ge/Si core/shell nanowire, the valence-band offset between the Ge core and the Si shell is large. Therefore, we assume hard-wall boundary conditions.  
\begin{figure}[t]
\centering
\includegraphics[width=\columnwidth]{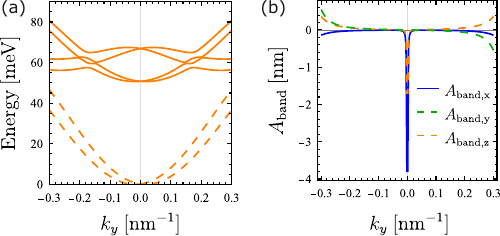}
\caption {\label{fig:CoreShelldispersion} \small{Calculated core/shell nanowire dispersion and Berry connections. The electric field is $E=10$ mV/nm, and $\phi_E=30\degree$. (a) Subbands of the core/shell nanowire, with zero magnetic field. The lowest two bands shown with dashed lines are used to construct the effective Hamiltonian. (b) Berry connections in the band basis with an out-of-plane magnetic field of $B_z$=20 mT.}}
\refstepcounter{subfigure}\label{fig:CoreShelldispersiona}
\refstepcounter{subfigure}\label{fig:CoreShelldispersionb}
\end{figure}
We used the 6-band LKBP Hamiltonian to describe the nanowire. We expand the LKBP Hamiltonian on the eigenfunctions of the infinite square-well potential to eliminate the directions $x$ and $z$. In this way, we arrive at the full Hamiltonian of the nanowire depending only on $k_y$, $H_\mathrm{full}^{\mathrm{(cs)}}(k_y)$. We note that the ansatz we use to expand the LKBP Hamiltonian is not a solution of the confinement problem, as it does not take into account the applied electric field. However, this is not an issue if we use a large enough basis set. More details can be found in the Appendix~\ref{App:cs}.

The numerical solution of $H_\mathrm{full}^{\mathrm{(cs)}}(k_y)$ produces subbands as a function of $k_y$, which can be seen in Fig.~\ref{fig:CoreShelldispersiona}. The lowest three subbands are plotted, with large spin-splittings at finite $k_y$. The lowest two spin-split subbands plotted with dashed lines are used to construct the nanowire effective Hamiltonian. The dispersion Hamiltonian in the band basis becomes of the same form as for the gate-defined nanowire, see Eq.~\eqref{eq:HbandGate}.

To calculate the Berry connections in the band basis, we apply the same gauge fixing technique as for gate-defined nanowires. We need to express the eigenstates at finite $k_y$ with the eigenstates at $k_y=0$. However, the eigenstates are represented in the basis of infinite square-well solutions. Therefore, we apply a unitary transformation to the eigenbasis of the eigenstates at $k_y=0$, then we set the corresponding vector components' phases to be zero, see Eq.~\eqref{eq:gaugeFixing}. The resulting Berry connections are shown in Fig.~\ref{fig:CoreShelldispersionb}. Interestingly, even though the SOI is large, the Berry connections are small even in the band basis. This highlights that the pseudospin basis is more suitable for studying the SOI in the system.

As a next step, we transform the dispersion Hamiltonian from Eq.~\eqref{eq:HbandGate} to the pseudospin basis as described in previous sections. The effective Hamiltonian allowed by the symmetries has the same form as in Eq.~\eqref{eq:gatedispersion}, with the symmetry properties described in Eq.~\eqref{eq:Gatesymmetries}.

In Eq.~\eqref{eq:gatedispersion}, there are two different spin-orbit functions $\alpha_x(k_y)$ and $\alpha_z(k_y)$. If the electric field is parallel to $x$ or $z$, we can redefine the pseudospin axes with a global unitary transformation such that $\alpha_z(k_y)=0$ for every $k_y$. However, if the electric field has an arbitrary direction in the $x$-$z$ plane, there is no such global transformation. Therefore, we keep both $\alpha_x(k_y)$ and $\alpha_z(k_y)$ functions. 

We choose the pseudospin axes in the following physically intuitive way. We apply an electric field with components $E_x=E_z=1$ mV/nm and prescribe that the two spin-orbit functions are equal at $k_y=0.1$ $\mathrm{nm^{-1}}$. In this way, the pseudospin axes are aligned to the crystal axes. For most calculations, we choose $\phi_E=30\degree$ for the direction of the electric field. 

\begin{figure}[tp!]
\centering
\includegraphics[width=\columnwidth]{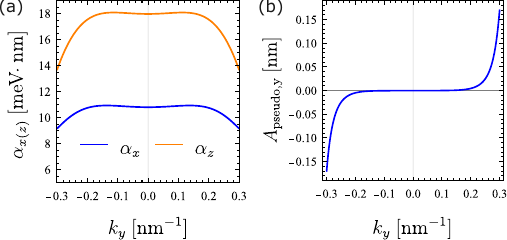}
\caption {\label{fig:CoreShellalpha} \small{Calculated core/shell nanowire spin-orbit functions and Berry connections. The electric field is $E=10$ mV/nm with angle $\phi_E=30\degree$. (a) The spin-orbit functions $\alpha_x(k_y)$ and $\alpha_z(k_y)$ in Eq.~\eqref{eq:gatedispersion}. (b) Berry connections in the pseudospin basis, with an out-of-plane magnetic field of $B_z$=20 mT. The $A_\mathrm{pseudo,x}$ and $A_\mathrm{pseudo,z}$ components are zero.}}
\refstepcounter{subfigure}\label{fig:CoreShellalphaa}
\refstepcounter{subfigure}\label{fig:CoreShellalphab}
\end{figure}

Figure~\ref{fig:CoreShellalphaa} shows the calculated spin-orbit functions $\alpha_x(k_y)$ and $\alpha_z(k_y)$, see Eq.~\eqref{eq:gatedispersion}. We can see that for small $k_y$ they are constant, which means that the SOI is linear. The spin-orbit functions are large, as expected for a core/shell nanowire and with an applied electric field of $E=10$ mV/nm. Figure~\ref{fig:CoreShellalphab} shows that the Berry connections in the pseudospin basis are negligible, therefore, we omit them.

\begin{figure}[tp!]
\centering
\includegraphics[width=\columnwidth]{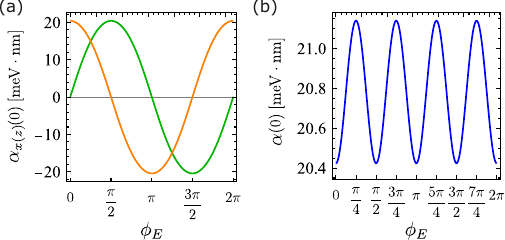}
\caption {\label{fig:CoreShellalphaRot} \small{Calculated spin-orbit functions from Eq.~\eqref{eq:gatedispersion} at $k_y=0$. The electric field is $E=10$ mV/nm. (a) The spin-orbit coefficients are plotted as functions of the in-plane angle of the electric field. (b) Total spin-orbit coefficient $\alpha(0)=\sqrt{\alpha_x(0)^2+\alpha_z(0)^2}$ as a function of the in-plane angle of the electric field.}}
\refstepcounter{subfigure}\label{fig:CoreShellalphaRota}
\refstepcounter{subfigure}\label{fig:CoreShellalphaRotb}
\end{figure}

\begin{figure}[bp!]
\centering
\includegraphics[width=\columnwidth]{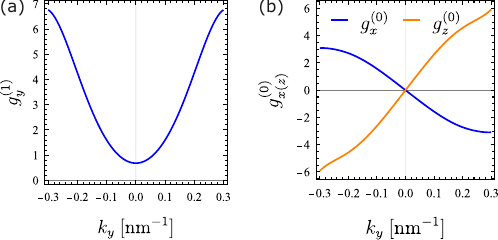}
\caption {\label{fig:CoreShellgzg0} \small{Calculated $g$-functions of the core/shell nanowire, see Eq.~\eqref{eq:gatedispersion}. The electric field is $E=10$ mV/nm with angle $\phi_E=30\degree$.}}
\refstepcounter{subfigure}\label{fig:CoreShellgzg0a}
\refstepcounter{subfigure}\label{fig:CoreShellgzg0b}
\end{figure}

In Fig.~\ref{fig:CoreShellalphaRota} we plot the spin-orbit functions $\alpha_x(k_y)$ and $\alpha_z(k_y)$ at $k_y=0$ for different in-plane angles $\phi_E$ of the electric field, see Fig.~\ref{fig:nanowiresb}. The spin-orbit coefficients approximately rotate with the electric field, $\alpha_x\approx \alpha(0) \sin{(\phi_E)}$ and $\alpha_z \approx \alpha(0) \cos{(\phi_E)}$. Figure~\ref{fig:CoreShellalphaRotb} shows the total SOI $\alpha(0)=\sqrt{\alpha_x(0)^2+\alpha_y(0)^2}$, which is approximately constant, devations come from the non-cylindrical geometry. The total SOI has a $\pi/4$ period, and it is maximal when the electric field is along the diagonal of the square. 

We remark that, for the core/shell nanowire, the pseudospin axes cannot be perfectly aligned to the crystal axes. When the electric field is along the square diagonal, $\phi_E=\pi/4$, the two spin-orbit coefficients are equal. However, when the electric field is parallel to the $x$-axis, $\phi_E=0$, we expect an exactly zero $\alpha_x$ coefficient. Instead, we obtain a value around 0.1 $\mathrm{meV}\cdot \mathrm{nm}$, which is negligible on the scale of Fig.~\ref{fig:CoreShellalphaRot}. We attribute this imperfection to the fact that the eigenstates of the core/shell nanowire, even at $k_y=0$, contain significant heavy-hole, light-hole, and split-off contributions, and these states have different rotational symmetry properties.

Figure \ref{fig:CoreShellgzg0a} shows the $g_y(k_y)$ $g$-function. As expected, for small $k_y$ it is close to 0, because the confinement along $y$ is not considered yet. Figure \ref{fig:CoreShellgzg0b} shows the spin-independent Zeeman functions $g_x^{(0)}(k_y)$ and $g_z^{(0)}(k_y)$, which for finite $k_y$ can be significant. 

\begin{figure}[htp!]
\centering
\includegraphics[width=\columnwidth]{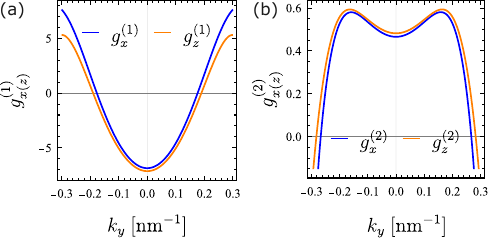}
\caption {\label{fig:CoreShellgxy12} \small{Calculated $g$-functions of the core/shell nanowire, see Eq.~\eqref{eq:gatedispersion}. The electric field is $E=10$ mV/nm with angle $\phi_E=30\degree$.}}
\refstepcounter{subfigure}\label{fig:CoreShellgxy12a}
\refstepcounter{subfigure}\label{fig:CoreShellgxy12b}
\end{figure}

\begin{figure}[b!]
\centering
\includegraphics[width=\columnwidth]{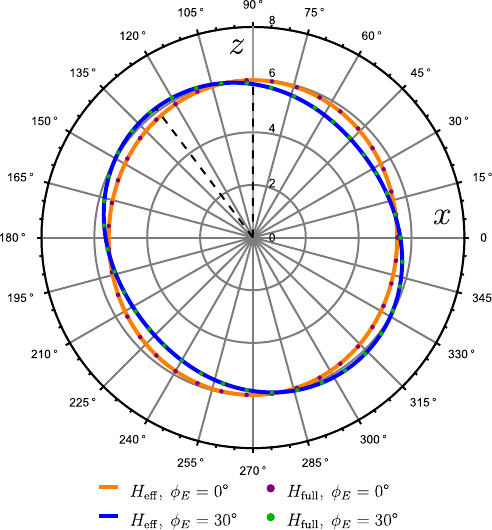}
\caption {\label{fig:CoreShellplanexy} \small{Effective $g$-factor of the core/shell nanowire quantum dot for different magnetic field orientations in the $x$-$z$ plane. The electric field is $E=10$ mV/nm, and $\hbar\omega_y=2$ meV. The orange and blue curves are calculated using the effective Hamiltonian in Eq.~\eqref{eq:Heff_cs}, the purple and green markers are calculated using the full Hamiltonian $H_\mathrm{full}^\mathrm{(cs)}$. The black dashed lines show the maximal $g$-factors.}}
\end{figure}

Figure \ref{fig:CoreShellgxy12a} shows the in-plane $g$-functions $g_x^{(1)}(k_y)$ and $g_z^{(1)}(k_y)$, while $g_x^{(2)}(k_y)$ and $g_z^{(2)}(k_y)$ are plotted in Fig.~\ref{fig:CoreShellgxy12b}. We can only see a small asymmetry between the $x$ and $z$ directions, arising from the electric field with direction $\phi_E=30\degree$. The in-plane $g$-functions are much larger than the $g$-function along the nanowire, because of the strong in-plane confinement. 

Now, we use the effective nanowire Hamiltonian from Eq.~\eqref{eq:gatedispersion} to describe a nanowire quantum dot confined by the potential
\begin{equation}\label{eq:csConf}
    V_\mathrm{cs}(y)=\frac{m^*}{2}\omega_y^2 y^2, 
\end{equation}
where $m^*$ is the effective mass. Due to the large spin-orbit splitting of the core/shell dispersion (Fig.~\ref{fig:CoreShelldispersion}), it is challenging to determine the effective mass. However, using the effective Hamiltonian from Eq.~\eqref{eq:gatedispersion}, we can calculate $m^*$ by fitting $E_0(k_z)$.

The effective Hamiltonian of the nanowire quantum dot can be written as 
\begin{equation}\label{eq:Heff_cs}
H_\mathrm{eff}^\mathrm{(cs)}=H_\mathrm{pseudo}^{(d,\mathrm{nw})}(k_y)+V_\mathrm{cs}(y+A_\mathrm{pseudo}(k_y)),
\end{equation}
where $H_\mathrm{pseudo}^{(d,\mathrm{nw})}$ is in Eq.~\eqref{eq:gatedispersion} and $V_\mathrm{cs}$ is in Eq.~\eqref{eq:csConf}. 

\begin{figure}[b!]
\centering
\includegraphics[width=\columnwidth]{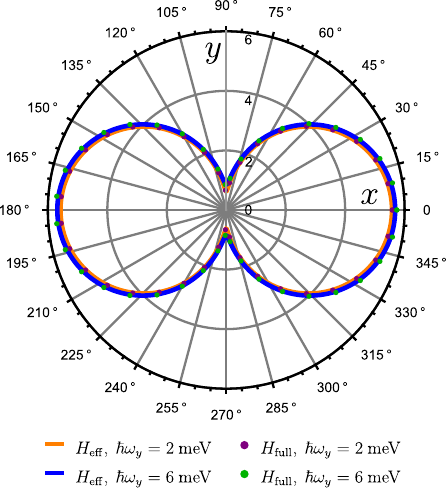}
\caption {\label{fig:CoreShellplanexz} \small{Effective $g$-factor of the core/shell nanowire for different magnetic field orientations in the $x$-$y$ plane, for different confinement strengths. The electric field is $E=10$ mV/nm with angle $\phi_E=30\degree$. The orange and blue curves are the results of the effective Hamiltonian from Eq.~\eqref{eq:Heff_cs}}, while the purple and green markers are calculated using the full Hamiltonian $H_\mathrm{full}^{(cs)}$.}
\end{figure}

As a test of our effective Hamiltonian in Eq.~\eqref{eq:Heff_cs}, we study the $g$-factor anisotropy of the nanowire quantum dot. To make the calculations more convenient and fast, we neglect the small Berry connection contribution $A_\mathrm{pseudo}$, see Fig.~\ref{fig:CoreShellalphab}.

\begin{figure}[htp!]
\centering
\includegraphics[width=\columnwidth]{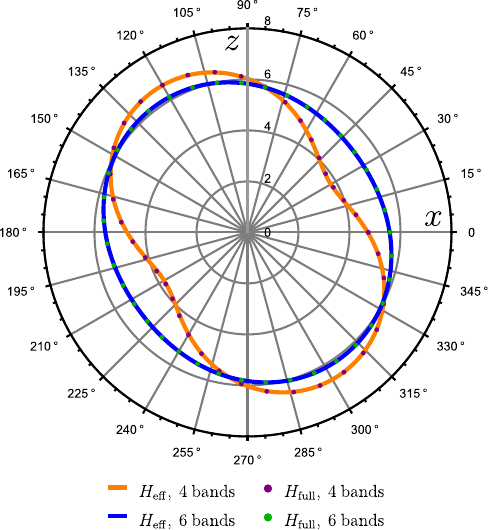}
\caption {\label{fig:CoreShell6vs4} \small{Effective $g$-factor of the core/shell nanowire for different magnetic field orientations in the $x$-$z$ plane. The electric field is $E=10$ mV/nm with angle $\phi_E=30\degree$ and $\hbar\omega_y=2$ meV. The blue curve (green markers) is the result of a calculation using the full (effective) 6-band LKBP Hamiltonian. The orange curve (purple markers) is the result of a calculation using the full (effective) 4-band LKBP Hamiltonian. }}
\end{figure}

Figure~\ref{fig:CoreShellplanexy} shows the in-plane $g$-factor as a function of the in-plane angle of the magnetic field. The blue curve represents the case when the angle of the electric field $E=10$ mV/nm is $\phi_E=30\degree$, while for the orange curve the angle is $\phi_E=0\degree$. The green and purple markers are the solutions of the full Hamiltonian $H_\mathrm{full}^{\mathrm{(cs)}}$. The agreement between the results of the effective Hamiltonian and the full Hamiltonian is excellent, which is expected, as the $z$-confinement is not strong, $\hbar\omega_z=2$ meV. The difference between the orange and blue curves is not large, even though the angle of the electric field differs by $30\degree$. This is a result of the fact that due to the strain, the eigenstates of the core/shell nanowire have predominantly light hole character, which have a much more isotropic $g$-tensor.

Next, we study the tunability of the $g$-factor in the $x$-$y$ plane with the confinement along the $y$-axis. Figure~\ref{fig:CoreShellplanexz} shows the $g$-factors for different directions in the $x$-$y$ plane for $\hbar\omega_y=2$ meV and $\hbar\omega_y=6$ meV. The difference between the two cases is small, as the $y$-confinement predominantly tunes the $g$-factor along $y$, which is small compared to the $g$-factors along $x$ and $z$. Figures~\ref{fig:CoreShellplanexy} and \ref{fig:CoreShellplanexz} overall suggest a small device-from-device variability of hole spin qubits confined in core/shell nanowires, contrary to hole spin qubits in planar heterostructures \cite{john2024two, seidler2025spatial}.

The small anisotropy of the $g$-tensor of the core/shell nanowire quantum dot is the result of the predominantly light-hole character of the states. 
Theoretical studies used the 4-band LKBP Hamiltonian to model nanowires \cite{kloeffel2011strong, kloeffel2018direct} or a 6-band LKBP Hamiltonian with isotropic approximation \cite{adelsberger2022hole}. Here, we employ the 6-band LKBP Hamiltonian including the anisotropies, $\gamma_2\neq \gamma_3$. Because of the strong confinement and the mixing between the split-off holes and light holes, the split-off holes can become important. In Fig.~\ref{fig:CoreShell6vs4} we compare the in-plane $g$-factors as a function of the magnetic field angle calculated with the 6-band LKBP Hamiltonian and with the 4-band Hamiltonian. We can see that the 6-band Hamiltonian yields much more isotropic results than the 4-band Hamiltonian. This is because the heavy-hole contribution of the states is larger when the 4-band Hamiltonian is used. When the 6-band Hamiltonian is used, some of the mixing with the heavy holes is replaced by mixing with split-off holes, which introduces more isotropy to the $g$-tensor.

\section{Conclusion}\label{sec:Conclusion}
In this paper, we constructed non-perturbative effective Hamiltonians of hole spin qubits in planar heterostructures and nanowires. We studied two different planar structures, a SiGe/Ge/SiGe hosting a heavy-hole qubit, and a SnGe/Ge hosting a light-hole qubit. First, we solved the strong $z$-confinement problem yielding the 2DHG subbands. Then, we constructed the effective Hamiltonian using the lowest two spin-split subbands. We have shown that the quantum geometry of the 2DHG naturally emerges in the effective Hamiltonian, the position operators in the in-plane confinement potential have to be shifted by Berry connections. We defined the pseudospin non-perturbatively by prescribing that the Berry connection contributions are small. This led to the discovery of a weak direct-Rashba SOI induced by the in-plane confinement. 

To demonstrate the broad applicability of our method, we also studied a gate-defined nanowire in SiGe/Ge/SiGe and a Ge/Si core/shell nanowire quantum dot. Similarly to planar heterostructures, we first solved the strong confinement problem to obtain the nanowire subbands, then we constructed the effective Hamiltonians on the lowest two spin-split subbands. The resulting effective Hamiltonians yielded results in excellent agreement with the full Hamiltonian for realistic confinement strengths. Our results also highlighted the importance of the split-off holes for the $g$-tensor anisotropy of core/shell nanowires. 

The SOI of light-hole spin qubits is highly tunable with electric fields. Using the non-perturbative effective Hamiltonian, we have shown that in the regime where the SOI is close to its minimum, the SOI is dominated by the in-plane confinement, not the SOI of the 2DHG. The total SOI of the light-hole qubit considered in this work cannot be completely turned off. Broadly speaking, to study the spin-orbit switch functionality of hole spin qubits, one has to consider the full 3D confinement problem. Even if the SOI originating from the strong confinement can be turned off, the weak confinement creates a residual SOI, leading to decoherence during idling. While relaxing the weak confinement mitigates the residual SOI, the dot size cannot be increased arbitrarily. A sufficiently strong confinement is required to maintain a large enough orbital level spacing, necessary to have a well-isolated qubit subspace.

Throughout the paper, we considered hole spin qubits in Ge. Our method does not assume an isotropic approximation; therefore, it could be directly applied to Si hole spin qubits using the 6-band LKBP Hamiltonian with the appropriate parameters for Si. Additionally, our method should apply to a wide variety of systems, in which the essential physics can be captured by an effective Hamiltonian constructed from the lowest two subbands. 

\begin{acknowledgments}
We thank Abhikbrata Sarkar, Even Thingstad, Zeb Krix and Patrick del Vecchio for useful discussions. 
This work was supported as part of NCCR SPIN, a National Center of Competence in Research, funded by the Swiss National Science Foundation (grant number 225153). This work has received funding from the Swiss State Secretariat for Education, Research and Innovation (SERI) under contract number M822.00078. D.L. acknowledges the Deanship of Research and the Quantum Center for the support received under Grant no. CUP25102 and no. INQC2500, respectively. The Gen-Q programme has received funding from the European Union’s Horizon Europe research and innovation programme under the Marie Skłodowska-Curie grant agreement number 101217386.
\end{acknowledgments}

\appendix

\section{Derivation of the effective Hamiltonian in the band basis}\label{App:HeffBand}
In this section of the Appendix, we derive the effective Hamiltonian in Eq.~\eqref{eq:Heff}. We start with the Hamiltonian describing the 2DHG and the in-plane confinement, see Eq.~\eqref{eq:Hfull}. We find the eigenstates and eigenvalues of the 2DHG Hamiltonian: 
\begin{equation}
H_0(\bm{k})\ket{\psi_n(\bm{k})}=E_n(\bm{k})\ket{\psi_n(\bm{k})}, 
\end{equation}
where $n \in \{1,\hdots , N\}$ integer, with $N$ being the number of $z$-confinement eigenstates that we use to expand the 3D Hamiltonian on. We represent the eigenstates $\ket{\psi_n(\bm{k})}$ in the basis of eigenstates at $\bm{k}=0$, we denote the representation of the states as $\psi_{n,i}(\bm{k})$, with $i \in\{1,\hdots, N\}$. 

We construct an $N\times N$ unitary matrix $U(\bm{k})$ with matrix elements $U_{ij}(\bm{k})=\psi_{i,j}(\bm{k})$, which we use to diagonalize $H_0(\bm{k})$: 
\begin{equation}
\begin{aligned}
    H_\mathrm{0,band}(\bm{k})&=U^\dag(\bm{k}) H_0(\bm{k})U(\bm{k})=\\&=\mathrm{Diag}(E_1(\bm{k}),\hdots, E_N(\bm{k})),
\end{aligned}
\end{equation}
which is the 2DHG Hamiltonian represented on its eigenbasis, on the band basis. The in-plane confinement potential $V(x,y)$ has to be transformed as well: 
\begin{equation}
    V_\mathrm{band}(x,y)=U^\dag(\bm{k})V(x,y)U(\bm{k}).
\end{equation}
We Taylor-expand the confinement potential around $x_0$, $y_0$ points: 
\begin{equation}
\begin{aligned}
    V(x,y)&=V(x_0,y_0)+x\partial_xV+y\partial_yV+\\&+\frac{1}{2}x^2 \partial^2_x V+xy \partial_x\partial_y V+\frac{1}{2}y^2 \partial_y^2 V+\hdots,
\end{aligned}
\end{equation}
where the derivatives are taken at $x=x_0$ and $y=y_0$. The transformed linear term in $x$ can be written as follows:
\begin{equation}\label{eq:transformedx}
    U^\dag(\bm{k})xU(\bm{k})=x+U^\dag(\bm{k})[x,U(\bm{k})]. 
\end{equation}
Assuming that $U(\bm{k})$ is a smooth function of $k_x$ and $k_y$, we can use the following identity: 
\begin{equation}
    [x,U(\bm{k})]=i\partial_{k_x}U(\bm{k}). 
\end{equation}
Using this identity, Eq.~\eqref{eq:transformedx} becomes: 
\begin{equation}
U^\dag(\bm{k})xU(\bm{k})=x+iU^\dag(k)\partial_{k_x}U(\bm{k})=x+\mathcal{A}^{(x)}(\bm{k}), 
\end{equation}
where we introduced the notation $\mathcal{A}^{(x)}(\bm{k})=iU^\dag(\bm{k})\partial_{k_y}U(\bm{k})$, which is an $N\times N$ matrix containing Berry connections between different bands. Similarly, $y$ transforms the following way: 
\begin{equation}
U^\dag(\bm{k})yU(\bm{k})=y+iU^\dag(k)\partial_{k_y}U(\bm{k})=y+\mathcal{A}^{(y)}(\bm{k}). 
\end{equation}
An arbitrary term $x^n y^m$ in the Taylor expansion can be written as: 
\begin{equation}
    U^\dag(\bm{k})x^ny^mU(\bm{k})=(x+\mathcal{A}^{(x)}(\bm{k}))^n(y+\mathcal{A}^{(y)}(\bm{k}))^m,
\end{equation}
which means that the transformed confinement potential becomes: 
\begin{equation}
    U^\dag(\bm{k})V(x,y)U(\bm{k})=V(x+\mathcal{A}^{(x)}(\bm{k}),y+\mathcal{A}^{(y)}(\bm{k})).
\end{equation}

The full Hamiltonian in the band basis: 
\begin{equation}\label{eq:Hfullband}
\begin{aligned}
    H_\mathrm{full,band}&=\mathrm{Diag}(E_1(\bm{k}),\hdots, E_N(\bm{k}))+\\&+V(x+\mathcal{A}^{(x)}(\bm{k}),y+\mathcal{A}^{(y)}(\bm{k})).
\end{aligned}
\end{equation}
This is still the exact Hamiltonian (aside from the truncation of the $z$-confinement basis to $N$ states), but represented in the band basis. We can derive an effective Hamiltonian in the lowest two spin-split subbands by applying a Schriffer-Wolff transformation. If we assume a large separation between the lowest two spin-split subbands and the rest, it is enough to apply a trivial first-order Schrieffer-Wolff transformation and take the corresponding $2\times 2$ block of Eq.~\eqref{eq:Hfullband}. The effective Hamiltonian becomes: 
\begin{equation}
\begin{aligned}
    H_\mathrm{eff,band}&=\mathrm{Diag}(E_1(\bm{k}),E_2(\bm{k}))+\\&+V(x+A^{(x)}(\bm{k}),y+A^{(y)}(\bm{k})),
\end{aligned}
\end{equation}
where $A^{(x)}(\bm{k})$ and $A^{(y)}(\bm{k})$ are the $2\times 2$ blocks of $\mathcal{A}^{(x)}(\bm{k})$ and $\mathcal{A}^{(y)}(\bm{k})$ corresponding the two lowest spin-split subbands. We also neglected contributions from Berry connections between the lowest two bands and bands outside the effective subspace. We justify this approximation by considering the quadratic term in the Taylor expansion of the potential:
\begin{equation}
\begin{aligned}
    (x+\mathcal{A}^{(x)}(\bm{k}))^2&=x^2+x\mathcal{A}^{(x)}(\bm{k})+\mathcal{A}^{(x)}(\bm{k})x+\\&+\mathcal{A}^{(x)}(\bm{k})\mathcal{A}^{(x)}(\bm{k}).
\end{aligned}
\end{equation}
When we project these down to the effective subspace, we obtain a matrix product from the last term: 
\begin{equation}\label{eq:BerryQuadValid}
\begin{aligned}
    &[\mathcal{A}^{(x)}(\bm{k})\mathcal{A}^{(x)}(\bm{k})]_{\sigma\sigma'}=\sum_{l=1}^N \mathcal{A}^{(x)}_{\sigma l}(\bm{k})\mathcal{A}^{(x)}_{l\sigma'}(\bm{k})=\\&=
    [A^{(x)}(\bm{k})A^{(x)}(\bm{k})]_{\sigma\sigma'}+\sum_{l=3}^N \mathcal{A}^{(x)}_{\sigma l}(\bm{k})\mathcal{A}^{(x)}_{l\sigma'}(\bm{k}).
\end{aligned}
\end{equation}
Berry connections between nondegenerate states can be written as follows: 
\begin{equation}
    \mathcal{A}^{(x)}_{mn}(\bm{k})=i\frac{\bra{\psi_m(\bm{k})}\partial_{k_x}H_0(\bm{k})\ket{\psi_n(\bm{k})}}{E_n-E_m}.
\end{equation}
Therefore, if the separation between the effective subspace and the higher bands is large, $\mathcal{A}^{(x)}_{\sigma l}(\bm{k})$ can be neglected in Eq.~\eqref{eq:BerryQuadValid}. This means that the $N\times N$ Berry connection matrix $\mathcal{A}^{(x)}(\bm{k})$ can be replaced with the $2\times 2$ block $A^{(x)}(\bm{k})$. Similarly, $\mathcal{A}^{(y)}(\bm{k})$ can be replaced with $A^{(y)}(\bm{k})$.
\section{Berry connections in the pseudospin basis}\label{App:Apseudo}
We use the matrix $X(\bm{k})$ to transform the dispersion Hamiltonian and the Berry connection, see Eqs.~\eqref{eq:dispersion_band} and ~\eqref{eq:AbandTransform}. Here, we prove that this transformation is indeed an isomorphism between the subspaces $S(\bm{k})$ and $S_0$, and approximately cancels out the Berry connections, making it the appropriate transformation to define pseudospin. 

We define two operators $\mathcal{U}_0$ and $\mathcal{U}(\bm{k})$, $\mathcal{U}_0:\mathbb{C}^2\rightarrow S_0$, and $\mathcal{U}(\bm{k}):\mathbb{C}^2\rightarrow S(\bm{k})$, the following way: 
\begin{subequations}
    \begin{align}
       &\mathcal{U}_0e_\sigma=\ket{\psi_\sigma(0)}, \\ 
       &\mathcal{U}(\bm{k})e_\sigma=\ket{\psi_\sigma(\bm{k})},
    \end{align}
\end{subequations}where $\sigma \in \{1,2\}$ and $e_\sigma$ is a basis vector in $\mathbb{C}^2$. These operators are isometries, $\mathcal{U}_0^\dag \mathcal{U}_0=\mathcal{U}^\dag(\bm{k}) \mathcal{U}(\bm{k})=I_2$. $X$ is a mapping between the representations of vectors from $S(\bm{k})$ and $S_0$ on $\mathbb{C}^2$, $X:\mathbb{C}^2_\mathrm{band}\rightarrow \mathbb{C}^2_\mathrm{pseudo}$, $X=M(M^\dag M)^{-1/2}$. Using that $X^\dag=(M^\dag M)^{-1/2}M^\dag$, we can see that $X$ is unitary, $X^\dag X=X X^\dag=I_2$. The $W$ isomorphism between the spaces $S(\bm{k})$ and $S_0$, $W:S(\bm{k})\rightarrow S_0$ is constructed the following way: 
\begin{equation}
    W(\bm{k})=\mathcal{U}_0X(\bm{k})\mathcal{U}^\dag(\bm{k}).
\end{equation}
This mapping is unitary, $W^\dag(\bm{k}) W(\bm{k})=\mathcal{U}(\bm{k})\mathcal{U}^\dag (\bm{k})$, $W(\bm{k}) W^\dag(\bm{k})=\mathcal{U}_0\mathcal{U}_0^\dag$, where $\mathcal{U}(\bm{k})\mathcal{U}^\dag (\bm{k})$ and $\mathcal{U}_0\mathcal{U}_0^\dag$ are projectors to subspaces $S(\bm{k})$ and $S_0$, therefore identity operators on the subspaces. This isomorphism is not unique, we could redefine $W(\bm{k})$ using a different $X(\bm{k})$, $\Tilde{X}(\bm{k})=X(\bm{k})F(\bm{k})$, where $F(\bm{k})$ is an arbitrary $2\times 2$ unitary matrix. 

Now we show how to transform operators from the band basis to the pseudospin basis. Let us take the dispersion operator $H^{(d)}$ which acts on the band basis, $H^{(d)}_\mathrm{band}:S(\bm{k})\rightarrow S(\bm{k})$. To obtain the form in the pseudospin basis, first we transform $H^{(d)}$ with the isomorphism $W(\bm{k})$, then we take the representation using $\mathcal{U}_0$: 
\begin{equation}
\begin{aligned}
    H^{(d)}_\mathrm{pseudo}&=\mathcal{U}_0^\dag W(\bm{k})H^{(d)}W^\dag(\bm{k})\mathcal{U}_0=\\
    &=\mathcal{U}_0^\dag \mathcal{U}_0 X(\bm{k}) \mathcal{U}^\dag(\bm{k}) H^{(d)} \mathcal{U}(k)X^\dag(\bm{k}) \mathcal{U}_0^\dag \mathcal{U}_0= \\&
    =X(\bm{k})H^{(d)}_\mathrm{band}(\bm{k})X^\dag(\bm{k}),
\end{aligned}
\end{equation}
where we used that $H^{(d)}_\mathrm{band}(\bm{k})=\mathcal{U}^\dag(\bm{k})H^{(d)}(\bm{k})\mathcal{U}(\bm{k})$. The Berry connection is the abstract operator $i\partial_{k_\mu}$ acting on $S(\bm{k})$, $i\partial_{k_\mu}:S(\bm{k})\rightarrow S(\bm{k})$. In the band basis it can be written as $A^{(\mu)}_\mathrm{band}(\bm{k})=\mathcal{U}^\dag (\bm{k})i \partial_{k_\mu}\mathcal{U}(\bm{k})$. The form in the pseudospin basis becomes: 
\begin{equation}\label{eq:APseudoDeriv}
\begin{aligned}
    A_\mathrm{pseudo}^{(\mu)}(\bm{k})&=\mathcal{U}_0^\dag W(\bm{k})i\partial_{k_\mu}(W^\dag(\bm{k})\mathcal{U}_0)=\\& =X(\bm{k})\mathcal{U}^\dag(\bm{k})i\partial_{k_\mu}(\mathcal{U}(\bm{k})X^\dag(\bm{k}))=\\&=
    A_\mathrm{band}^{(\mu)}(\bm{k})+X(\bm{k})i\partial_{k_\mu}X^\dag(\bm{k}).
\end{aligned}
\end{equation}
Now we show that the Berry connections in the pseudospin basis, Eq.~\eqref{eq:APseudoDeriv}, are approximately canceled out by the transformation $X(\bm{k})$. Let us introduce $U_\mathrm{eff}(\bm{k})$ $N\times 2$ matrix with matrix elements $U_\mathrm{eff,l\sigma }(\bm{k})=\psi_{\sigma,l}(\bm{k})$, with $\sigma \in\{1,2\}$ and $l \in \{1,\hdots ,N\}$. $U_\mathrm{eff}(\bm{k})$ is the matrix representation of $\mathcal{U}(\bm{k})$, using the canonical basis vectors $e_1=(1,0)$ and $e_2=(0,1)$ on $\mathbb{C}^2$. Using this matrix, the $2\times 2$ Berry connection matrix $A^{(\mu)}_\mathrm{band}(\bm{k})$ can be written as: 
\begin{equation}
    A^{(\mu)}(\bm{k})=iU_\mathrm{eff}^\dag(\bm{k})\partial_{k_\mu}U_\mathrm{eff}(\bm{k}), 
\end{equation}
from which follows that:
\begin{equation}\label{eq:UeffIdentity}
    iU_\mathrm{eff}^\dag(\bm{0})\partial_{k_\mu}U_\mathrm{eff}(\bm{k})=U_\mathrm{eff}^{\dag}(\bm{0})U_\mathrm{eff}(\bm{k})A^{(\mu)}_\mathrm{band}(\bm{k}).
\end{equation}
Here we can recognize the overlap matrix $M(\bm{k})=U_\mathrm{eff}^\dag(\bm{0})U_\mathrm{eff}(\bm{k})$, which can be used to rewrite Eq.~\eqref{eq:UeffIdentity}: 
\begin{equation}\label{eq:MArelation}
    \partial_{k_\mu}M(\bm{k})^\dag=i A_\mathrm{band}^{(\mu)}(\bm{k}) M^\dag(\bm{k}).
\end{equation}
We take the derivative of $M^\dag(\bm{k} )M(\bm{k})$ and use Eq.~\eqref{eq:MArelation}:
\begin{equation}
\begin{aligned}
   &\partial_{k_\mu}(M^\dag(\bm{k})M(\bm{k}))=(\partial_{k_\mu} M^\dag(\bm{k}))M(\bm{k})+\\&+M^\dag(\bm{k})\partial_{k_\mu}(\bm{k})M(\bm{k})=i[A_\mathrm{band}^{(\mu)}(\bm{k}),M^\dag(\bm{k})M(\bm{k})].
\end{aligned}
\end{equation}
$M(\bm{0})$ is the identity matrix, therefore $M^\dag(\bm{k})M(\bm{k})$ has a zero derivative at the $\Gamma$-point. This means that $M^\dag(\bm{k})M(\bm{k})=I_2 +\mathcal{O}(k^2)$, from which $[M^\dag(\bm{k})M(\bm{k})]^{-1/2}=I_2 +\mathcal{O}(k^2)$. The transformation matrix $X(\bm{k})$ has the following form: 
\begin{equation}
    X(\bm{k})=M(\bm{k})+\mathcal{O}(k^2).  
\end{equation}
The Berry connection in the pseudospin basis becomes: 
\begin{equation}
\begin{aligned}
    A_\mathrm{pseudo}^{(\mu)}(\bm{k})&=M(\bm{k})A_\mathrm{band}^{(\mu)}M^\dag(\bm{k})+\\ &+iM(\bm{k})\partial_{k_\mu}M^\dag(\bm{k})+\mathcal{O}(k^2).
\end{aligned}
\end{equation}
Using Eq.~\eqref{eq:MArelation}, we can see that the two terms cancel out and yield: 
\begin{equation}
    A_\mathrm{pseudo}^{(\mu)}(\bm{k})=0+\mathcal{O}(k^2), 
\end{equation}
the linear contributions of the Berry connections become zero in the pseudospin representation.

\section{6-band Luttinger-Kohn-Bir-Pikus Hamiltonian}\label{app:Ham}
We use the 6-band Luttinger-Kohn-Bir-Pikus Hamiltonian \cite{luttinger1956quantum,bir1974symmetry,winkler2001spin,del2025fully} to model the system:
\begin{equation}\label{eq:HLKBP}
    H_\mathrm{LKBP}=H_\mathrm{LK}+H_\mathrm{BP}, 
\end{equation}
where $H_\mathrm{LK}$ is the Luttinger-Kohn Hamiltonian and $H_\mathrm{BP}$ is the Bir-Pikus Hamiltonian. $H_\mathrm{LK}$ contains the kinetic contributions: 
\begin{widetext}
\begin{equation}\label{eq:HLK}
  H_\mathrm{LK}=\begin{pmatrix}
      P+Q & -S & R & 0 & S/\sqrt{2} & -\sqrt{2}R \\ 
      -S^* & P-Q & 0 & R & \sqrt{2}Q & -\sqrt{3/2} S\\ 
      R^* & 0 & P-Q & S & -\sqrt{3/2} S^* & -\sqrt{2}Q \\
      0 & R^* & S^* & P+Q & \sqrt{2}R^* & S^*/\sqrt{2} \\ 
      S^*/\sqrt{2} & \sqrt{2} Q^* & -\sqrt{3/2}S & \sqrt{2}R & P+\Delta_\mathrm{SO} & 0 \\ 
      -\sqrt{2}R^* & -\sqrt{3/2}S^* & -\sqrt{2}Q^* & S/\sqrt{2} & 0 & P+\Delta_\mathrm{SO}
  \end{pmatrix},
\end{equation}
With elements: 
\begin{equation}
\begin{aligned}
    &P=\frac{\hbar^2}{2m_0}\gamma_1(k_x^2+k_y^2+k_z^2), \hspace{5mm} 
    Q=-\frac{\hbar^2}{2m_0}\gamma_2(2k_z^2-k_x^2-k_y^2), \\
    &R=\sqrt{3}\frac{\hbar^2}{2m_0}[-\gamma_2(k_x^2-k_y^2)+2i\gamma_3 k_xk_y], \hspace{5mm} S=\sqrt{3}\frac{\hbar^2}{m_0}\gamma_3(k_x-ik_y)k_z,  
\end{aligned}
\end{equation}
where $m_0$ is the bare electron mass, $\gamma_1$, $\gamma_2$ and $\gamma_3$ are the Luttinger-parameters, and $\Delta_\mathrm{SO}$ is the spin split-off gap. The Luttinger-Kohn Hamiltonian from Eq.~\eqref{eq:HLK} is represented on the basis of heavy holes, light holes and split-off holes, the states in order are $\ket{\frac{3}{2},\frac{3}{2}}$, $\ket{\frac{3}{2},\frac{1}{2}}$, $\ket{\frac{3}{2},-\frac{1}{2}}$, $\ket{\frac{3}{2},-\frac{3}{2}}$, $\ket{\frac{1}{2},\frac{1}{2}}$ and $\ket{\frac{1}{2},-\frac{1}{2}}$. Assuming biaxial strain, i.e. $\epsilon_{xy}=\epsilon_{xz}=\epsilon_{yz}=0$ and $\epsilon_{xx}=\epsilon_{yy}$, the Bir-Pikus Hamiltonian can be written as: 
\begin{equation}\label{eq:HBPmatrix}
    H_\mathrm{BP}=-a_v\mathrm{Tr\epsilon}-b_v(\epsilon_{xx}-\epsilon_{zz})\begin{pmatrix}
        1 & 0 & 0 & 0 & 0 & 0 \\ 
        0 & -1 & 0 & 0 & \sqrt{2} & 0 \\ 
        0 & 0 & -1 & 0 & 0 & -\sqrt{2} \\
        0 & 0 & 0 & 1 &  0& 0 \\
        0 & \sqrt{2} & 0 & 0 & 0 & 0 \\
        0 & 0 & -\sqrt{2} &0 & 0 & 0
    \end{pmatrix},
\end{equation}
where $a_v$ and $b_v$ are deformation potentials.

\end{widetext}

Magnetic fields are taken into account with the substitution: 
\begin{equation}
    \bm{k}=-i\nabla+\frac{e}{\hbar}\bm{A},
\end{equation}
where $\bm{A}$ is the vector potential. Additionally, we have to include the direct Zeeman coupling Hamiltonian $H_Z$: 
\begin{equation}
    H_Z=\begin{pmatrix}
        H_Z^\mathrm{(HL)} & H_Z^{(C)} \\[2mm] 
        H_Z^{(C)\dag} & H_Z^\mathrm{(SO)}
    \end{pmatrix}. 
\end{equation}
Here, $H_Z^\mathrm{(HL)}$ is the Zeeman term in the heavy-hole and light-hole block: 
\begin{equation}
    H_Z^\mathrm{(HL)}=2\mu_B (\kappa \bm{J}\cdot \bm{B}+q\bm{\mathcal{J}}\cdot \bm{B}), 
\end{equation}
where $\kappa$ and $q$ are constants, $\bm{J}=(J_x,J_y,J_z)$, and $\bm{\mathcal{J}}=(J_x^3,J_y^3,J_z^3)$, where $J_i$ are the 3/2 spin matrices. For split-off holes, $H_Z^\mathrm{(SO)}$ is:
\begin{equation}
    H_Z^\mathrm{(SO)}=
        2\mu_B \kappa \bm{\sigma}\cdot \bm{B}. 
\end{equation}
The off-diagonal block $H_Z^{(C)}$ can be written as:
\begin{equation}
    H_Z^{(C)}=3 \mu_B \kappa\bm{U}\cdot \bm{B}, 
\end{equation}
where $\bm{U}=(U_x,U_y,U_z)$ with the components: 
\begin{subequations}
\begin{align}
    U_x&=\frac{1}{3\sqrt{2}}\begin{pmatrix}
        -\sqrt{3} & 0 \\ 
        0 & -1\\
        1 & 0\\
        0 & \sqrt{3}
    \end{pmatrix},\\[2mm]
    U_y&=\frac{i}{3\sqrt{2}}\begin{pmatrix}
        \sqrt{3} & 0 \\
        0 & 1 \\ 
        1 & 0 \\
        0 & \sqrt{3}
    \end{pmatrix},\\[2mm]
    U_z&=\frac{\sqrt{2}}{3} \begin{pmatrix}
        0 & 0 \\ 
        1 & 0 \\ 
        0 & 1 \\ 
        0 & 0
    \end{pmatrix}.
\end{align}
\end{subequations}
For the planar heterostructures, we use the $\bm{A}=(zB_y,-zB_x,0)$ gauge for the vector potential. We neglect orbital contributions in the $z$-direction because we apply only small out-of-plane magnetic fields. 
For the core/shell nanowire, we use the $\bm{A}=(-yB_z/2,xB_z/2,yB_x-xB_y)$ gauge.
\section{Solution of the $z$-confinement for planar heterostructures}\label{app:conf}
Here, we describe the solution of the $z$-confinement problem for planar heterostructures. We follow the procedure presented in Refs.~\cite{del2024light,del2025fully,rotaru2025hole}. We note that the material parameters depend on position $z$, they do not commute with $k_z$. Considering zero magnetic fields and $k_x=k_y=0$, the quantum well has the following Hamiltonian \cite{del2023light,del2024light,del2025fully}: 
\begin{equation}\label{eq:QW}
    H_\mathrm{QW}=H_k+H_\mathrm{BP}+V_\mathrm{qw},
\end{equation}
where $H_k$ is the kinetic term: 
\begin{widetext}
\begin{equation}
H_k=\frac{\hbar^2}{2m_0}\begin{pmatrix}
    k_z\gamma_-k_z & 0 & 0 & 0 & 0 & 0\\
    0 & k_z\gamma_+k_z & 0 & 0 &-2\sqrt{2}k_z\gamma_2k_z&0  \\
    0 & 0 & k_z\gamma_+k_z & 0 & 0&2\sqrt{2}k_z\gamma_2k_z \\ 
    0 & 0 & 0 & k_z\gamma_-k_z & 0 & 0 \\ 
    0 & -2\sqrt{2}k_z\gamma_2k_z & 0 & 0 & k_z\gamma_1k_z & 0 \\
    0 & 0 & 2\sqrt{2}k_z\gamma_2k_z & 0 & 0 & k_z\gamma_1 k_z
\end{pmatrix}, 
\end{equation}
\end{widetext}
with $\gamma_\pm=\gamma_1\pm 2\gamma_2$. Here, we have position dependent Luttinger-parameters. In addition, $H_\mathrm{BP}$, appearing in Eq.~\eqref{eq:QW}, is given in Eq.~\eqref{eq:HBPmatrix}, and $V_\mathrm{qw}$ has the following form: 
\begin{equation}
    V_\mathrm{qw}=-E_V-eE_zz-V_\mathrm{SO}, 
\end{equation}
where $E_V$ is the valence band edge energy without spin-orbit coupling, and $V_\mathrm{SO}$ is:
\begin{equation}
    V_\mathrm{SO}=\frac{\Delta_\mathrm{SO}}{3}\mathrm{Diag}\left(1,1,1,1,-2,-2\right).
\end{equation}

Note that the heavy-hole part of the quantum well Hamiltonian from Eq.~\eqref{eq:QW} is decoupled from the rest, while light holes and split-off holes are coupled. This means that we get pure heavy-hole subbands and mixed light-hole and split-off subbands. The mixed states with not too large energy predominantly have light-hole contributions. Therefore, we refer to them simply as the light hole states, keeping in mind that they also contain split-off contributions. 

For planar heterostructures, the strain tensor elements can be calculated in the following way: 
\begin{equation}
    \epsilon_{xx}=\epsilon_{yy}=\frac{a_x}{a_\mathrm{Ge}}-1, \hspace{2mm} \epsilon_{zz}=-2\frac{c_{12}}{c_{11}}\epsilon_{xx},
\end{equation}
where $a_x$ is the lattice constant of $\mathrm{Si_{x}Ge_{1-x}}$ or $\mathrm{Sn_x Ge_{1-x}}$, $a_\mathrm{Ge}$ is the lattice constant of Ge, and $c_{12}$ and $c_{11}$ are elastic constants of Ge.  

We solve the Hamiltonian in Eq.~\eqref{eq:QW} numerically using finite difference discretization. We construct the Hamiltonian of the 2DHG $H_0(k_x,k_y)$ from Eq.~\eqref{eq:Hfull} by expanding the LKBP Hamiltonian, Eq.~\eqref{eq:HLKBP}, on the solutions of Eq.~\eqref{eq:QW}, keeping terms $[q,k_z]\neq 0, [\kappa,k_z]\neq0$, see Refs. \cite{del2024light,del2025fully,rotaru2025hole}. 
Contrary to previous studies that considered an infinite quantum well \cite{miserev2017dimensional, terrazos2021theory, sarkar2023electrical}, here we assume a finite quantum well, similarly to Refs. \cite{wang2024modeling, del2025fully}. For infinite quantum wells, the energy of the eigenstates scales quadratically with the number of the eigenstate, $E_\mathrm{inf}\propto n^2$, while for a triangular well, in the presence of an electric field, the energy scales approximately as $E_\mathrm{tri}\propto n^{2/3}$ \cite{ando1982electronic}. This means that achieving convergence when expanding the Hamiltonian from Eq.~\eqref{eq:HLKBP} is more difficult. We choose the thickness of the lower SiGe (SnGe) layer (see Fig.~\ref{fig:devices}) to be large enough to contain all eigenfunctions required for convergence, without introducing artificial boundary effects. 

We solve the full quantum dot Hamiltonian $H_0(k_x,k_y)+V(x,y)$ numerically by expanding it on a 2D oscillator basis, characterized by the effective mass $m^*$ and confinement frequencies $\omega_x$ and $\omega_y$. We obtain the effective mass by fitting a parabola to the lowest subband of the 2DHG, see Fig.~\ref{fig:dispersion2DHG}. 

\begin{figure}[h!]
\centering
\includegraphics[width=\columnwidth]{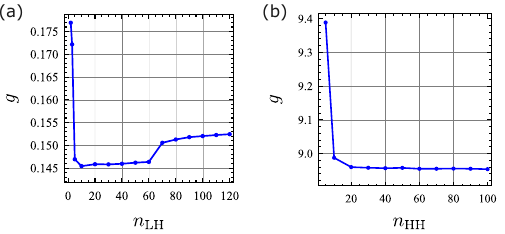}
\caption {\label{fig:converge} \small{Convergence of the in-plane $g$-factor with the basis size. The magnetic field is along the $x$ direction. (a) Convergence for a heavy-hole qubit, $n_\mathrm{HH}=60$, and $\hbar\omega_x=\hbar\omega_y=1$ meV. (b) Convergence for a light-hole qubit, $n_\mathrm{LH}=100$ and $\hbar\omega_x=\hbar\omega_y=1$ meV.}}
\refstepcounter{subfigure}\label{fig:convergea}
\refstepcounter{subfigure}\label{fig:convergeb}
\end{figure}
In Fig.~\ref{fig:converge} we show the convergence of the in-plane $g$-factor by including different numbers of eigenstates from the $z$-confinement problem, calculated using the full Hamiltonian. Throughout this work, we always consider the g-factors in the linear response regime, assuming weak magnetic fields. Figure~\ref{fig:convergea} shows the in-plane $g$-factor for the heavy hole spin qubit, see Fig.~\ref{fig:devicesa}. The magnetic field is applied in-plane, along the $x$ direction. The number of heavy-hole states is $n_\mathrm{HH}=60$, and the number of light-hole states is varied. We see that around $n_\mathrm{LH}=40$, the $g$-factor seems to be converged. However, states above $n_\mathrm{LH}=60$ have strong contributions to the $g$-factor. This is because around $n_\mathrm{LH}=60$-$70$, the light-hole eigenstates have a large overlap with the heavy-hole ground state. We note that convergence was observed in Ref.~\cite{wang2024modeling} around $n_\mathrm{LH}=50$. However, in our work, we use a 6-band Hamiltonian compared to the 4-band Hamiltonian used in~\cite{wang2024modeling}, furthermore, in our case, the quantum well is shallower. This means that it is possible that in~\cite{wang2024modeling} full convergence was achieved around $n_\mathrm{LH}=50$, while in our case we need to go above $n_\mathrm{LH}=70$. 

Figure~\ref{fig:convergeb} shows the $g$-factor along $x$ direction for a light-hole qubit from Fig.~\ref{fig:devicesb}. The number of light-hole states is kept fixed at $n_\mathrm{LH}=100$, and the number of heavy-hole states is varied. In both Figs.~\ref{fig:convergea} and~\ref{fig:convergeb}, we can see that the $g$-factors are not completely converged. However, our main focus is not to obtain the most precise results for the qubits but to compare the effective Hamiltonians with the full Hamiltonians. Furthermore, we neglected more important effects, such as inhomogeneous and shear strains \cite{abadillo2023hole}, interface effects \cite{rodriguez2023linear, sarkar2025effect}, and inhomogeneous or nonseparable confinements in the plane \cite{martinez2022hole}, which would have a much larger impact on the results compared to the small error we make stopping at $n_\mathrm{LH}=120$ in Fig.~\ref{fig:convergea} and at $n_\mathrm{HH}=100$ in Fig.~\ref{fig:convergeb}. Even the uncertainties of the different material parameters, e.g. the valence band offsets or the deformation potentials, are expected to introduce a larger uncertainty in the results.

\section{Solution of the $x$-$z$ confinement for core/shell nanowires}\label{App:cs}
In the main text, we also considered a Ge/Si core/shell nanowire. Compared with the SiGe/Ge/SiGe and SnGe/Ge heterostructures, the valence band offset between Si and Ge is much larger. Therefore, in this case, we assume hard-wall boundary conditions: 
\begin{equation}
    V_\mathrm{square}(x, z) =
\begin{cases}
  0, & |x| < \frac{L}{2} \text{ and } |z| < \frac{L}{2}, \\
  \infty, & \text{otherwise,}
\end{cases}
\end{equation}
and we use the following ansatz, which satisfies the boundary conditions:
\begin{equation}
f_{n_x, n_z}(x, z) = \frac{2 \sin \left[ n_x \pi \left( \frac{x}{L} + \frac{1}{2} \right) \right] \sin \left[ n_z \pi \left( \frac{z}{L} + \frac{1}{2} \right) \right]}{L}
\end{equation}
The nanowire is described using the Hamiltonian in Eq.~\eqref{eq:HLKBP}.
The Bir-Pikus Hamiltonian in Eq.~\eqref{eq:HBPmatrix} depends only on the strain tensor elements $\epsilon_{xx}=\epsilon_{zz}$ and $\epsilon_{yy}$, assuming no shear strain. The strain tensor elements depend on the parameter $\gamma$ \cite{kloeffel2011strong,kloeffel2014acoustic,kloeffel2018direct}. In our calculations we take $\gamma=0.4$, which results in $\epsilon_{xx}=-5.8\cdot 10^{-3}$ and $\epsilon_{yy}=-21.8 \cdot 10^{-3}$. 

\section{Material parameters}\label{App:parameters}
The material parameters used in the calculations can be found in Tab.~\ref{tab:parameters}. To calculate the material parameters of the alloys, in some cases we used a linear interpolation between Ge and Si (Sn). In the remaining cases, we used bowing parameters, e.g., the lattice constant can be calculated as: 
\begin{equation}
    a_\mathrm{Si_xGe_{1-x}}=a_\mathrm{Si} x+a_\mathrm{Ge}(1-x)-bx(1-x),
\end{equation}
where $b$ is the corresponding bowing parameter. The values in the columns $\mathrm{Si_x Ge_{1-x}}$ and $\mathrm{Sn_x Ge_{1-x}}$ in Tab.~\ref{tab:parameters} are the bowing parameters. If the value is missing, it means a linear interpolation was used to calculate the parameter of the respective alloy. 

The Luttinger parameters $\gamma_1$, $\gamma_2$, $\gamma_3$ and the Zeeman parameter $\kappa$ for a $\mathrm{Si_{0.2}Ge_{0.8}}$ alloy are \cite{fraj2007band}:
\begin{equation}\label{eq:gammaSiGe}
\begin{aligned}
    \gamma_{1,\mathrm{SiGe}}^{(0.2)}&=8.447, \hspace{3mm} \gamma_{2,\mathrm{SiGe}}^{(0.2)}=1.947, \\
    \gamma_{3,\mathrm{SiGe}}^{(0.2)}&=3.338, \hspace{3mm} \kappa_\mathrm{SiGe}^{(0.2)}=1.153. 
\end{aligned}
\end{equation}
For the $\mathrm{Sn_x Ge_{1-x}}$ alloy, the Luttinger-parameters can be calculated in the [0,0.2] range as found in Ref.~\cite{del2025fully}: 
\begin{equation}\label{eq:gammaSnGe}
    \gamma_i=\left(1-\frac{x}{0.2}\right)\gamma_{i,\mathrm{Ge}}+\frac{x}{0.2}\gamma_i^{(0.2)}-\frac{x}{0.2}\left(1-\frac{x}{0.2}\right)b_{\gamma_i}, 
\end{equation}
with $\gamma_1^{(0.2)}=29.2108$, $\gamma_2^{(0.2)}=12.2413$, $\gamma_3^{(0.3)}=13.7387$, $b_{\gamma_1}=20.3391$, $b_{\gamma_2}=9.6609$ and $b_{\gamma_3}=9.8187$.

\begin{widetext}
\begin{table}[H]
\centering
\caption{Material parameters and bowings used in the calculations.}\label{tab:parameters}
\renewcommand{\arraystretch}{1.3} 
\begin{tabular}{lccccc}
\toprule
 & \textbf{Ge} & \textbf{Si$_x$Ge$_{1-x}$}& \textbf{Si}& \textbf{Sn$_x$Ge$_{1-x}$} & \textbf{Sn} \\ 
\midrule

\multicolumn{4}{l}{Lattice constant} \\
$a_0$ \quad [\AA] & $5.65235^{\mathrm{a}}$ & $0.0188^{\mathrm{i}}$&$5.43^{\mathrm{i}}$&$-0.083^{\mathrm{g}}$ & $6.480117^{\mathrm{b}}$ \\
\hline

\multicolumn{4}{l}{Bulk band energies} \\
$E_V$ [eV] & 0 & & $-0.49^{\mathrm{j}*}$& &$0.69^{\mathrm{h}*}$ \\
$\Delta_\mathrm{SO}$ \quad [eV] & $0.290^{\mathrm{b}}$& & $0.044^{\mathrm{i}}$ & $-0.100^{\mathrm{g}}$ & $0.770^{\mathrm{f}}$ \\
\hline

\multicolumn{4}{l}{Elastic constants} \\
$c_{12}/c_{11}$ & $0.333^{\mathrm{b}}$ &- & - & - & -\\
\hline

\multicolumn{4}{l}{Deformation potentials} \\
$a_v$ \quad [eV] & $1.24^{\mathrm{c}}$ &- & -& - & - \\
$b_v$ \quad [eV] & $-2.86^{\mathrm{d}}$ &- & - & - &- \\
\hline

\multicolumn{4}{l}{Effective mass and spin parameters} \\
$\gamma_1$ & $13.38^{\mathrm{e}}$ & $\dagger$ & $\dagger$ & $\ddagger$ & $\ddagger$\\
$\gamma_2$ & $4.24^{\mathrm{e}}$ & $\dagger$ & $\dagger$ & $\ddagger$ & $\ddagger$\\
$\gamma_3$ & $5.69^{\mathrm{e}}$ & $\dagger$ & $\dagger$ & $\ddagger$ & $\ddagger$\\
$\kappa$ & $3.41^{\mathrm{f}}$& $\dagger$& $\dagger$& &$-11.84^{\mathrm{f}}$ \\
$q$ & $0.06^{\mathrm{e}}$& &$0.01^{\mathrm{f}}$ & & $0.30^{\mathrm{f}}$ \\
\bottomrule

\multicolumn{6}{l}{
  \begin{tabular}[t]{@{}l@{}}
  References: $^{\mathrm{a}}$:\cite{reeber1996thermal}, $^{\mathrm{b}}$:\cite{madelung2012semiconductors}, $^{\mathrm{c}}$:\cite{van1989band}, $^{\mathrm{d}}$:\cite{van1986theoretical}, 
  $^{\mathrm{e}}$:\cite{winkler2001spin}, $^{\mathrm{f}}$:\cite{lawaetz1971valence}, $^{\mathrm{g}}$:\cite{polak2017electronic},
  $^{\mathrm{h}}$:\cite{menendez2004type}, \\$^{\mathrm{i}}$:\cite{fischetti1996band},$^{\mathrm{j}}$:\cite{edward1990measurement} \\
  $^*$ Relative to Ge \\
  $^\dagger$ See Eq.~\eqref{eq:gammaSiGe}\\ 
  $\ddagger$ See Eq.~\eqref{eq:gammaSnGe}
  \end{tabular}
}
\end{tabular}
\end{table}
\end{widetext}

\bibliography{references} 

\begin{thebibliography}{95}%
\makeatletter
\providecommand \@ifxundefined [1]{%
 \@ifx{#1\undefined}
}%
\providecommand \@ifnum [1]{%
 \ifnum #1\expandafter \@firstoftwo
 \else \expandafter \@secondoftwo
 \fi
}%
\providecommand \@ifx [1]{%
 \ifx #1\expandafter \@firstoftwo
 \else \expandafter \@secondoftwo
 \fi
}%
\providecommand \natexlab [1]{#1}%
\providecommand \enquote  [1]{``#1''}%
\providecommand \bibnamefont  [1]{#1}%
\providecommand \bibfnamefont [1]{#1}%
\providecommand \citenamefont [1]{#1}%
\providecommand \href@noop [0]{\@secondoftwo}%
\providecommand \href [0]{\begingroup \@sanitize@url \@href}%
\providecommand \@href[1]{\@@startlink{#1}\@@href}%
\providecommand \@@href[1]{\endgroup#1\@@endlink}%
\providecommand \@sanitize@url [0]{\catcode `\\12\catcode `\$12\catcode `\&12\catcode `\#12\catcode `\^12\catcode `\_12\catcode `\%12\relax}%
\providecommand \@@startlink[1]{}%
\providecommand \@@endlink[0]{}%
\providecommand \url  [0]{\begingroup\@sanitize@url \@url }%
\providecommand \@url [1]{\endgroup\@href {#1}{\urlprefix }}%
\providecommand \urlprefix  [0]{URL }%
\providecommand \Eprint [0]{\href }%
\providecommand \doibase [0]{https://doi.org/}%
\providecommand \selectlanguage [0]{\@gobble}%
\providecommand \bibinfo  [0]{\@secondoftwo}%
\providecommand \bibfield  [0]{\@secondoftwo}%
\providecommand \translation [1]{[#1]}%
\providecommand \BibitemOpen [0]{}%
\providecommand \bibitemStop [0]{}%
\providecommand \bibitemNoStop [0]{.\EOS\space}%
\providecommand \EOS [0]{\spacefactor3000\relax}%
\providecommand \BibitemShut  [1]{\csname bibitem#1\endcsname}%
\let\auto@bib@innerbib\@empty
\bibitem [{\citenamefont {Loss}\ and\ \citenamefont {DiVincenzo}(1998)}]{PhysRevA.57.120}%
  \BibitemOpen
  \bibfield  {author} {\bibinfo {author} {\bibfnamefont {D.}~\bibnamefont {Loss}}\ and\ \bibinfo {author} {\bibfnamefont {D.~P.}\ \bibnamefont {DiVincenzo}},\ }\bibfield  {title} {\bibinfo {title} {Quantum computation with quantum dots},\ }\href {https://link.aps.org/doi/10.1103/PhysRevA.57.120} {\bibfield  {journal} {\bibinfo  {journal} {Phys. Rev. A}\ }\textbf {\bibinfo {volume} {57}},\ \bibinfo {pages} {120} (\bibinfo {year} {1998})}\BibitemShut {NoStop}%
\bibitem [{\citenamefont {Fang}\ \emph {et~al.}(2023)\citenamefont {Fang}, \citenamefont {Philippopoulos}, \citenamefont {Culcer}, \citenamefont {Coish},\ and\ \citenamefont {Chesi}}]{Fang_2023}%
  \BibitemOpen
  \bibfield  {author} {\bibinfo {author} {\bibfnamefont {Y.}~\bibnamefont {Fang}}, \bibinfo {author} {\bibfnamefont {P.}~\bibnamefont {Philippopoulos}}, \bibinfo {author} {\bibfnamefont {D.}~\bibnamefont {Culcer}}, \bibinfo {author} {\bibfnamefont {W.~A.}\ \bibnamefont {Coish}},\ and\ \bibinfo {author} {\bibfnamefont {S.}~\bibnamefont {Chesi}},\ }\bibfield  {title} {\bibinfo {title} {Recent advances in hole-spin qubits},\ }\href {https://doi.org/10.1088/2633-4356/acb87e} {\bibfield  {journal} {\bibinfo  {journal} {Materials for Quantum Technology}\ }\textbf {\bibinfo {volume} {3}},\ \bibinfo {pages} {012003} (\bibinfo {year} {2023})}\BibitemShut {NoStop}%
\bibitem [{\citenamefont {Burkard}\ \emph {et~al.}(2023)\citenamefont {Burkard}, \citenamefont {Ladd}, \citenamefont {Pan}, \citenamefont {Nichol},\ and\ \citenamefont {Petta}}]{RevModPhys.95.025003}%
  \BibitemOpen
  \bibfield  {author} {\bibinfo {author} {\bibfnamefont {G.}~\bibnamefont {Burkard}}, \bibinfo {author} {\bibfnamefont {T.~D.}\ \bibnamefont {Ladd}}, \bibinfo {author} {\bibfnamefont {A.}~\bibnamefont {Pan}}, \bibinfo {author} {\bibfnamefont {J.~M.}\ \bibnamefont {Nichol}},\ and\ \bibinfo {author} {\bibfnamefont {J.~R.}\ \bibnamefont {Petta}},\ }\bibfield  {title} {\bibinfo {title} {Semiconductor spin qubits},\ }\href {https://doi.org/10.1103/RevModPhys.95.025003} {\bibfield  {journal} {\bibinfo  {journal} {Rev. Mod. Phys.}\ }\textbf {\bibinfo {volume} {95}},\ \bibinfo {pages} {025003} (\bibinfo {year} {2023})}\BibitemShut {NoStop}%
\bibitem [{\citenamefont {Philips}\ \emph {et~al.}(2022)\citenamefont {Philips}, \citenamefont {M{\k{a}}dzik}, \citenamefont {Amitonov}, \citenamefont {de~Snoo}, \citenamefont {Russ}, \citenamefont {Kalhor}, \citenamefont {Volk}, \citenamefont {Lawrie}, \citenamefont {Brousse}, \citenamefont {Tryputen} \emph {et~al.}}]{philips2022universal}%
  \BibitemOpen
  \bibfield  {author} {\bibinfo {author} {\bibfnamefont {S.~G.}\ \bibnamefont {Philips}}, \bibinfo {author} {\bibfnamefont {M.~T.}\ \bibnamefont {M{\k{a}}dzik}}, \bibinfo {author} {\bibfnamefont {S.~V.}\ \bibnamefont {Amitonov}}, \bibinfo {author} {\bibfnamefont {S.~L.}\ \bibnamefont {de~Snoo}}, \bibinfo {author} {\bibfnamefont {M.}~\bibnamefont {Russ}}, \bibinfo {author} {\bibfnamefont {N.}~\bibnamefont {Kalhor}}, \bibinfo {author} {\bibfnamefont {C.}~\bibnamefont {Volk}}, \bibinfo {author} {\bibfnamefont {W.~I.}\ \bibnamefont {Lawrie}}, \bibinfo {author} {\bibfnamefont {D.}~\bibnamefont {Brousse}}, \bibinfo {author} {\bibfnamefont {L.}~\bibnamefont {Tryputen}}, \emph {et~al.},\ }\bibfield  {title} {\bibinfo {title} {Universal control of a six-qubit quantum processor in silicon},\ }\href {https://doi.org/10.1038/s41586-022-05117-x} {\bibfield  {journal} {\bibinfo  {journal} {Nature}\ }\textbf {\bibinfo {volume} {609}},\ \bibinfo {pages} {919} (\bibinfo {year} {2022})}\BibitemShut {NoStop}%
\bibitem [{\citenamefont {Hendrickx}\ \emph {et~al.}(2021)\citenamefont {Hendrickx}, \citenamefont {Lawrie}, \citenamefont {Russ}, \citenamefont {van Riggelen}, \citenamefont {de~Snoo}, \citenamefont {Schouten}, \citenamefont {Sammak}, \citenamefont {Scappucci},\ and\ \citenamefont {Veldhorst}}]{hendrickx2021four}%
  \BibitemOpen
  \bibfield  {author} {\bibinfo {author} {\bibfnamefont {N.~W.}\ \bibnamefont {Hendrickx}}, \bibinfo {author} {\bibfnamefont {W.~I.}\ \bibnamefont {Lawrie}}, \bibinfo {author} {\bibfnamefont {M.}~\bibnamefont {Russ}}, \bibinfo {author} {\bibfnamefont {F.}~\bibnamefont {van Riggelen}}, \bibinfo {author} {\bibfnamefont {S.~L.}\ \bibnamefont {de~Snoo}}, \bibinfo {author} {\bibfnamefont {R.~N.}\ \bibnamefont {Schouten}}, \bibinfo {author} {\bibfnamefont {A.}~\bibnamefont {Sammak}}, \bibinfo {author} {\bibfnamefont {G.}~\bibnamefont {Scappucci}},\ and\ \bibinfo {author} {\bibfnamefont {M.}~\bibnamefont {Veldhorst}},\ }\bibfield  {title} {\bibinfo {title} {A four-qubit germanium quantum processor},\ }\href {https://doi.org/10.1038/s41586-021-03332-6} {\bibfield  {journal} {\bibinfo  {journal} {Nature}\ }\textbf {\bibinfo {volume} {591}},\ \bibinfo {pages} {580} (\bibinfo {year} {2021})}\BibitemShut {NoStop}%
\bibitem [{\citenamefont {Borsoi}\ \emph {et~al.}(2024)\citenamefont {Borsoi}, \citenamefont {Hendrickx}, \citenamefont {John}, \citenamefont {Meyer}, \citenamefont {Motz}, \citenamefont {van Riggelen}, \citenamefont {Sammak}, \citenamefont {de~Snoo}, \citenamefont {Scappucci},\ and\ \citenamefont {Veldhorst}}]{borsoi2024shared}%
  \BibitemOpen
  \bibfield  {author} {\bibinfo {author} {\bibfnamefont {F.}~\bibnamefont {Borsoi}}, \bibinfo {author} {\bibfnamefont {N.~W.}\ \bibnamefont {Hendrickx}}, \bibinfo {author} {\bibfnamefont {V.}~\bibnamefont {John}}, \bibinfo {author} {\bibfnamefont {M.}~\bibnamefont {Meyer}}, \bibinfo {author} {\bibfnamefont {S.}~\bibnamefont {Motz}}, \bibinfo {author} {\bibfnamefont {F.}~\bibnamefont {van Riggelen}}, \bibinfo {author} {\bibfnamefont {A.}~\bibnamefont {Sammak}}, \bibinfo {author} {\bibfnamefont {S.~L.}\ \bibnamefont {de~Snoo}}, \bibinfo {author} {\bibfnamefont {G.}~\bibnamefont {Scappucci}},\ and\ \bibinfo {author} {\bibfnamefont {M.}~\bibnamefont {Veldhorst}},\ }\bibfield  {title} {\bibinfo {title} {Shared control of a 16 semiconductor quantum dot crossbar array},\ }\href {https://doi.org/10.1038/s41565-023-01491-3} {\bibfield  {journal} {\bibinfo  {journal} {Nature Nanotechnology}\ }\textbf {\bibinfo {volume} {19}},\ \bibinfo {pages} {21} (\bibinfo {year} {2024})}\BibitemShut {NoStop}%
\bibitem [{\citenamefont {Zhang}\ \emph {et~al.}(2025)\citenamefont {Zhang}, \citenamefont {Morozova}, \citenamefont {Rimbach-Russ}, \citenamefont {Jirovec}, \citenamefont {Hsiao}, \citenamefont {Fari{\~n}a}, \citenamefont {Wang}, \citenamefont {Oosterhout}, \citenamefont {Sammak}, \citenamefont {Scappucci} \emph {et~al.}}]{zhang2025universal}%
  \BibitemOpen
  \bibfield  {author} {\bibinfo {author} {\bibfnamefont {X.}~\bibnamefont {Zhang}}, \bibinfo {author} {\bibfnamefont {E.}~\bibnamefont {Morozova}}, \bibinfo {author} {\bibfnamefont {M.}~\bibnamefont {Rimbach-Russ}}, \bibinfo {author} {\bibfnamefont {D.}~\bibnamefont {Jirovec}}, \bibinfo {author} {\bibfnamefont {T.-K.}\ \bibnamefont {Hsiao}}, \bibinfo {author} {\bibfnamefont {P.~C.}\ \bibnamefont {Fari{\~n}a}}, \bibinfo {author} {\bibfnamefont {C.-A.}\ \bibnamefont {Wang}}, \bibinfo {author} {\bibfnamefont {S.~D.}\ \bibnamefont {Oosterhout}}, \bibinfo {author} {\bibfnamefont {A.}~\bibnamefont {Sammak}}, \bibinfo {author} {\bibfnamefont {G.}~\bibnamefont {Scappucci}}, \emph {et~al.},\ }\bibfield  {title} {\bibinfo {title} {Universal control of four singlet--triplet qubits},\ }\href {https://doi.org/10.1038/s41565-024-01817-9} {\bibfield  {journal} {\bibinfo  {journal} {Nature Nanotechnology}\ }\textbf {\bibinfo {volume} {20}},\ \bibinfo {pages} {209} (\bibinfo {year} {2025})}\BibitemShut {NoStop}%
\bibitem [{\citenamefont {Wang}\ \emph {et~al.}(2024{\natexlab{a}})\citenamefont {Wang}, \citenamefont {John}, \citenamefont {Tidjani}, \citenamefont {Yu}, \citenamefont {Ivlev}, \citenamefont {D{\'e}prez}, \citenamefont {van Riggelen-Doelman}, \citenamefont {Woods}, \citenamefont {Hendrickx}, \citenamefont {Lawrie} \emph {et~al.}}]{wang2024operating}%
  \BibitemOpen
  \bibfield  {author} {\bibinfo {author} {\bibfnamefont {C.-A.}\ \bibnamefont {Wang}}, \bibinfo {author} {\bibfnamefont {V.}~\bibnamefont {John}}, \bibinfo {author} {\bibfnamefont {H.}~\bibnamefont {Tidjani}}, \bibinfo {author} {\bibfnamefont {C.~X.}\ \bibnamefont {Yu}}, \bibinfo {author} {\bibfnamefont {A.~S.}\ \bibnamefont {Ivlev}}, \bibinfo {author} {\bibfnamefont {C.}~\bibnamefont {D{\'e}prez}}, \bibinfo {author} {\bibfnamefont {F.}~\bibnamefont {van Riggelen-Doelman}}, \bibinfo {author} {\bibfnamefont {B.~D.}\ \bibnamefont {Woods}}, \bibinfo {author} {\bibfnamefont {N.~W.}\ \bibnamefont {Hendrickx}}, \bibinfo {author} {\bibfnamefont {W.~I.}\ \bibnamefont {Lawrie}}, \emph {et~al.},\ }\bibfield  {title} {\bibinfo {title} {Operating semiconductor quantum processors with hopping spins},\ }\href {https://www.science.org/doi/abs/10.1126/science.ado5915} {\bibfield  {journal} {\bibinfo  {journal} {Science}\ }\textbf {\bibinfo {volume} {385}},\ \bibinfo {pages} {447} (\bibinfo {year}
  {2024}{\natexlab{a}})}\BibitemShut {NoStop}%
\bibitem [{\citenamefont {John}\ \emph {et~al.}(2025)\citenamefont {John}, \citenamefont {Yu}, \citenamefont {van Straaten}, \citenamefont {Rodr{\'\i}guez-Mena}, \citenamefont {Rodr{\'\i}guez}, \citenamefont {Oosterhout}, \citenamefont {Stehouwer}, \citenamefont {Scappucci}, \citenamefont {Rimbach-Russ}, \citenamefont {Bosco} \emph {et~al.}}]{john2024two}%
  \BibitemOpen
  \bibfield  {author} {\bibinfo {author} {\bibfnamefont {V.}~\bibnamefont {John}}, \bibinfo {author} {\bibfnamefont {C.~X.}\ \bibnamefont {Yu}}, \bibinfo {author} {\bibfnamefont {B.}~\bibnamefont {van Straaten}}, \bibinfo {author} {\bibfnamefont {E.~A.}\ \bibnamefont {Rodr{\'\i}guez-Mena}}, \bibinfo {author} {\bibfnamefont {M.}~\bibnamefont {Rodr{\'\i}guez}}, \bibinfo {author} {\bibfnamefont {S.~D.}\ \bibnamefont {Oosterhout}}, \bibinfo {author} {\bibfnamefont {L.~E.}\ \bibnamefont {Stehouwer}}, \bibinfo {author} {\bibfnamefont {G.}~\bibnamefont {Scappucci}}, \bibinfo {author} {\bibfnamefont {M.}~\bibnamefont {Rimbach-Russ}}, \bibinfo {author} {\bibfnamefont {S.}~\bibnamefont {Bosco}}, \emph {et~al.},\ }\bibfield  {title} {\bibinfo {title} {Robust and localised control of a 10-spin qubit array in germanium},\ }\href {https://doi.org/10.1038/s41467-025-65577-3} {\bibfield  {journal} {\bibinfo  {journal} {Nature Communications}\ }\textbf {\bibinfo {volume} {16}},\ \bibinfo {pages} {10560} (\bibinfo {year}
  {2025})}\BibitemShut {NoStop}%
\bibitem [{\citenamefont {George}\ \emph {et~al.}(2025)\citenamefont {George}, \citenamefont {Mądzik}, \citenamefont {Henry}, \citenamefont {Wagner}, \citenamefont {Islam}, \citenamefont {Borjans}, \citenamefont {Connors}, \citenamefont {Corrigan}, \citenamefont {Curry}, \citenamefont {Harper} \emph {et~al.}}]{george202412}%
  \BibitemOpen
  \bibfield  {author} {\bibinfo {author} {\bibfnamefont {H.~C.}\ \bibnamefont {George}}, \bibinfo {author} {\bibfnamefont {M.~T.}\ \bibnamefont {Mądzik}}, \bibinfo {author} {\bibfnamefont {E.~M.}\ \bibnamefont {Henry}}, \bibinfo {author} {\bibfnamefont {A.~J.}\ \bibnamefont {Wagner}}, \bibinfo {author} {\bibfnamefont {M.~M.}\ \bibnamefont {Islam}}, \bibinfo {author} {\bibfnamefont {F.}~\bibnamefont {Borjans}}, \bibinfo {author} {\bibfnamefont {E.~J.}\ \bibnamefont {Connors}}, \bibinfo {author} {\bibfnamefont {J.}~\bibnamefont {Corrigan}}, \bibinfo {author} {\bibfnamefont {M.}~\bibnamefont {Curry}}, \bibinfo {author} {\bibfnamefont {M.~K.}\ \bibnamefont {Harper}}, \emph {et~al.},\ }\bibfield  {title} {\bibinfo {title} {12-spin-qubit arrays fabricated on a 300 mm semiconductor manufacturing line},\ }\href {https://doi.org/10.1021/acs.nanolett.4c05205} {\bibfield  {journal} {\bibinfo  {journal} {Nano Letters}\ }\textbf {\bibinfo {volume} {25}},\ \bibinfo {pages} {793} (\bibinfo {year} {2025})}\BibitemShut
  {NoStop}%
\bibitem [{\citenamefont {Han~Lim}\ \emph {et~al.}(2025)\citenamefont {Han~Lim}, \citenamefont {Tanttu}, \citenamefont {Youn}, \citenamefont {Huang}, \citenamefont {Serrano}, \citenamefont {Dickie}, \citenamefont {Yianni}, \citenamefont {Hudson}, \citenamefont {Escott}, \citenamefont {Yang} \emph {et~al.}}]{lim20242x2}%
  \BibitemOpen
  \bibfield  {author} {\bibinfo {author} {\bibfnamefont {W.}~\bibnamefont {Han~Lim}}, \bibinfo {author} {\bibfnamefont {T.}~\bibnamefont {Tanttu}}, \bibinfo {author} {\bibfnamefont {T.}~\bibnamefont {Youn}}, \bibinfo {author} {\bibfnamefont {J.~Y.}\ \bibnamefont {Huang}}, \bibinfo {author} {\bibfnamefont {S.}~\bibnamefont {Serrano}}, \bibinfo {author} {\bibfnamefont {A.}~\bibnamefont {Dickie}}, \bibinfo {author} {\bibfnamefont {S.}~\bibnamefont {Yianni}}, \bibinfo {author} {\bibfnamefont {F.~E.}\ \bibnamefont {Hudson}}, \bibinfo {author} {\bibfnamefont {C.~C.}\ \bibnamefont {Escott}}, \bibinfo {author} {\bibfnamefont {C.~H.}\ \bibnamefont {Yang}}, \emph {et~al.},\ }\bibfield  {title} {\bibinfo {title} {A 2$\times$2 quantum dot array in silicon with fully tunable pairwise interdot coupling},\ }\href {https://doi.org/10.1021/acs.nanolett.4c06264} {\bibfield  {journal} {\bibinfo  {journal} {Nano Letters}\ }\textbf {\bibinfo {volume} {25}},\ \bibinfo {pages} {10263} (\bibinfo {year} {2025})}\BibitemShut {NoStop}%
\bibitem [{\citenamefont {Veldhorst}\ \emph {et~al.}(2014)\citenamefont {Veldhorst}, \citenamefont {Hwang}, \citenamefont {Yang}, \citenamefont {Leenstra}, \citenamefont {de~Ronde}, \citenamefont {Dehollain}, \citenamefont {Muhonen}, \citenamefont {Hudson}, \citenamefont {Itoh}, \citenamefont {Morello} \emph {et~al.}}]{veldhorst2014addressable}%
  \BibitemOpen
  \bibfield  {author} {\bibinfo {author} {\bibfnamefont {M.}~\bibnamefont {Veldhorst}}, \bibinfo {author} {\bibfnamefont {J.}~\bibnamefont {Hwang}}, \bibinfo {author} {\bibfnamefont {C.}~\bibnamefont {Yang}}, \bibinfo {author} {\bibfnamefont {A.}~\bibnamefont {Leenstra}}, \bibinfo {author} {\bibfnamefont {B.}~\bibnamefont {de~Ronde}}, \bibinfo {author} {\bibfnamefont {J.}~\bibnamefont {Dehollain}}, \bibinfo {author} {\bibfnamefont {J.}~\bibnamefont {Muhonen}}, \bibinfo {author} {\bibfnamefont {F.}~\bibnamefont {Hudson}}, \bibinfo {author} {\bibfnamefont {K.~M.}\ \bibnamefont {Itoh}}, \bibinfo {author} {\bibfnamefont {A.~t.}\ \bibnamefont {Morello}}, \emph {et~al.},\ }\bibfield  {title} {\bibinfo {title} {An addressable quantum dot qubit with fault-tolerant control-fidelity},\ }\href {https://doi.org/10.1038/nnano.2014.216} {\bibfield  {journal} {\bibinfo  {journal} {Nature Nanotechnology}\ }\textbf {\bibinfo {volume} {9}},\ \bibinfo {pages} {981} (\bibinfo {year} {2014})}\BibitemShut {NoStop}%
\bibitem [{\citenamefont {Yoneda}\ \emph {et~al.}(2018)\citenamefont {Yoneda}, \citenamefont {Takeda}, \citenamefont {Otsuka}, \citenamefont {Nakajima}, \citenamefont {Delbecq}, \citenamefont {Allison}, \citenamefont {Honda}, \citenamefont {Kodera}, \citenamefont {Oda}, \citenamefont {Hoshi} \emph {et~al.}}]{yoneda2018quantum}%
  \BibitemOpen
  \bibfield  {author} {\bibinfo {author} {\bibfnamefont {J.}~\bibnamefont {Yoneda}}, \bibinfo {author} {\bibfnamefont {K.}~\bibnamefont {Takeda}}, \bibinfo {author} {\bibfnamefont {T.}~\bibnamefont {Otsuka}}, \bibinfo {author} {\bibfnamefont {T.}~\bibnamefont {Nakajima}}, \bibinfo {author} {\bibfnamefont {M.~R.}\ \bibnamefont {Delbecq}}, \bibinfo {author} {\bibfnamefont {G.}~\bibnamefont {Allison}}, \bibinfo {author} {\bibfnamefont {T.}~\bibnamefont {Honda}}, \bibinfo {author} {\bibfnamefont {T.}~\bibnamefont {Kodera}}, \bibinfo {author} {\bibfnamefont {S.}~\bibnamefont {Oda}}, \bibinfo {author} {\bibfnamefont {Y.}~\bibnamefont {Hoshi}}, \emph {et~al.},\ }\bibfield  {title} {\bibinfo {title} {A quantum-dot spin qubit with coherence limited by charge noise and fidelity higher than 99.9\%},\ }\href {https://doi.org/10.1038/s41565-017-0014-x} {\bibfield  {journal} {\bibinfo  {journal} {Nature nanotechnology}\ }\textbf {\bibinfo {volume} {13}},\ \bibinfo {pages} {102} (\bibinfo {year} {2018})}\BibitemShut
  {NoStop}%
\bibitem [{\citenamefont {Lawrie}\ \emph {et~al.}(2023)\citenamefont {Lawrie}, \citenamefont {Rimbach-Russ}, \citenamefont {Riggelen}, \citenamefont {Hendrickx}, \citenamefont {Snoo}, \citenamefont {Sammak}, \citenamefont {Scappucci}, \citenamefont {Helsen},\ and\ \citenamefont {Veldhorst}}]{lawrie2023simultaneous}%
  \BibitemOpen
  \bibfield  {author} {\bibinfo {author} {\bibfnamefont {W.}~\bibnamefont {Lawrie}}, \bibinfo {author} {\bibfnamefont {M.}~\bibnamefont {Rimbach-Russ}}, \bibinfo {author} {\bibfnamefont {F.~v.}\ \bibnamefont {Riggelen}}, \bibinfo {author} {\bibfnamefont {N.}~\bibnamefont {Hendrickx}}, \bibinfo {author} {\bibfnamefont {S.~d.}\ \bibnamefont {Snoo}}, \bibinfo {author} {\bibfnamefont {A.}~\bibnamefont {Sammak}}, \bibinfo {author} {\bibfnamefont {G.}~\bibnamefont {Scappucci}}, \bibinfo {author} {\bibfnamefont {J.}~\bibnamefont {Helsen}},\ and\ \bibinfo {author} {\bibfnamefont {M.}~\bibnamefont {Veldhorst}},\ }\bibfield  {title} {\bibinfo {title} {Simultaneous single-qubit driving of semiconductor spin qubits at the fault-tolerant threshold},\ }\href {https://doi.org/10.1038/s41467-023-39334-3} {\bibfield  {journal} {\bibinfo  {journal} {Nature Communications}\ }\textbf {\bibinfo {volume} {14}},\ \bibinfo {pages} {3617} (\bibinfo {year} {2023})}\BibitemShut {NoStop}%
\bibitem [{\citenamefont {Mills}\ \emph {et~al.}(2022{\natexlab{a}})\citenamefont {Mills}, \citenamefont {Guinn}, \citenamefont {Feldman}, \citenamefont {Sigillito}, \citenamefont {Gullans}, \citenamefont {Rakher}, \citenamefont {Kerckhoff}, \citenamefont {Jackson},\ and\ \citenamefont {Petta}}]{mills2022high}%
  \BibitemOpen
  \bibfield  {author} {\bibinfo {author} {\bibfnamefont {A.}~\bibnamefont {Mills}}, \bibinfo {author} {\bibfnamefont {C.}~\bibnamefont {Guinn}}, \bibinfo {author} {\bibfnamefont {M.}~\bibnamefont {Feldman}}, \bibinfo {author} {\bibfnamefont {A.}~\bibnamefont {Sigillito}}, \bibinfo {author} {\bibfnamefont {M.}~\bibnamefont {Gullans}}, \bibinfo {author} {\bibfnamefont {M.}~\bibnamefont {Rakher}}, \bibinfo {author} {\bibfnamefont {J.}~\bibnamefont {Kerckhoff}}, \bibinfo {author} {\bibfnamefont {C.}~\bibnamefont {Jackson}},\ and\ \bibinfo {author} {\bibfnamefont {J.}~\bibnamefont {Petta}},\ }\bibfield  {title} {\bibinfo {title} {High-fidelity state preparation, quantum control, and readout of an isotopically enriched silicon spin qubit},\ }\href {https://link.aps.org/doi/10.1103/PhysRevApplied.18.064028} {\bibfield  {journal} {\bibinfo  {journal} {Phys. Rev. Appl.}\ }\textbf {\bibinfo {volume} {18}},\ \bibinfo {pages} {064028} (\bibinfo {year} {2022}{\natexlab{a}})}\BibitemShut {NoStop}%
\bibitem [{\citenamefont {Huang}\ \emph {et~al.}(2024)\citenamefont {Huang}, \citenamefont {Su}, \citenamefont {Lim}, \citenamefont {Feng}, \citenamefont {van Straaten}, \citenamefont {Severin}, \citenamefont {Gilbert}, \citenamefont {Dumoulin~Stuyck}, \citenamefont {Tanttu}, \citenamefont {Serrano} \emph {et~al.}}]{huang2024high}%
  \BibitemOpen
  \bibfield  {author} {\bibinfo {author} {\bibfnamefont {J.~Y.}\ \bibnamefont {Huang}}, \bibinfo {author} {\bibfnamefont {R.~Y.}\ \bibnamefont {Su}}, \bibinfo {author} {\bibfnamefont {W.~H.}\ \bibnamefont {Lim}}, \bibinfo {author} {\bibfnamefont {M.}~\bibnamefont {Feng}}, \bibinfo {author} {\bibfnamefont {B.}~\bibnamefont {van Straaten}}, \bibinfo {author} {\bibfnamefont {B.}~\bibnamefont {Severin}}, \bibinfo {author} {\bibfnamefont {W.}~\bibnamefont {Gilbert}}, \bibinfo {author} {\bibfnamefont {N.}~\bibnamefont {Dumoulin~Stuyck}}, \bibinfo {author} {\bibfnamefont {T.}~\bibnamefont {Tanttu}}, \bibinfo {author} {\bibfnamefont {S.}~\bibnamefont {Serrano}}, \emph {et~al.},\ }\bibfield  {title} {\bibinfo {title} {High-fidelity spin qubit operation and algorithmic initialization above 1 {K}},\ }\href {https://doi.org/10.1038/s41586-024-07160-2} {\bibfield  {journal} {\bibinfo  {journal} {Nature}\ }\textbf {\bibinfo {volume} {627}},\ \bibinfo {pages} {772} (\bibinfo {year} {2024})}\BibitemShut {NoStop}%
\bibitem [{\citenamefont {Wu}\ \emph {et~al.}(2025)\citenamefont {Wu}, \citenamefont {Camenzind}, \citenamefont {B{\"u}tler}, \citenamefont {Jin}, \citenamefont {Noiri}, \citenamefont {Takeda}, \citenamefont {Nakajima}, \citenamefont {Kobayashi}, \citenamefont {Scappucci}, \citenamefont {Goan} \emph {et~al.}}]{wu2025simultaneous}%
  \BibitemOpen
  \bibfield  {author} {\bibinfo {author} {\bibfnamefont {Y.-H.}\ \bibnamefont {Wu}}, \bibinfo {author} {\bibfnamefont {L.~C.}\ \bibnamefont {Camenzind}}, \bibinfo {author} {\bibfnamefont {P.}~\bibnamefont {B{\"u}tler}}, \bibinfo {author} {\bibfnamefont {I.~K.}\ \bibnamefont {Jin}}, \bibinfo {author} {\bibfnamefont {A.}~\bibnamefont {Noiri}}, \bibinfo {author} {\bibfnamefont {K.}~\bibnamefont {Takeda}}, \bibinfo {author} {\bibfnamefont {T.}~\bibnamefont {Nakajima}}, \bibinfo {author} {\bibfnamefont {T.}~\bibnamefont {Kobayashi}}, \bibinfo {author} {\bibfnamefont {G.}~\bibnamefont {Scappucci}}, \bibinfo {author} {\bibfnamefont {H.-S.}\ \bibnamefont {Goan}}, \emph {et~al.},\ }\bibfield  {title} {\bibinfo {title} {Simultaneous high-fidelity single-qubit gates in a spin qubit array},\ }\href {https://arxiv.org/abs/2507.11918} {\bibfield  {journal} {\bibinfo  {journal} {arXiv preprint arXiv:2507.11918}\ } (\bibinfo {year} {2025})}\BibitemShut {NoStop}%
\bibitem [{\citenamefont {Xue}\ \emph {et~al.}(2022)\citenamefont {Xue}, \citenamefont {Russ}, \citenamefont {Samkharadze}, \citenamefont {Undseth}, \citenamefont {Sammak}, \citenamefont {Scappucci},\ and\ \citenamefont {Vandersypen}}]{xue2022quantum}%
  \BibitemOpen
  \bibfield  {author} {\bibinfo {author} {\bibfnamefont {X.}~\bibnamefont {Xue}}, \bibinfo {author} {\bibfnamefont {M.}~\bibnamefont {Russ}}, \bibinfo {author} {\bibfnamefont {N.}~\bibnamefont {Samkharadze}}, \bibinfo {author} {\bibfnamefont {B.}~\bibnamefont {Undseth}}, \bibinfo {author} {\bibfnamefont {A.}~\bibnamefont {Sammak}}, \bibinfo {author} {\bibfnamefont {G.}~\bibnamefont {Scappucci}},\ and\ \bibinfo {author} {\bibfnamefont {L.~M.}\ \bibnamefont {Vandersypen}},\ }\bibfield  {title} {\bibinfo {title} {Quantum logic with spin qubits crossing the surface code threshold},\ }\href {https://doi.org/10.1038/s41586-021-04273-w} {\bibfield  {journal} {\bibinfo  {journal} {Nature}\ }\textbf {\bibinfo {volume} {601}},\ \bibinfo {pages} {343} (\bibinfo {year} {2022})}\BibitemShut {NoStop}%
\bibitem [{\citenamefont {Mills}\ \emph {et~al.}(2022{\natexlab{b}})\citenamefont {Mills}, \citenamefont {Guinn}, \citenamefont {Gullans}, \citenamefont {Sigillito}, \citenamefont {Feldman}, \citenamefont {Nielsen},\ and\ \citenamefont {Petta}}]{mills2022two}%
  \BibitemOpen
  \bibfield  {author} {\bibinfo {author} {\bibfnamefont {A.~R.}\ \bibnamefont {Mills}}, \bibinfo {author} {\bibfnamefont {C.~R.}\ \bibnamefont {Guinn}}, \bibinfo {author} {\bibfnamefont {M.~J.}\ \bibnamefont {Gullans}}, \bibinfo {author} {\bibfnamefont {A.~J.}\ \bibnamefont {Sigillito}}, \bibinfo {author} {\bibfnamefont {M.~M.}\ \bibnamefont {Feldman}}, \bibinfo {author} {\bibfnamefont {E.}~\bibnamefont {Nielsen}},\ and\ \bibinfo {author} {\bibfnamefont {J.~R.}\ \bibnamefont {Petta}},\ }\bibfield  {title} {\bibinfo {title} {Two-qubit silicon quantum processor with operation fidelity exceeding 99\%},\ }\href {https://www.science.org/doi/abs/10.1126/sciadv.abn5130} {\bibfield  {journal} {\bibinfo  {journal} {Science Advances}\ }\textbf {\bibinfo {volume} {8}},\ \bibinfo {pages} {eabn5130} (\bibinfo {year} {2022}{\natexlab{b}})}\BibitemShut {NoStop}%
\bibitem [{\citenamefont {Noiri}\ \emph {et~al.}(2022)\citenamefont {Noiri}, \citenamefont {Takeda}, \citenamefont {Nakajima}, \citenamefont {Kobayashi}, \citenamefont {Sammak}, \citenamefont {Scappucci},\ and\ \citenamefont {Tarucha}}]{noiri2022fast}%
  \BibitemOpen
  \bibfield  {author} {\bibinfo {author} {\bibfnamefont {A.}~\bibnamefont {Noiri}}, \bibinfo {author} {\bibfnamefont {K.}~\bibnamefont {Takeda}}, \bibinfo {author} {\bibfnamefont {T.}~\bibnamefont {Nakajima}}, \bibinfo {author} {\bibfnamefont {T.}~\bibnamefont {Kobayashi}}, \bibinfo {author} {\bibfnamefont {A.}~\bibnamefont {Sammak}}, \bibinfo {author} {\bibfnamefont {G.}~\bibnamefont {Scappucci}},\ and\ \bibinfo {author} {\bibfnamefont {S.}~\bibnamefont {Tarucha}},\ }\bibfield  {title} {\bibinfo {title} {Fast universal quantum gate above the fault-tolerance threshold in silicon},\ }\href {https://doi.org/10.1038/s41586-021-04182-y} {\bibfield  {journal} {\bibinfo  {journal} {Nature}\ }\textbf {\bibinfo {volume} {601}},\ \bibinfo {pages} {338} (\bibinfo {year} {2022})}\BibitemShut {NoStop}%
\bibitem [{\citenamefont {Zwerver}\ \emph {et~al.}(2022)\citenamefont {Zwerver}, \citenamefont {Kr{\"a}henmann}, \citenamefont {Watson}, \citenamefont {Lampert}, \citenamefont {George}, \citenamefont {Pillarisetty}, \citenamefont {Bojarski}, \citenamefont {Amin}, \citenamefont {Amitonov}, \citenamefont {Boter} \emph {et~al.}}]{zwerver2022qubits}%
  \BibitemOpen
  \bibfield  {author} {\bibinfo {author} {\bibfnamefont {A.}~\bibnamefont {Zwerver}}, \bibinfo {author} {\bibfnamefont {T.}~\bibnamefont {Kr{\"a}henmann}}, \bibinfo {author} {\bibfnamefont {T.}~\bibnamefont {Watson}}, \bibinfo {author} {\bibfnamefont {L.}~\bibnamefont {Lampert}}, \bibinfo {author} {\bibfnamefont {H.~C.}\ \bibnamefont {George}}, \bibinfo {author} {\bibfnamefont {R.}~\bibnamefont {Pillarisetty}}, \bibinfo {author} {\bibfnamefont {S.}~\bibnamefont {Bojarski}}, \bibinfo {author} {\bibfnamefont {P.}~\bibnamefont {Amin}}, \bibinfo {author} {\bibfnamefont {S.}~\bibnamefont {Amitonov}}, \bibinfo {author} {\bibfnamefont {J.}~\bibnamefont {Boter}}, \emph {et~al.},\ }\bibfield  {title} {\bibinfo {title} {Qubits made by advanced semiconductor manufacturing},\ }\href {https://doi.org/10.1038/s41928-022-00727-9} {\bibfield  {journal} {\bibinfo  {journal} {Nature Electronics}\ }\textbf {\bibinfo {volume} {5}},\ \bibinfo {pages} {184} (\bibinfo {year} {2022})}\BibitemShut {NoStop}%
\bibitem [{\citenamefont {Steinacker}\ \emph {et~al.}(2025)\citenamefont {Steinacker}, \citenamefont {Dumoulin~Stuyck}, \citenamefont {Lim}, \citenamefont {Tanttu}, \citenamefont {Feng}, \citenamefont {Serrano}, \citenamefont {Nickl}, \citenamefont {Candido}, \citenamefont {Cifuentes}, \citenamefont {Vahapoglu} \emph {et~al.}}]{steinacker2025industry}%
  \BibitemOpen
  \bibfield  {author} {\bibinfo {author} {\bibfnamefont {P.}~\bibnamefont {Steinacker}}, \bibinfo {author} {\bibfnamefont {N.}~\bibnamefont {Dumoulin~Stuyck}}, \bibinfo {author} {\bibfnamefont {W.~H.}\ \bibnamefont {Lim}}, \bibinfo {author} {\bibfnamefont {T.}~\bibnamefont {Tanttu}}, \bibinfo {author} {\bibfnamefont {M.}~\bibnamefont {Feng}}, \bibinfo {author} {\bibfnamefont {S.}~\bibnamefont {Serrano}}, \bibinfo {author} {\bibfnamefont {A.}~\bibnamefont {Nickl}}, \bibinfo {author} {\bibfnamefont {M.}~\bibnamefont {Candido}}, \bibinfo {author} {\bibfnamefont {J.~D.}\ \bibnamefont {Cifuentes}}, \bibinfo {author} {\bibfnamefont {E.}~\bibnamefont {Vahapoglu}}, \emph {et~al.},\ }\bibfield  {title} {\bibinfo {title} {Industry-compatible silicon spin-qubit unit cells exceeding 99\% fidelity},\ }\href {https://doi.org/10.1038/s41586-025-09531-9} {\bibfield  {journal} {\bibinfo  {journal} {Nature}\ ,\ \bibinfo {pages} {1}} (\bibinfo {year} {2025})}\BibitemShut {NoStop}%
\bibitem [{\citenamefont {Eggli}\ \emph {et~al.}(2025)\citenamefont {Eggli}, \citenamefont {Patlatiuk}, \citenamefont {Kelly}, \citenamefont {Orekhov}, \citenamefont {Salis}, \citenamefont {Warburton}, \citenamefont {Zumb{\"u}hl},\ and\ \citenamefont {Kuhlmann}}]{eggli2025coupling}%
  \BibitemOpen
  \bibfield  {author} {\bibinfo {author} {\bibfnamefont {R.~S.}\ \bibnamefont {Eggli}}, \bibinfo {author} {\bibfnamefont {T.}~\bibnamefont {Patlatiuk}}, \bibinfo {author} {\bibfnamefont {E.~G.}\ \bibnamefont {Kelly}}, \bibinfo {author} {\bibfnamefont {A.}~\bibnamefont {Orekhov}}, \bibinfo {author} {\bibfnamefont {G.}~\bibnamefont {Salis}}, \bibinfo {author} {\bibfnamefont {R.~J.}\ \bibnamefont {Warburton}}, \bibinfo {author} {\bibfnamefont {D.~M.}\ \bibnamefont {Zumb{\"u}hl}},\ and\ \bibinfo {author} {\bibfnamefont {A.~V.}\ \bibnamefont {Kuhlmann}},\ }\bibfield  {title} {\bibinfo {title} {Coupling a high-{Q} resonator to a spin qubit with all-electrical control},\ }\href {https://link.aps.org/doi/10.1103/PhysRevResearch.7.013197} {\bibfield  {journal} {\bibinfo  {journal} {Physical Review Research}\ }\textbf {\bibinfo {volume} {7}},\ \bibinfo {pages} {013197} (\bibinfo {year} {2025})}\BibitemShut {NoStop}%
\bibitem [{\citenamefont {B{\o}ttcher}\ \emph {et~al.}(2022)\citenamefont {B{\o}ttcher}, \citenamefont {Harvey}, \citenamefont {Fallahi}, \citenamefont {Gardner}, \citenamefont {Manfra}, \citenamefont {Vool}, \citenamefont {Bartlett},\ and\ \citenamefont {Yacoby}}]{bottcher2022parametric}%
  \BibitemOpen
  \bibfield  {author} {\bibinfo {author} {\bibfnamefont {C.}~\bibnamefont {B{\o}ttcher}}, \bibinfo {author} {\bibfnamefont {S.}~\bibnamefont {Harvey}}, \bibinfo {author} {\bibfnamefont {S.}~\bibnamefont {Fallahi}}, \bibinfo {author} {\bibfnamefont {G.}~\bibnamefont {Gardner}}, \bibinfo {author} {\bibfnamefont {M.}~\bibnamefont {Manfra}}, \bibinfo {author} {\bibfnamefont {U.}~\bibnamefont {Vool}}, \bibinfo {author} {\bibfnamefont {S.}~\bibnamefont {Bartlett}},\ and\ \bibinfo {author} {\bibfnamefont {A.}~\bibnamefont {Yacoby}},\ }\bibfield  {title} {\bibinfo {title} {Parametric longitudinal coupling between a high-impedance superconducting resonator and a semiconductor quantum dot singlet-triplet spin qubit},\ }\href {https://doi.org/10.1038/s41467-022-32236-w} {\bibfield  {journal} {\bibinfo  {journal} {Nature Communications}\ }\textbf {\bibinfo {volume} {13}},\ \bibinfo {pages} {4773} (\bibinfo {year} {2022})}\BibitemShut {NoStop}%
\bibitem [{\citenamefont {Noirot}\ \emph {et~al.}(2026)\citenamefont {Noirot}, \citenamefont {Yu}, \citenamefont {Abadillo-Uriel}, \citenamefont {Dumur}, \citenamefont {Niebojewski}, \citenamefont {Bertrand}, \citenamefont {Maurand},\ and\ \citenamefont {Zihlmann}}]{noirot2025coherence}%
  \BibitemOpen
  \bibfield  {author} {\bibinfo {author} {\bibfnamefont {L.}~\bibnamefont {Noirot}}, \bibinfo {author} {\bibfnamefont {C.~X.}\ \bibnamefont {Yu}}, \bibinfo {author} {\bibfnamefont {J.~C.}\ \bibnamefont {Abadillo-Uriel}}, \bibinfo {author} {\bibfnamefont {{\'E}.}~\bibnamefont {Dumur}}, \bibinfo {author} {\bibfnamefont {H.}~\bibnamefont {Niebojewski}}, \bibinfo {author} {\bibfnamefont {B.}~\bibnamefont {Bertrand}}, \bibinfo {author} {\bibfnamefont {R.}~\bibnamefont {Maurand}},\ and\ \bibinfo {author} {\bibfnamefont {S.}~\bibnamefont {Zihlmann}},\ }\bibfield  {title} {\bibinfo {title} {Coherence of a hole-spin flopping-mode qubit in a circuit quantum electrodynamics environment},\ }\href {https://doi.org/10.1038/s41567-026-03262-y} {\bibfield  {journal} {\bibinfo  {journal} {Nature Physics}\ ,\ \bibinfo {pages} {1}} (\bibinfo {year} {2026})}\BibitemShut {NoStop}%
\bibitem [{\citenamefont {De~Palma}\ \emph {et~al.}(2024)\citenamefont {De~Palma}, \citenamefont {Oppliger}, \citenamefont {Jang}, \citenamefont {Bosco}, \citenamefont {Jan{\'\i}k}, \citenamefont {Calcaterra}, \citenamefont {Katsaros}, \citenamefont {Isella}, \citenamefont {Loss},\ and\ \citenamefont {Scarlino}}]{de2024strong}%
  \BibitemOpen
  \bibfield  {author} {\bibinfo {author} {\bibfnamefont {F.}~\bibnamefont {De~Palma}}, \bibinfo {author} {\bibfnamefont {F.}~\bibnamefont {Oppliger}}, \bibinfo {author} {\bibfnamefont {W.}~\bibnamefont {Jang}}, \bibinfo {author} {\bibfnamefont {S.}~\bibnamefont {Bosco}}, \bibinfo {author} {\bibfnamefont {M.}~\bibnamefont {Jan{\'\i}k}}, \bibinfo {author} {\bibfnamefont {S.}~\bibnamefont {Calcaterra}}, \bibinfo {author} {\bibfnamefont {G.}~\bibnamefont {Katsaros}}, \bibinfo {author} {\bibfnamefont {G.}~\bibnamefont {Isella}}, \bibinfo {author} {\bibfnamefont {D.}~\bibnamefont {Loss}},\ and\ \bibinfo {author} {\bibfnamefont {P.}~\bibnamefont {Scarlino}},\ }\bibfield  {title} {\bibinfo {title} {Strong hole-photon coupling in planar {G}e for probing charge degree and strongly correlated states},\ }\href {https://doi.org/10.1038/s41467-024-54520-7} {\bibfield  {journal} {\bibinfo  {journal} {Nature Communications}\ }\textbf {\bibinfo {volume} {15}},\ \bibinfo {pages} {10177} (\bibinfo {year} {2024})}\BibitemShut
  {NoStop}%
\bibitem [{\citenamefont {Mi}\ \emph {et~al.}(2018)\citenamefont {Mi}, \citenamefont {Benito}, \citenamefont {Putz}, \citenamefont {Zajac}, \citenamefont {Taylor}, \citenamefont {Burkard},\ and\ \citenamefont {Petta}}]{mi2018coherent}%
  \BibitemOpen
  \bibfield  {author} {\bibinfo {author} {\bibfnamefont {X.}~\bibnamefont {Mi}}, \bibinfo {author} {\bibfnamefont {M.}~\bibnamefont {Benito}}, \bibinfo {author} {\bibfnamefont {S.}~\bibnamefont {Putz}}, \bibinfo {author} {\bibfnamefont {D.~M.}\ \bibnamefont {Zajac}}, \bibinfo {author} {\bibfnamefont {J.~M.}\ \bibnamefont {Taylor}}, \bibinfo {author} {\bibfnamefont {G.}~\bibnamefont {Burkard}},\ and\ \bibinfo {author} {\bibfnamefont {J.~R.}\ \bibnamefont {Petta}},\ }\bibfield  {title} {\bibinfo {title} {A coherent spin--photon interface in silicon},\ }\href {https://doi.org/10.1038/nature25769} {\bibfield  {journal} {\bibinfo  {journal} {Nature}\ }\textbf {\bibinfo {volume} {555}},\ \bibinfo {pages} {599} (\bibinfo {year} {2018})}\BibitemShut {NoStop}%
\bibitem [{\citenamefont {Bosco}\ \emph {et~al.}(2024)\citenamefont {Bosco}, \citenamefont {Zou},\ and\ \citenamefont {Loss}}]{bosco2024high}%
  \BibitemOpen
  \bibfield  {author} {\bibinfo {author} {\bibfnamefont {S.}~\bibnamefont {Bosco}}, \bibinfo {author} {\bibfnamefont {J.}~\bibnamefont {Zou}},\ and\ \bibinfo {author} {\bibfnamefont {D.}~\bibnamefont {Loss}},\ }\bibfield  {title} {\bibinfo {title} {High-fidelity spin qubit shuttling via large spin-orbit interactions},\ }\href {https://link.aps.org/doi/10.1103/PRXQuantum.5.020353} {\bibfield  {journal} {\bibinfo  {journal} {PRX Quantum}\ }\textbf {\bibinfo {volume} {5}},\ \bibinfo {pages} {020353} (\bibinfo {year} {2024})}\BibitemShut {NoStop}%
\bibitem [{\citenamefont {K{\"u}nne}\ \emph {et~al.}(2024)\citenamefont {K{\"u}nne}, \citenamefont {Willmes}, \citenamefont {Oberl{\"a}nder}, \citenamefont {Gorjaew}, \citenamefont {Teske}, \citenamefont {Bhardwaj}, \citenamefont {Beer}, \citenamefont {Kammerloher}, \citenamefont {Otten}, \citenamefont {Seidler} \emph {et~al.}}]{kunne2024spinbus}%
  \BibitemOpen
  \bibfield  {author} {\bibinfo {author} {\bibfnamefont {M.}~\bibnamefont {K{\"u}nne}}, \bibinfo {author} {\bibfnamefont {A.}~\bibnamefont {Willmes}}, \bibinfo {author} {\bibfnamefont {M.}~\bibnamefont {Oberl{\"a}nder}}, \bibinfo {author} {\bibfnamefont {C.}~\bibnamefont {Gorjaew}}, \bibinfo {author} {\bibfnamefont {J.~D.}\ \bibnamefont {Teske}}, \bibinfo {author} {\bibfnamefont {H.}~\bibnamefont {Bhardwaj}}, \bibinfo {author} {\bibfnamefont {M.}~\bibnamefont {Beer}}, \bibinfo {author} {\bibfnamefont {E.}~\bibnamefont {Kammerloher}}, \bibinfo {author} {\bibfnamefont {R.}~\bibnamefont {Otten}}, \bibinfo {author} {\bibfnamefont {I.}~\bibnamefont {Seidler}}, \emph {et~al.},\ }\bibfield  {title} {\bibinfo {title} {The spinbus architecture for scaling spin qubits with electron shuttling},\ }\href {https://doi.org/10.1038/s41467-024-49182-4} {\bibfield  {journal} {\bibinfo  {journal} {Nature Communications}\ }\textbf {\bibinfo {volume} {15}},\ \bibinfo {pages} {4977} (\bibinfo {year} {2024})}\BibitemShut {NoStop}%
\bibitem [{\citenamefont {van Riggelen-Doelman}\ \emph {et~al.}(2024)\citenamefont {van Riggelen-Doelman}, \citenamefont {Wang}, \citenamefont {de~Snoo}, \citenamefont {Lawrie}, \citenamefont {Hendrickx}, \citenamefont {Rimbach-Russ}, \citenamefont {Sammak}, \citenamefont {Scappucci}, \citenamefont {D{\'e}prez},\ and\ \citenamefont {Veldhorst}}]{van2024coherent}%
  \BibitemOpen
  \bibfield  {author} {\bibinfo {author} {\bibfnamefont {F.}~\bibnamefont {van Riggelen-Doelman}}, \bibinfo {author} {\bibfnamefont {C.-A.}\ \bibnamefont {Wang}}, \bibinfo {author} {\bibfnamefont {S.~L.}\ \bibnamefont {de~Snoo}}, \bibinfo {author} {\bibfnamefont {W.~I.}\ \bibnamefont {Lawrie}}, \bibinfo {author} {\bibfnamefont {N.~W.}\ \bibnamefont {Hendrickx}}, \bibinfo {author} {\bibfnamefont {M.}~\bibnamefont {Rimbach-Russ}}, \bibinfo {author} {\bibfnamefont {A.}~\bibnamefont {Sammak}}, \bibinfo {author} {\bibfnamefont {G.}~\bibnamefont {Scappucci}}, \bibinfo {author} {\bibfnamefont {C.}~\bibnamefont {D{\'e}prez}},\ and\ \bibinfo {author} {\bibfnamefont {M.}~\bibnamefont {Veldhorst}},\ }\bibfield  {title} {\bibinfo {title} {Coherent spin qubit shuttling through germanium quantum dots},\ }\href {https://doi.org/10.1038/s41467-024-49358-y} {\bibfield  {journal} {\bibinfo  {journal} {Nature Communications}\ }\textbf {\bibinfo {volume} {15}},\ \bibinfo {pages} {5716} (\bibinfo {year} {2024})}\BibitemShut
  {NoStop}%
\bibitem [{\citenamefont {De~Smet}\ \emph {et~al.}(2025)\citenamefont {De~Smet}, \citenamefont {Matsumoto}, \citenamefont {Zwerver}, \citenamefont {Tryputen}, \citenamefont {de~Snoo}, \citenamefont {Amitonov}, \citenamefont {Katiraee-Far}, \citenamefont {Sammak}, \citenamefont {Samkharadze}, \citenamefont {G{\"u}l} \emph {et~al.}}]{de2025high}%
  \BibitemOpen
  \bibfield  {author} {\bibinfo {author} {\bibfnamefont {M.}~\bibnamefont {De~Smet}}, \bibinfo {author} {\bibfnamefont {Y.}~\bibnamefont {Matsumoto}}, \bibinfo {author} {\bibfnamefont {A.-M.~J.}\ \bibnamefont {Zwerver}}, \bibinfo {author} {\bibfnamefont {L.}~\bibnamefont {Tryputen}}, \bibinfo {author} {\bibfnamefont {S.~L.}\ \bibnamefont {de~Snoo}}, \bibinfo {author} {\bibfnamefont {S.~V.}\ \bibnamefont {Amitonov}}, \bibinfo {author} {\bibfnamefont {S.~R.}\ \bibnamefont {Katiraee-Far}}, \bibinfo {author} {\bibfnamefont {A.}~\bibnamefont {Sammak}}, \bibinfo {author} {\bibfnamefont {N.}~\bibnamefont {Samkharadze}}, \bibinfo {author} {\bibfnamefont {{\"O}.}~\bibnamefont {G{\"u}l}}, \emph {et~al.},\ }\bibfield  {title} {\bibinfo {title} {High-fidelity single-spin shuttling in silicon},\ }\href {https://doi.org/10.1038/s41565-025-01920-5} {\bibfield  {journal} {\bibinfo  {journal} {Nature Nanotechnology}\ ,\ \bibinfo {pages} {1}} (\bibinfo {year} {2025})}\BibitemShut {NoStop}%
\bibitem [{\citenamefont {Yoneda}\ \emph {et~al.}(2021)\citenamefont {Yoneda}, \citenamefont {Huang}, \citenamefont {Feng}, \citenamefont {Yang}, \citenamefont {Chan}, \citenamefont {Tanttu}, \citenamefont {Gilbert}, \citenamefont {Leon}, \citenamefont {Hudson}, \citenamefont {Itoh} \emph {et~al.}}]{yoneda2021coherent}%
  \BibitemOpen
  \bibfield  {author} {\bibinfo {author} {\bibfnamefont {J.}~\bibnamefont {Yoneda}}, \bibinfo {author} {\bibfnamefont {W.}~\bibnamefont {Huang}}, \bibinfo {author} {\bibfnamefont {M.}~\bibnamefont {Feng}}, \bibinfo {author} {\bibfnamefont {C.~H.}\ \bibnamefont {Yang}}, \bibinfo {author} {\bibfnamefont {K.~W.}\ \bibnamefont {Chan}}, \bibinfo {author} {\bibfnamefont {T.}~\bibnamefont {Tanttu}}, \bibinfo {author} {\bibfnamefont {W.}~\bibnamefont {Gilbert}}, \bibinfo {author} {\bibfnamefont {R.}~\bibnamefont {Leon}}, \bibinfo {author} {\bibfnamefont {F.}~\bibnamefont {Hudson}}, \bibinfo {author} {\bibfnamefont {K.}~\bibnamefont {Itoh}}, \emph {et~al.},\ }\bibfield  {title} {\bibinfo {title} {Coherent spin qubit transport in silicon},\ }\href {https://doi.org/10.1038/s41467-021-24371-7} {\bibfield  {journal} {\bibinfo  {journal} {Nature Communications}\ }\textbf {\bibinfo {volume} {12}},\ \bibinfo {pages} {4114} (\bibinfo {year} {2021})}\BibitemShut {NoStop}%
\bibitem [{\citenamefont {Ademi}\ \emph {et~al.}(2025)\citenamefont {Ademi}, \citenamefont {Bassi}, \citenamefont {Yu}, \citenamefont {de~Snoo}, \citenamefont {Oosterhout}, \citenamefont {Sammak}, \citenamefont {Vandersypen}, \citenamefont {Scappucci}, \citenamefont {D{\'e}prez},\ and\ \citenamefont {Veldhorst}}]{ademi2025distributing}%
  \BibitemOpen
  \bibfield  {author} {\bibinfo {author} {\bibfnamefont {Z.}~\bibnamefont {Ademi}}, \bibinfo {author} {\bibfnamefont {M.}~\bibnamefont {Bassi}}, \bibinfo {author} {\bibfnamefont {C.~X.}\ \bibnamefont {Yu}}, \bibinfo {author} {\bibfnamefont {S.~L.}\ \bibnamefont {de~Snoo}}, \bibinfo {author} {\bibfnamefont {S.~D.}\ \bibnamefont {Oosterhout}}, \bibinfo {author} {\bibfnamefont {A.}~\bibnamefont {Sammak}}, \bibinfo {author} {\bibfnamefont {L.~M.}\ \bibnamefont {Vandersypen}}, \bibinfo {author} {\bibfnamefont {G.}~\bibnamefont {Scappucci}}, \bibinfo {author} {\bibfnamefont {C.}~\bibnamefont {D{\'e}prez}},\ and\ \bibinfo {author} {\bibfnamefont {M.}~\bibnamefont {Veldhorst}},\ }\bibfield  {title} {\bibinfo {title} {Distributing entanglement between distant semiconductor qubit registers using a shared-control shuttling link},\ }\href {https://arxiv.org/abs/2510.26860} {\bibfield  {journal} {\bibinfo  {journal} {arXiv preprint arXiv:2510.26860}\ } (\bibinfo {year} {2025})}\BibitemShut {NoStop}%
\bibitem [{\citenamefont {Scappucci}\ \emph {et~al.}(2021)\citenamefont {Scappucci}, \citenamefont {Kloeffel}, \citenamefont {Zwanenburg}, \citenamefont {Loss}, \citenamefont {Myronov}, \citenamefont {Zhang}, \citenamefont {De~Franceschi}, \citenamefont {Katsaros},\ and\ \citenamefont {Veldhorst}}]{scappucci2021germanium}%
  \BibitemOpen
  \bibfield  {author} {\bibinfo {author} {\bibfnamefont {G.}~\bibnamefont {Scappucci}}, \bibinfo {author} {\bibfnamefont {C.}~\bibnamefont {Kloeffel}}, \bibinfo {author} {\bibfnamefont {F.~A.}\ \bibnamefont {Zwanenburg}}, \bibinfo {author} {\bibfnamefont {D.}~\bibnamefont {Loss}}, \bibinfo {author} {\bibfnamefont {M.}~\bibnamefont {Myronov}}, \bibinfo {author} {\bibfnamefont {J.-J.}\ \bibnamefont {Zhang}}, \bibinfo {author} {\bibfnamefont {S.}~\bibnamefont {De~Franceschi}}, \bibinfo {author} {\bibfnamefont {G.}~\bibnamefont {Katsaros}},\ and\ \bibinfo {author} {\bibfnamefont {M.}~\bibnamefont {Veldhorst}},\ }\bibfield  {title} {\bibinfo {title} {The germanium quantum information route},\ }\href {https://doi.org/10.1038/s41578-020-00262-z} {\bibfield  {journal} {\bibinfo  {journal} {Nature Reviews Materials}\ }\textbf {\bibinfo {volume} {6}},\ \bibinfo {pages} {926} (\bibinfo {year} {2021})}\BibitemShut {NoStop}%
\bibitem [{\citenamefont {Hendrickx}\ \emph {et~al.}(2020{\natexlab{a}})\citenamefont {Hendrickx}, \citenamefont {Lawrie}, \citenamefont {Petit}, \citenamefont {Sammak}, \citenamefont {Scappucci},\ and\ \citenamefont {Veldhorst}}]{hendrickx2020single}%
  \BibitemOpen
  \bibfield  {author} {\bibinfo {author} {\bibfnamefont {N.}~\bibnamefont {Hendrickx}}, \bibinfo {author} {\bibfnamefont {W.}~\bibnamefont {Lawrie}}, \bibinfo {author} {\bibfnamefont {L.}~\bibnamefont {Petit}}, \bibinfo {author} {\bibfnamefont {A.}~\bibnamefont {Sammak}}, \bibinfo {author} {\bibfnamefont {G.}~\bibnamefont {Scappucci}},\ and\ \bibinfo {author} {\bibfnamefont {M.}~\bibnamefont {Veldhorst}},\ }\bibfield  {title} {\bibinfo {title} {A single-hole spin qubit},\ }\href {https://doi.org/10.1038/s41467-020-17211-7} {\bibfield  {journal} {\bibinfo  {journal} {Nature Communications}\ }\textbf {\bibinfo {volume} {11}},\ \bibinfo {pages} {3478} (\bibinfo {year} {2020}{\natexlab{a}})}\BibitemShut {NoStop}%
\bibitem [{\citenamefont {Watzinger}\ \emph {et~al.}(2018)\citenamefont {Watzinger}, \citenamefont {Kuku{\v{c}}ka}, \citenamefont {Vuku{\v{s}}i{\'c}}, \citenamefont {Gao}, \citenamefont {Wang}, \citenamefont {Sch{\"a}ffler}, \citenamefont {Zhang},\ and\ \citenamefont {Katsaros}}]{watzinger2018germanium}%
  \BibitemOpen
  \bibfield  {author} {\bibinfo {author} {\bibfnamefont {H.}~\bibnamefont {Watzinger}}, \bibinfo {author} {\bibfnamefont {J.}~\bibnamefont {Kuku{\v{c}}ka}}, \bibinfo {author} {\bibfnamefont {L.}~\bibnamefont {Vuku{\v{s}}i{\'c}}}, \bibinfo {author} {\bibfnamefont {F.}~\bibnamefont {Gao}}, \bibinfo {author} {\bibfnamefont {T.}~\bibnamefont {Wang}}, \bibinfo {author} {\bibfnamefont {F.}~\bibnamefont {Sch{\"a}ffler}}, \bibinfo {author} {\bibfnamefont {J.-J.}\ \bibnamefont {Zhang}},\ and\ \bibinfo {author} {\bibfnamefont {G.}~\bibnamefont {Katsaros}},\ }\bibfield  {title} {\bibinfo {title} {A germanium hole spin qubit},\ }\href {https://doi.org/10.1038/s41467-018-06418-4} {\bibfield  {journal} {\bibinfo  {journal} {Nature communications}\ }\textbf {\bibinfo {volume} {9}},\ \bibinfo {pages} {3902} (\bibinfo {year} {2018})}\BibitemShut {NoStop}%
\bibitem [{\citenamefont {Jirovec}\ \emph {et~al.}(2021)\citenamefont {Jirovec}, \citenamefont {Hofmann}, \citenamefont {Ballabio}, \citenamefont {Mutter}, \citenamefont {Tavani}, \citenamefont {Botifoll}, \citenamefont {Crippa}, \citenamefont {Kukucka}, \citenamefont {Sagi}, \citenamefont {Martins} \emph {et~al.}}]{jirovec2021singlet}%
  \BibitemOpen
  \bibfield  {author} {\bibinfo {author} {\bibfnamefont {D.}~\bibnamefont {Jirovec}}, \bibinfo {author} {\bibfnamefont {A.}~\bibnamefont {Hofmann}}, \bibinfo {author} {\bibfnamefont {A.}~\bibnamefont {Ballabio}}, \bibinfo {author} {\bibfnamefont {P.~M.}\ \bibnamefont {Mutter}}, \bibinfo {author} {\bibfnamefont {G.}~\bibnamefont {Tavani}}, \bibinfo {author} {\bibfnamefont {M.}~\bibnamefont {Botifoll}}, \bibinfo {author} {\bibfnamefont {A.}~\bibnamefont {Crippa}}, \bibinfo {author} {\bibfnamefont {J.}~\bibnamefont {Kukucka}}, \bibinfo {author} {\bibfnamefont {O.}~\bibnamefont {Sagi}}, \bibinfo {author} {\bibfnamefont {F.}~\bibnamefont {Martins}}, \emph {et~al.},\ }\bibfield  {title} {\bibinfo {title} {A singlet-triplet hole spin qubit in planar ge},\ }\href {https://doi.org/10.1038/s41563-021-01022-2} {\bibfield  {journal} {\bibinfo  {journal} {Nature Materials}\ }\textbf {\bibinfo {volume} {20}},\ \bibinfo {pages} {1106} (\bibinfo {year} {2021})}\BibitemShut {NoStop}%
\bibitem [{\citenamefont {Hendrickx}\ \emph {et~al.}(2020{\natexlab{b}})\citenamefont {Hendrickx}, \citenamefont {Franke}, \citenamefont {Sammak}, \citenamefont {Scappucci},\ and\ \citenamefont {Veldhorst}}]{hendrickx2020fast}%
  \BibitemOpen
  \bibfield  {author} {\bibinfo {author} {\bibfnamefont {N.}~\bibnamefont {Hendrickx}}, \bibinfo {author} {\bibfnamefont {D.}~\bibnamefont {Franke}}, \bibinfo {author} {\bibfnamefont {A.}~\bibnamefont {Sammak}}, \bibinfo {author} {\bibfnamefont {G.}~\bibnamefont {Scappucci}},\ and\ \bibinfo {author} {\bibfnamefont {M.}~\bibnamefont {Veldhorst}},\ }\bibfield  {title} {\bibinfo {title} {Fast two-qubit logic with holes in germanium},\ }\href {https://doi.org/10.1038/s41586-019-1919-3} {\bibfield  {journal} {\bibinfo  {journal} {Nature}\ }\textbf {\bibinfo {volume} {577}},\ \bibinfo {pages} {487} (\bibinfo {year} {2020}{\natexlab{b}})}\BibitemShut {NoStop}%
\bibitem [{\citenamefont {Ivlev}\ \emph {et~al.}(2025)\citenamefont {Ivlev}, \citenamefont {Crielaard}, \citenamefont {Meyer}, \citenamefont {Lawrie}, \citenamefont {Hendrickx}, \citenamefont {Sammak}, \citenamefont {Matsumoto}, \citenamefont {Vandersypen}, \citenamefont {Scappucci}, \citenamefont {D{\'e}prez} \emph {et~al.}}]{ivlev2025operating}%
  \BibitemOpen
  \bibfield  {author} {\bibinfo {author} {\bibfnamefont {A.~S.}\ \bibnamefont {Ivlev}}, \bibinfo {author} {\bibfnamefont {D.~R.}\ \bibnamefont {Crielaard}}, \bibinfo {author} {\bibfnamefont {M.}~\bibnamefont {Meyer}}, \bibinfo {author} {\bibfnamefont {W.~I.}\ \bibnamefont {Lawrie}}, \bibinfo {author} {\bibfnamefont {N.~W.}\ \bibnamefont {Hendrickx}}, \bibinfo {author} {\bibfnamefont {A.}~\bibnamefont {Sammak}}, \bibinfo {author} {\bibfnamefont {Y.}~\bibnamefont {Matsumoto}}, \bibinfo {author} {\bibfnamefont {L.~M.}\ \bibnamefont {Vandersypen}}, \bibinfo {author} {\bibfnamefont {G.}~\bibnamefont {Scappucci}}, \bibinfo {author} {\bibfnamefont {C.}~\bibnamefont {D{\'e}prez}}, \emph {et~al.},\ }\bibfield  {title} {\bibinfo {title} {Operating semiconductor qubits without individual barrier gates},\ }\href {https://link.aps.org/doi/10.1103/xhq3-4jxz} {\bibfield  {journal} {\bibinfo  {journal} {Physical Review X}\ }\textbf {\bibinfo {volume} {15}},\ \bibinfo {pages} {031042} (\bibinfo {year} {2025})}\BibitemShut
  {NoStop}%
\bibitem [{\citenamefont {Bulaev}\ and\ \citenamefont {Loss}(2007)}]{bulaev2007electric}%
  \BibitemOpen
  \bibfield  {author} {\bibinfo {author} {\bibfnamefont {D.~V.}\ \bibnamefont {Bulaev}}\ and\ \bibinfo {author} {\bibfnamefont {D.}~\bibnamefont {Loss}},\ }\bibfield  {title} {\bibinfo {title} {Electric dipole spin resonance for heavy holes in quantum dots},\ }\href {https://link.aps.org/doi/10.1103/PhysRevLett.98.097202} {\bibfield  {journal} {\bibinfo  {journal} {Physical Review Letters}\ }\textbf {\bibinfo {volume} {98}},\ \bibinfo {pages} {097202} (\bibinfo {year} {2007})}\BibitemShut {NoStop}%
\bibitem [{\citenamefont {Wang}\ \emph {et~al.}(2022)\citenamefont {Wang}, \citenamefont {Xu}, \citenamefont {Gao}, \citenamefont {Liu}, \citenamefont {Ma}, \citenamefont {Zhang}, \citenamefont {Wang}, \citenamefont {Cao}, \citenamefont {Wang}, \citenamefont {Zhang} \emph {et~al.}}]{wang2022ultrafast}%
  \BibitemOpen
  \bibfield  {author} {\bibinfo {author} {\bibfnamefont {K.}~\bibnamefont {Wang}}, \bibinfo {author} {\bibfnamefont {G.}~\bibnamefont {Xu}}, \bibinfo {author} {\bibfnamefont {F.}~\bibnamefont {Gao}}, \bibinfo {author} {\bibfnamefont {H.}~\bibnamefont {Liu}}, \bibinfo {author} {\bibfnamefont {R.-L.}\ \bibnamefont {Ma}}, \bibinfo {author} {\bibfnamefont {X.}~\bibnamefont {Zhang}}, \bibinfo {author} {\bibfnamefont {Z.}~\bibnamefont {Wang}}, \bibinfo {author} {\bibfnamefont {G.}~\bibnamefont {Cao}}, \bibinfo {author} {\bibfnamefont {T.}~\bibnamefont {Wang}}, \bibinfo {author} {\bibfnamefont {J.-J.}\ \bibnamefont {Zhang}}, \emph {et~al.},\ }\bibfield  {title} {\bibinfo {title} {Ultrafast coherent control of a hole spin qubit in a germanium quantum dot},\ }\href {https://doi.org/10.1038/s41467-021-27880-7} {\bibfield  {journal} {\bibinfo  {journal} {Nature Communications}\ }\textbf {\bibinfo {volume} {13}},\ \bibinfo {pages} {206} (\bibinfo {year} {2022})}\BibitemShut {NoStop}%
\bibitem [{\citenamefont {Froning}\ \emph {et~al.}(2021)\citenamefont {Froning}, \citenamefont {Camenzind}, \citenamefont {van~der Molen}, \citenamefont {Li}, \citenamefont {Bakkers}, \citenamefont {Zumb{\"u}hl},\ and\ \citenamefont {Braakman}}]{froning2021ultrafast}%
  \BibitemOpen
  \bibfield  {author} {\bibinfo {author} {\bibfnamefont {F.~N.}\ \bibnamefont {Froning}}, \bibinfo {author} {\bibfnamefont {L.~C.}\ \bibnamefont {Camenzind}}, \bibinfo {author} {\bibfnamefont {O.~A.}\ \bibnamefont {van~der Molen}}, \bibinfo {author} {\bibfnamefont {A.}~\bibnamefont {Li}}, \bibinfo {author} {\bibfnamefont {E.~P.}\ \bibnamefont {Bakkers}}, \bibinfo {author} {\bibfnamefont {D.~M.}\ \bibnamefont {Zumb{\"u}hl}},\ and\ \bibinfo {author} {\bibfnamefont {F.~R.}\ \bibnamefont {Braakman}},\ }\bibfield  {title} {\bibinfo {title} {Ultrafast hole spin qubit with gate-tunable spin--orbit switch functionality},\ }\href {https://doi.org/10.1038/s41565-020-00828-6} {\bibfield  {journal} {\bibinfo  {journal} {Nature Nanotechnology}\ }\textbf {\bibinfo {volume} {16}},\ \bibinfo {pages} {308} (\bibinfo {year} {2021})}\BibitemShut {NoStop}%
\bibitem [{\citenamefont {Bosco}\ \emph {et~al.}(2021)\citenamefont {Bosco}, \citenamefont {Het{\'e}nyi},\ and\ \citenamefont {Loss}}]{bosco2021hole}%
  \BibitemOpen
  \bibfield  {author} {\bibinfo {author} {\bibfnamefont {S.}~\bibnamefont {Bosco}}, \bibinfo {author} {\bibfnamefont {B.}~\bibnamefont {Het{\'e}nyi}},\ and\ \bibinfo {author} {\bibfnamefont {D.}~\bibnamefont {Loss}},\ }\bibfield  {title} {\bibinfo {title} {Hole spin qubits in {S}i {F}in{FET}s with fully tunable spin-orbit coupling and sweet spots for charge noise},\ }\href {https://link.aps.org/doi/10.1103/PRXQuantum.2.010348} {\bibfield  {journal} {\bibinfo  {journal} {PRX Quantum}\ }\textbf {\bibinfo {volume} {2}},\ \bibinfo {pages} {010348} (\bibinfo {year} {2021})}\BibitemShut {NoStop}%
\bibitem [{\citenamefont {Abadillo-Uriel}\ \emph {et~al.}(2023)\citenamefont {Abadillo-Uriel}, \citenamefont {Rodr{\'\i}guez-Mena}, \citenamefont {Martinez},\ and\ \citenamefont {Niquet}}]{abadillo2023hole}%
  \BibitemOpen
  \bibfield  {author} {\bibinfo {author} {\bibfnamefont {J.~C.}\ \bibnamefont {Abadillo-Uriel}}, \bibinfo {author} {\bibfnamefont {E.~A.}\ \bibnamefont {Rodr{\'\i}guez-Mena}}, \bibinfo {author} {\bibfnamefont {B.}~\bibnamefont {Martinez}},\ and\ \bibinfo {author} {\bibfnamefont {Y.-M.}\ \bibnamefont {Niquet}},\ }\bibfield  {title} {\bibinfo {title} {Hole-spin driving by strain-induced spin-orbit interactions},\ }\href {https://link.aps.org/doi/10.1103/PhysRevLett.131.097002} {\bibfield  {journal} {\bibinfo  {journal} {Physical Review Letters}\ }\textbf {\bibinfo {volume} {131}},\ \bibinfo {pages} {097002} (\bibinfo {year} {2023})}\BibitemShut {NoStop}%
\bibitem [{\citenamefont {Martinez}\ \emph {et~al.}(2022)\citenamefont {Martinez}, \citenamefont {Abadillo-Uriel}, \citenamefont {Rodr{\'\i}guez-Mena},\ and\ \citenamefont {Niquet}}]{martinez2022hole}%
  \BibitemOpen
  \bibfield  {author} {\bibinfo {author} {\bibfnamefont {B.}~\bibnamefont {Martinez}}, \bibinfo {author} {\bibfnamefont {J.~C.}\ \bibnamefont {Abadillo-Uriel}}, \bibinfo {author} {\bibfnamefont {E.~A.}\ \bibnamefont {Rodr{\'\i}guez-Mena}},\ and\ \bibinfo {author} {\bibfnamefont {Y.-M.}\ \bibnamefont {Niquet}},\ }\bibfield  {title} {\bibinfo {title} {Hole spin manipulation in inhomogeneous and nonseparable electric fields},\ }\href {https://link.aps.org/doi/10.1103/PhysRevB.106.235426} {\bibfield  {journal} {\bibinfo  {journal} {Physical Review B}\ }\textbf {\bibinfo {volume} {106}},\ \bibinfo {pages} {235426} (\bibinfo {year} {2022})}\BibitemShut {NoStop}%
\bibitem [{\citenamefont {Terrazos}\ \emph {et~al.}(2021)\citenamefont {Terrazos}, \citenamefont {Marcellina}, \citenamefont {Wang}, \citenamefont {Coppersmith}, \citenamefont {Friesen}, \citenamefont {Hamilton}, \citenamefont {Hu}, \citenamefont {Koiller}, \citenamefont {Saraiva}, \citenamefont {Culcer} \emph {et~al.}}]{terrazos2021theory}%
  \BibitemOpen
  \bibfield  {author} {\bibinfo {author} {\bibfnamefont {L.}~\bibnamefont {Terrazos}}, \bibinfo {author} {\bibfnamefont {E.}~\bibnamefont {Marcellina}}, \bibinfo {author} {\bibfnamefont {Z.}~\bibnamefont {Wang}}, \bibinfo {author} {\bibfnamefont {S.}~\bibnamefont {Coppersmith}}, \bibinfo {author} {\bibfnamefont {M.}~\bibnamefont {Friesen}}, \bibinfo {author} {\bibfnamefont {A.}~\bibnamefont {Hamilton}}, \bibinfo {author} {\bibfnamefont {X.}~\bibnamefont {Hu}}, \bibinfo {author} {\bibfnamefont {B.}~\bibnamefont {Koiller}}, \bibinfo {author} {\bibfnamefont {A.}~\bibnamefont {Saraiva}}, \bibinfo {author} {\bibfnamefont {D.}~\bibnamefont {Culcer}}, \emph {et~al.},\ }\bibfield  {title} {\bibinfo {title} {Theory of hole-spin qubits in strained germanium quantum dots},\ }\href {https://link.aps.org/doi/10.1103/PhysRevB.103.125201} {\bibfield  {journal} {\bibinfo  {journal} {Physical Review B}\ }\textbf {\bibinfo {volume} {103}},\ \bibinfo {pages} {125201} (\bibinfo {year} {2021})}\BibitemShut {NoStop}%
\bibitem [{\citenamefont {Liles}\ \emph {et~al.}(2021)\citenamefont {Liles}, \citenamefont {Martins}, \citenamefont {Miserev}, \citenamefont {Kiselev}, \citenamefont {Thorvaldson}, \citenamefont {Rendell}, \citenamefont {Jin}, \citenamefont {Hudson}, \citenamefont {Veldhorst}, \citenamefont {Itoh} \emph {et~al.}}]{liles2021electrical}%
  \BibitemOpen
  \bibfield  {author} {\bibinfo {author} {\bibfnamefont {S.}~\bibnamefont {Liles}}, \bibinfo {author} {\bibfnamefont {F.}~\bibnamefont {Martins}}, \bibinfo {author} {\bibfnamefont {D.}~\bibnamefont {Miserev}}, \bibinfo {author} {\bibfnamefont {A.}~\bibnamefont {Kiselev}}, \bibinfo {author} {\bibfnamefont {I.}~\bibnamefont {Thorvaldson}}, \bibinfo {author} {\bibfnamefont {M.}~\bibnamefont {Rendell}}, \bibinfo {author} {\bibfnamefont {I.}~\bibnamefont {Jin}}, \bibinfo {author} {\bibfnamefont {F.}~\bibnamefont {Hudson}}, \bibinfo {author} {\bibfnamefont {M.}~\bibnamefont {Veldhorst}}, \bibinfo {author} {\bibfnamefont {K.}~\bibnamefont {Itoh}}, \emph {et~al.},\ }\bibfield  {title} {\bibinfo {title} {Electrical control of the g tensor of the first hole in a silicon {MOS} quantum dot},\ }\href {https://link.aps.org/doi/10.1103/PhysRevB.104.235303} {\bibfield  {journal} {\bibinfo  {journal} {Physical Review B}\ }\textbf {\bibinfo {volume} {104}},\ \bibinfo {pages} {235303} (\bibinfo {year} {2021})}\BibitemShut
  {NoStop}%
\bibitem [{\citenamefont {Bassi}\ \emph {et~al.}(2025)\citenamefont {Bassi}, \citenamefont {Rodr{\'\i}guez-Mena}, \citenamefont {Brun}, \citenamefont {Zihlmann}, \citenamefont {Nguyen}, \citenamefont {Champain}, \citenamefont {Abadillo-Uriel}, \citenamefont {Bertrand}, \citenamefont {Niebojewski}, \citenamefont {Maurand} \emph {et~al.}}]{bassi2024optimal}%
  \BibitemOpen
  \bibfield  {author} {\bibinfo {author} {\bibfnamefont {M.}~\bibnamefont {Bassi}}, \bibinfo {author} {\bibfnamefont {E.}~\bibnamefont {Rodr{\'\i}guez-Mena}}, \bibinfo {author} {\bibfnamefont {B.}~\bibnamefont {Brun}}, \bibinfo {author} {\bibfnamefont {S.}~\bibnamefont {Zihlmann}}, \bibinfo {author} {\bibfnamefont {T.}~\bibnamefont {Nguyen}}, \bibinfo {author} {\bibfnamefont {V.}~\bibnamefont {Champain}}, \bibinfo {author} {\bibfnamefont {J.~C.}\ \bibnamefont {Abadillo-Uriel}}, \bibinfo {author} {\bibfnamefont {B.}~\bibnamefont {Bertrand}}, \bibinfo {author} {\bibfnamefont {H.}~\bibnamefont {Niebojewski}}, \bibinfo {author} {\bibfnamefont {R.}~\bibnamefont {Maurand}}, \emph {et~al.},\ }\bibfield  {title} {\bibinfo {title} {Optimal operation of hole spin qubits},\ }\href {https://doi.org/10.1038/s41567-025-03106-1} {\bibfield  {journal} {\bibinfo  {journal} {Nature Physics}\ ,\ \bibinfo {pages} {1}} (\bibinfo {year} {2025})}\BibitemShut {NoStop}%
\bibitem [{\citenamefont {Mauro}\ \emph {et~al.}(2025)\citenamefont {Mauro}, \citenamefont {Rodr{\'\i}guez}, \citenamefont {Rodr{\'\i}guez-Mena},\ and\ \citenamefont {Niquet}}]{mauro2025hole}%
  \BibitemOpen
  \bibfield  {author} {\bibinfo {author} {\bibfnamefont {L.}~\bibnamefont {Mauro}}, \bibinfo {author} {\bibfnamefont {M.~J.}\ \bibnamefont {Rodr{\'\i}guez}}, \bibinfo {author} {\bibfnamefont {E.~A.}\ \bibnamefont {Rodr{\'\i}guez-Mena}},\ and\ \bibinfo {author} {\bibfnamefont {Y.-M.}\ \bibnamefont {Niquet}},\ }\bibfield  {title} {\bibinfo {title} {Hole spin qubits in unstrained germanium layers},\ }\href {https://doi.org/10.1038/s41534-025-01108-8} {\bibfield  {journal} {\bibinfo  {journal} {npj Quantum Information}\ }\textbf {\bibinfo {volume} {11}},\ \bibinfo {pages} {167} (\bibinfo {year} {2025})}\BibitemShut {NoStop}%
\bibitem [{\citenamefont {Seidler}\ \emph {et~al.}(2025)\citenamefont {Seidler}, \citenamefont {Het{\'e}nyi}, \citenamefont {Sommer}, \citenamefont {Massai}, \citenamefont {Tsoukalas}, \citenamefont {Kelly}, \citenamefont {Orekhov}, \citenamefont {Aldeghi}, \citenamefont {Bedell}, \citenamefont {Paredes} \emph {et~al.}}]{seidler2025spatial}%
  \BibitemOpen
  \bibfield  {author} {\bibinfo {author} {\bibfnamefont {I.}~\bibnamefont {Seidler}}, \bibinfo {author} {\bibfnamefont {B.}~\bibnamefont {Het{\'e}nyi}}, \bibinfo {author} {\bibfnamefont {L.}~\bibnamefont {Sommer}}, \bibinfo {author} {\bibfnamefont {L.}~\bibnamefont {Massai}}, \bibinfo {author} {\bibfnamefont {K.}~\bibnamefont {Tsoukalas}}, \bibinfo {author} {\bibfnamefont {E.~G.}\ \bibnamefont {Kelly}}, \bibinfo {author} {\bibfnamefont {A.}~\bibnamefont {Orekhov}}, \bibinfo {author} {\bibfnamefont {M.}~\bibnamefont {Aldeghi}}, \bibinfo {author} {\bibfnamefont {S.~W.}\ \bibnamefont {Bedell}}, \bibinfo {author} {\bibfnamefont {S.}~\bibnamefont {Paredes}}, \emph {et~al.},\ }\bibfield  {title} {\bibinfo {title} {Spatial uniformity of g-tensor and spin-orbit interaction in germanium hole spin qubits},\ }\href {https://arxiv.org/abs/2510.03125} {\bibfield  {journal} {\bibinfo  {journal} {arXiv preprint arXiv:2510.03125}\ } (\bibinfo {year} {2025})}\BibitemShut {NoStop}%
\bibitem [{\citenamefont {Sommer}\ \emph {et~al.}(2026)\citenamefont {Sommer}, \citenamefont {Seidler}, \citenamefont {Schupp}, \citenamefont {Paredes}, \citenamefont {Hendrickx}, \citenamefont {Massai}, \citenamefont {Bedell}, \citenamefont {Salis}, \citenamefont {Mergenthaler}, \citenamefont {Harvey-Collard} \emph {et~al.}}]{sommer2026disentangling}%
  \BibitemOpen
  \bibfield  {author} {\bibinfo {author} {\bibfnamefont {L.}~\bibnamefont {Sommer}}, \bibinfo {author} {\bibfnamefont {I.}~\bibnamefont {Seidler}}, \bibinfo {author} {\bibfnamefont {F.}~\bibnamefont {Schupp}}, \bibinfo {author} {\bibfnamefont {S.}~\bibnamefont {Paredes}}, \bibinfo {author} {\bibfnamefont {N.}~\bibnamefont {Hendrickx}}, \bibinfo {author} {\bibfnamefont {L.}~\bibnamefont {Massai}}, \bibinfo {author} {\bibfnamefont {S.}~\bibnamefont {Bedell}}, \bibinfo {author} {\bibfnamefont {G.}~\bibnamefont {Salis}}, \bibinfo {author} {\bibfnamefont {M.}~\bibnamefont {Mergenthaler}}, \bibinfo {author} {\bibfnamefont {P.}~\bibnamefont {Harvey-Collard}}, \emph {et~al.},\ }\bibfield  {title} {\bibinfo {title} {Disentangling orbital and confinement contributions to $ g $-factor in {G}e/{S}i{G}e hole quantum dots},\ }\href {https://arxiv.org/abs/2602.09913} {\bibfield  {journal} {\bibinfo  {journal} {arXiv preprint arXiv:2602.09913}\ } (\bibinfo {year} {2026})}\BibitemShut {NoStop}%
\bibitem [{\citenamefont {Bulaev}\ and\ \citenamefont {Loss}(2005)}]{bulaev2005spin}%
  \BibitemOpen
  \bibfield  {author} {\bibinfo {author} {\bibfnamefont {D.~V.}\ \bibnamefont {Bulaev}}\ and\ \bibinfo {author} {\bibfnamefont {D.}~\bibnamefont {Loss}},\ }\bibfield  {title} {\bibinfo {title} {Spin relaxation and decoherence of holes in quantum dots},\ }\href {https://link.aps.org/doi/10.1103/PhysRevLett.95.076805} {\bibfield  {journal} {\bibinfo  {journal} {Physical Review Letters}\ }\textbf {\bibinfo {volume} {95}},\ \bibinfo {pages} {076805} (\bibinfo {year} {2005})}\BibitemShut {NoStop}%
\bibitem [{\citenamefont {Winkler}\ \emph {et~al.}(2001)\citenamefont {Winkler}, \citenamefont {Papadakis}, \citenamefont {De~Poortere},\ and\ \citenamefont {Shayegan}}]{winkler2001spin}%
  \BibitemOpen
  \bibfield  {author} {\bibinfo {author} {\bibfnamefont {R.}~\bibnamefont {Winkler}}, \bibinfo {author} {\bibfnamefont {S.}~\bibnamefont {Papadakis}}, \bibinfo {author} {\bibfnamefont {E.}~\bibnamefont {De~Poortere}},\ and\ \bibinfo {author} {\bibfnamefont {M.}~\bibnamefont {Shayegan}},\ }\bibfield  {title} {\bibinfo {title} {Spin-orbit coupling in two-dimensional electron and hole systems},\ }in\ \href {https://link.springer.com/10.1007/b13586} {\emph {\bibinfo {booktitle} {Advances in Solid State Physics}}}\ (\bibinfo  {publisher} {Springer},\ \bibinfo {year} {2001})\ pp.\ \bibinfo {pages} {211--223}\BibitemShut {NoStop}%
\bibitem [{\citenamefont {Marcellina}\ \emph {et~al.}(2017)\citenamefont {Marcellina}, \citenamefont {Hamilton}, \citenamefont {Winkler},\ and\ \citenamefont {Culcer}}]{marcellina2017spin}%
  \BibitemOpen
  \bibfield  {author} {\bibinfo {author} {\bibfnamefont {E.}~\bibnamefont {Marcellina}}, \bibinfo {author} {\bibfnamefont {A.}~\bibnamefont {Hamilton}}, \bibinfo {author} {\bibfnamefont {R.}~\bibnamefont {Winkler}},\ and\ \bibinfo {author} {\bibfnamefont {D.}~\bibnamefont {Culcer}},\ }\bibfield  {title} {\bibinfo {title} {Spin-orbit interactions in inversion-asymmetric two-dimensional hole systems: A variational analysis},\ }\href {https://link.aps.org/doi/10.1103/PhysRevB.95.075305} {\bibfield  {journal} {\bibinfo  {journal} {Physical Review B}\ }\textbf {\bibinfo {volume} {95}},\ \bibinfo {pages} {075305} (\bibinfo {year} {2017})}\BibitemShut {NoStop}%
\bibitem [{\citenamefont {Rodr{\'\i}guez-Mena}\ \emph {et~al.}(2023)\citenamefont {Rodr{\'\i}guez-Mena}, \citenamefont {Abadillo-Uriel}, \citenamefont {Veste}, \citenamefont {Martinez}, \citenamefont {Li}, \citenamefont {Skl{\'e}nard},\ and\ \citenamefont {Niquet}}]{rodriguez2023linear}%
  \BibitemOpen
  \bibfield  {author} {\bibinfo {author} {\bibfnamefont {E.~A.}\ \bibnamefont {Rodr{\'\i}guez-Mena}}, \bibinfo {author} {\bibfnamefont {J.~C.}\ \bibnamefont {Abadillo-Uriel}}, \bibinfo {author} {\bibfnamefont {G.}~\bibnamefont {Veste}}, \bibinfo {author} {\bibfnamefont {B.}~\bibnamefont {Martinez}}, \bibinfo {author} {\bibfnamefont {J.}~\bibnamefont {Li}}, \bibinfo {author} {\bibfnamefont {B.}~\bibnamefont {Skl{\'e}nard}},\ and\ \bibinfo {author} {\bibfnamefont {Y.-M.}\ \bibnamefont {Niquet}},\ }\bibfield  {title} {\bibinfo {title} {Linear-in-momentum spin orbit interactions in planar {G}e/{G}e{S}i heterostructures and spin qubits},\ }\href {https://link.aps.org/doi/10.1103/PhysRevB.108.205416} {\bibfield  {journal} {\bibinfo  {journal} {Physical Review B}\ }\textbf {\bibinfo {volume} {108}},\ \bibinfo {pages} {205416} (\bibinfo {year} {2023})}\BibitemShut {NoStop}%
\bibitem [{\citenamefont {Sarkar}\ \emph {et~al.}(2025)\citenamefont {Sarkar}, \citenamefont {Chowdhury}, \citenamefont {Hu}, \citenamefont {Saraiva}, \citenamefont {Dzurak}, \citenamefont {Hamilton}, \citenamefont {Rahman},\ and\ \citenamefont {Culcer}}]{sarkar2025effect}%
  \BibitemOpen
  \bibfield  {author} {\bibinfo {author} {\bibfnamefont {A.}~\bibnamefont {Sarkar}}, \bibinfo {author} {\bibfnamefont {P.}~\bibnamefont {Chowdhury}}, \bibinfo {author} {\bibfnamefont {X.}~\bibnamefont {Hu}}, \bibinfo {author} {\bibfnamefont {A.}~\bibnamefont {Saraiva}}, \bibinfo {author} {\bibfnamefont {A.}~\bibnamefont {Dzurak}}, \bibinfo {author} {\bibfnamefont {A.}~\bibnamefont {Hamilton}}, \bibinfo {author} {\bibfnamefont {R.}~\bibnamefont {Rahman}},\ and\ \bibinfo {author} {\bibfnamefont {D.}~\bibnamefont {Culcer}},\ }\bibfield  {title} {\bibinfo {title} {Effect of disorder and strain on the operation of planar {G}e hole spin qubits},\ }\href {https://doi.org/10.1038/s41534-025-01130-w} {\bibfield  {journal} {\bibinfo  {journal} {npj Quantum Information}\ }\textbf {\bibinfo {volume} {11}},\ \bibinfo {pages} {185} (\bibinfo {year} {2025})}\BibitemShut {NoStop}%
\bibitem [{\citenamefont {Assali}\ \emph {et~al.}(2022)\citenamefont {Assali}, \citenamefont {Attiaoui}, \citenamefont {Vecchio}, \citenamefont {Mukherjee}, \citenamefont {Nicolas},\ and\ \citenamefont {Moutanabbir}}]{assali2022light}%
  \BibitemOpen
  \bibfield  {author} {\bibinfo {author} {\bibfnamefont {S.}~\bibnamefont {Assali}}, \bibinfo {author} {\bibfnamefont {A.}~\bibnamefont {Attiaoui}}, \bibinfo {author} {\bibfnamefont {P.~D.}\ \bibnamefont {Vecchio}}, \bibinfo {author} {\bibfnamefont {S.}~\bibnamefont {Mukherjee}}, \bibinfo {author} {\bibfnamefont {J.}~\bibnamefont {Nicolas}},\ and\ \bibinfo {author} {\bibfnamefont {O.}~\bibnamefont {Moutanabbir}},\ }\bibfield  {title} {\bibinfo {title} {A light-hole germanium quantum well on silicon},\ }\href {https://advanced.onlinelibrary.wiley.com/doi/abs/10.1002/adma.202201192} {\bibfield  {journal} {\bibinfo  {journal} {Advanced Materials}\ }\textbf {\bibinfo {volume} {34}},\ \bibinfo {pages} {2201192} (\bibinfo {year} {2022})}\BibitemShut {NoStop}%
\bibitem [{\citenamefont {Del~Vecchio}\ and\ \citenamefont {Moutanabbir}(2023)}]{del2023light}%
  \BibitemOpen
  \bibfield  {author} {\bibinfo {author} {\bibfnamefont {P.}~\bibnamefont {Del~Vecchio}}\ and\ \bibinfo {author} {\bibfnamefont {O.}~\bibnamefont {Moutanabbir}},\ }\bibfield  {title} {\bibinfo {title} {Light-hole gate-defined spin-orbit qubit},\ }\href {https://link.aps.org/doi/10.1103/PhysRevB.107.L161406} {\bibfield  {journal} {\bibinfo  {journal} {Physical Review B}\ }\textbf {\bibinfo {volume} {107}},\ \bibinfo {pages} {L161406} (\bibinfo {year} {2023})}\BibitemShut {NoStop}%
\bibitem [{\citenamefont {Del~Vecchio}\ and\ \citenamefont {Moutanabbir}(2024)}]{del2024light}%
  \BibitemOpen
  \bibfield  {author} {\bibinfo {author} {\bibfnamefont {P.}~\bibnamefont {Del~Vecchio}}\ and\ \bibinfo {author} {\bibfnamefont {O.}~\bibnamefont {Moutanabbir}},\ }\bibfield  {title} {\bibinfo {title} {Light-hole spin confined in germanium},\ }\href {https://link.aps.org/doi/10.1103/PhysRevB.110.045409} {\bibfield  {journal} {\bibinfo  {journal} {Physical Review B}\ }\textbf {\bibinfo {volume} {110}},\ \bibinfo {pages} {045409} (\bibinfo {year} {2024})}\BibitemShut {NoStop}%
\bibitem [{\citenamefont {Kloeffel}\ \emph {et~al.}(2011)\citenamefont {Kloeffel}, \citenamefont {Trif},\ and\ \citenamefont {Loss}}]{kloeffel2011strong}%
  \BibitemOpen
  \bibfield  {author} {\bibinfo {author} {\bibfnamefont {C.}~\bibnamefont {Kloeffel}}, \bibinfo {author} {\bibfnamefont {M.}~\bibnamefont {Trif}},\ and\ \bibinfo {author} {\bibfnamefont {D.}~\bibnamefont {Loss}},\ }\bibfield  {title} {\bibinfo {title} {Strong spin-orbit interaction and helical hole states in {G}e/{S}i nanowires},\ }\href {https://link.aps.org/doi/10.1103/PhysRevB.84.195314} {\bibfield  {journal} {\bibinfo  {journal} {Physical Review B}\ }\textbf {\bibinfo {volume} {84}},\ \bibinfo {pages} {195314} (\bibinfo {year} {2011})}\BibitemShut {NoStop}%
\bibitem [{\citenamefont {Kloeffel}\ \emph {et~al.}(2018)\citenamefont {Kloeffel}, \citenamefont {Ran{\v{c}}i{\'c}},\ and\ \citenamefont {Loss}}]{kloeffel2018direct}%
  \BibitemOpen
  \bibfield  {author} {\bibinfo {author} {\bibfnamefont {C.}~\bibnamefont {Kloeffel}}, \bibinfo {author} {\bibfnamefont {M.~J.}\ \bibnamefont {Ran{\v{c}}i{\'c}}},\ and\ \bibinfo {author} {\bibfnamefont {D.}~\bibnamefont {Loss}},\ }\bibfield  {title} {\bibinfo {title} {Direct rashba spin-orbit interaction in {S}i and {G}e nanowires with different growth directions},\ }\href {https://link.aps.org/doi/10.1103/PhysRevB.97.235422} {\bibfield  {journal} {\bibinfo  {journal} {Physical Review B}\ }\textbf {\bibinfo {volume} {97}},\ \bibinfo {pages} {235422} (\bibinfo {year} {2018})}\BibitemShut {NoStop}%
\bibitem [{\citenamefont {Carballido}\ \emph {et~al.}(2025)\citenamefont {Carballido}, \citenamefont {Svab}, \citenamefont {Eggli}, \citenamefont {Patlatiuk}, \citenamefont {Chevalier~Kwon}, \citenamefont {Schuff}, \citenamefont {Kaiser}, \citenamefont {Camenzind}, \citenamefont {Li}, \citenamefont {Ares} \emph {et~al.}}]{carballido2025compromise}%
  \BibitemOpen
  \bibfield  {author} {\bibinfo {author} {\bibfnamefont {M.~J.}\ \bibnamefont {Carballido}}, \bibinfo {author} {\bibfnamefont {S.}~\bibnamefont {Svab}}, \bibinfo {author} {\bibfnamefont {R.~S.}\ \bibnamefont {Eggli}}, \bibinfo {author} {\bibfnamefont {T.}~\bibnamefont {Patlatiuk}}, \bibinfo {author} {\bibfnamefont {P.}~\bibnamefont {Chevalier~Kwon}}, \bibinfo {author} {\bibfnamefont {J.}~\bibnamefont {Schuff}}, \bibinfo {author} {\bibfnamefont {R.~M.}\ \bibnamefont {Kaiser}}, \bibinfo {author} {\bibfnamefont {L.~C.}\ \bibnamefont {Camenzind}}, \bibinfo {author} {\bibfnamefont {A.}~\bibnamefont {Li}}, \bibinfo {author} {\bibfnamefont {N.}~\bibnamefont {Ares}}, \emph {et~al.},\ }\bibfield  {title} {\bibinfo {title} {Compromise-free scaling of qubit speed and coherence},\ }\href {https://doi.org/10.1038/s41467-025-62614-z} {\bibfield  {journal} {\bibinfo  {journal} {Nature Communications}\ }\textbf {\bibinfo {volume} {16}},\ \bibinfo {pages} {7616} (\bibinfo {year} {2025})}\BibitemShut {NoStop}%
\bibitem [{\citenamefont {Sarkar}\ \emph {et~al.}(2023)\citenamefont {Sarkar}, \citenamefont {Wang}, \citenamefont {Rendell}, \citenamefont {Hendrickx}, \citenamefont {Veldhorst}, \citenamefont {Scappucci}, \citenamefont {Khalifa}, \citenamefont {Salfi}, \citenamefont {Saraiva}, \citenamefont {Dzurak} \emph {et~al.}}]{sarkar2023electrical}%
  \BibitemOpen
  \bibfield  {author} {\bibinfo {author} {\bibfnamefont {A.}~\bibnamefont {Sarkar}}, \bibinfo {author} {\bibfnamefont {Z.}~\bibnamefont {Wang}}, \bibinfo {author} {\bibfnamefont {M.}~\bibnamefont {Rendell}}, \bibinfo {author} {\bibfnamefont {N.~W.}\ \bibnamefont {Hendrickx}}, \bibinfo {author} {\bibfnamefont {M.}~\bibnamefont {Veldhorst}}, \bibinfo {author} {\bibfnamefont {G.}~\bibnamefont {Scappucci}}, \bibinfo {author} {\bibfnamefont {M.}~\bibnamefont {Khalifa}}, \bibinfo {author} {\bibfnamefont {J.}~\bibnamefont {Salfi}}, \bibinfo {author} {\bibfnamefont {A.}~\bibnamefont {Saraiva}}, \bibinfo {author} {\bibfnamefont {A.}~\bibnamefont {Dzurak}}, \emph {et~al.},\ }\bibfield  {title} {\bibinfo {title} {Electrical operation of planar {G}e hole spin qubits in an in-plane magnetic field},\ }\href {https://link.aps.org/doi/10.1103/PhysRevB.108.245301} {\bibfield  {journal} {\bibinfo  {journal} {Physical Review B}\ }\textbf {\bibinfo {volume} {108}},\ \bibinfo {pages} {245301} (\bibinfo {year} {2023})}\BibitemShut
  {NoStop}%
\bibitem [{\citenamefont {Wang}\ \emph {et~al.}(2024{\natexlab{b}})\citenamefont {Wang}, \citenamefont {Ercan}, \citenamefont {Gyure}, \citenamefont {Scappucci}, \citenamefont {Veldhorst},\ and\ \citenamefont {Rimbach-Russ}}]{wang2024modeling}%
  \BibitemOpen
  \bibfield  {author} {\bibinfo {author} {\bibfnamefont {C.-A.}\ \bibnamefont {Wang}}, \bibinfo {author} {\bibfnamefont {H.~E.}\ \bibnamefont {Ercan}}, \bibinfo {author} {\bibfnamefont {M.~F.}\ \bibnamefont {Gyure}}, \bibinfo {author} {\bibfnamefont {G.}~\bibnamefont {Scappucci}}, \bibinfo {author} {\bibfnamefont {M.}~\bibnamefont {Veldhorst}},\ and\ \bibinfo {author} {\bibfnamefont {M.}~\bibnamefont {Rimbach-Russ}},\ }\bibfield  {title} {\bibinfo {title} {Modeling of planar germanium hole qubits in electric and magnetic fields},\ }\href {https://doi.org/10.1038/s41534-024-00897-8} {\bibfield  {journal} {\bibinfo  {journal} {npj Quantum Information}\ }\textbf {\bibinfo {volume} {10}},\ \bibinfo {pages} {102} (\bibinfo {year} {2024}{\natexlab{b}})}\BibitemShut {NoStop}%
\bibitem [{\citenamefont {Miles}\ \emph {et~al.}(2025)\citenamefont {Miles}, \citenamefont {Bozkurt}, \citenamefont {Varjas},\ and\ \citenamefont {Wimmer}}]{miles2025effective}%
  \BibitemOpen
  \bibfield  {author} {\bibinfo {author} {\bibfnamefont {S.}~\bibnamefont {Miles}}, \bibinfo {author} {\bibfnamefont {A.~M.}\ \bibnamefont {Bozkurt}}, \bibinfo {author} {\bibfnamefont {D.}~\bibnamefont {Varjas}},\ and\ \bibinfo {author} {\bibfnamefont {M.}~\bibnamefont {Wimmer}},\ }\bibfield  {title} {\bibinfo {title} {Effective hamiltonians for {G}e/{S}i core/shell nanowires from higher order perturbation theory},\ }\href {https://arxiv.org/abs/2511.11809} {\bibfield  {journal} {\bibinfo  {journal} {arXiv preprint arXiv:2511.11809}\ } (\bibinfo {year} {2025})}\BibitemShut {NoStop}%
\bibitem [{\citenamefont {Adelsberger}\ \emph {et~al.}(2022)\citenamefont {Adelsberger}, \citenamefont {Benito}, \citenamefont {Bosco}, \citenamefont {Klinovaja},\ and\ \citenamefont {Loss}}]{adelsberger2022hole}%
  \BibitemOpen
  \bibfield  {author} {\bibinfo {author} {\bibfnamefont {C.}~\bibnamefont {Adelsberger}}, \bibinfo {author} {\bibfnamefont {M.}~\bibnamefont {Benito}}, \bibinfo {author} {\bibfnamefont {S.}~\bibnamefont {Bosco}}, \bibinfo {author} {\bibfnamefont {J.}~\bibnamefont {Klinovaja}},\ and\ \bibinfo {author} {\bibfnamefont {D.}~\bibnamefont {Loss}},\ }\bibfield  {title} {\bibinfo {title} {Hole-spin qubits in {G}e nanowire quantum dots: Interplay of orbital magnetic field, strain, and growth direction},\ }\href {https://link.aps.org/doi/10.1103/PhysRevB.105.075308} {\bibfield  {journal} {\bibinfo  {journal} {Physical Review B}\ }\textbf {\bibinfo {volume} {105}},\ \bibinfo {pages} {075308} (\bibinfo {year} {2022})}\BibitemShut {NoStop}%
\bibitem [{\citenamefont {Hendrickx}\ \emph {et~al.}(2024)\citenamefont {Hendrickx}, \citenamefont {Massai}, \citenamefont {Mergenthaler}, \citenamefont {Schupp}, \citenamefont {Paredes}, \citenamefont {Bedell}, \citenamefont {Salis},\ and\ \citenamefont {Fuhrer}}]{hendrickx2024sweet}%
  \BibitemOpen
  \bibfield  {author} {\bibinfo {author} {\bibfnamefont {N.}~\bibnamefont {Hendrickx}}, \bibinfo {author} {\bibfnamefont {L.}~\bibnamefont {Massai}}, \bibinfo {author} {\bibfnamefont {M.}~\bibnamefont {Mergenthaler}}, \bibinfo {author} {\bibfnamefont {F.~J.}\ \bibnamefont {Schupp}}, \bibinfo {author} {\bibfnamefont {S.}~\bibnamefont {Paredes}}, \bibinfo {author} {\bibfnamefont {S.}~\bibnamefont {Bedell}}, \bibinfo {author} {\bibfnamefont {G.}~\bibnamefont {Salis}},\ and\ \bibinfo {author} {\bibfnamefont {A.}~\bibnamefont {Fuhrer}},\ }\bibfield  {title} {\bibinfo {title} {Sweet-spot operation of a germanium hole spin qubit with highly anisotropic noise sensitivity},\ }\href {https://doi.org/10.1038/s41563-024-01857-5} {\bibfield  {journal} {\bibinfo  {journal} {Nature Materials}\ }\textbf {\bibinfo {volume} {23}},\ \bibinfo {pages} {920} (\bibinfo {year} {2024})}\BibitemShut {NoStop}%
\bibitem [{\citenamefont {Brauns}\ \emph {et~al.}(2016)\citenamefont {Brauns}, \citenamefont {Ridderbos}, \citenamefont {Li}, \citenamefont {Bakkers}, \citenamefont {Van Der~Wiel},\ and\ \citenamefont {Zwanenburg}}]{brauns2016anisotropic}%
  \BibitemOpen
  \bibfield  {author} {\bibinfo {author} {\bibfnamefont {M.}~\bibnamefont {Brauns}}, \bibinfo {author} {\bibfnamefont {J.}~\bibnamefont {Ridderbos}}, \bibinfo {author} {\bibfnamefont {A.}~\bibnamefont {Li}}, \bibinfo {author} {\bibfnamefont {E.~P.}\ \bibnamefont {Bakkers}}, \bibinfo {author} {\bibfnamefont {W.~G.}\ \bibnamefont {Van Der~Wiel}},\ and\ \bibinfo {author} {\bibfnamefont {F.~A.}\ \bibnamefont {Zwanenburg}},\ }\bibfield  {title} {\bibinfo {title} {Anisotropic {P}auli spin blockade in hole quantum dots},\ }\href {https://link.aps.org/doi/10.1103/PhysRevB.94.041411} {\bibfield  {journal} {\bibinfo  {journal} {Physical Review B}\ }\textbf {\bibinfo {volume} {94}},\ \bibinfo {pages} {041411} (\bibinfo {year} {2016})}\BibitemShut {NoStop}%
\bibitem [{\citenamefont {Froning}\ \emph {et~al.}(2018)\citenamefont {Froning}, \citenamefont {Rehmann}, \citenamefont {Ridderbos}, \citenamefont {Brauns}, \citenamefont {Zwanenburg}, \citenamefont {Li}, \citenamefont {Bakkers}, \citenamefont {Zumb{\"u}hl},\ and\ \citenamefont {Braakman}}]{froning2018single}%
  \BibitemOpen
  \bibfield  {author} {\bibinfo {author} {\bibfnamefont {F.}~\bibnamefont {Froning}}, \bibinfo {author} {\bibfnamefont {M.}~\bibnamefont {Rehmann}}, \bibinfo {author} {\bibfnamefont {J.}~\bibnamefont {Ridderbos}}, \bibinfo {author} {\bibfnamefont {M.}~\bibnamefont {Brauns}}, \bibinfo {author} {\bibfnamefont {F.}~\bibnamefont {Zwanenburg}}, \bibinfo {author} {\bibfnamefont {A.}~\bibnamefont {Li}}, \bibinfo {author} {\bibfnamefont {E.}~\bibnamefont {Bakkers}}, \bibinfo {author} {\bibfnamefont {D.}~\bibnamefont {Zumb{\"u}hl}},\ and\ \bibinfo {author} {\bibfnamefont {F.}~\bibnamefont {Braakman}},\ }\bibfield  {title} {\bibinfo {title} {Single, double, and triple quantum dots in {G}e/{S}i nanowires},\ }\href {https://doi.org/10.1063/1.5042501} {\bibfield  {journal} {\bibinfo  {journal} {Applied Physics Letters}\ }\textbf {\bibinfo {volume} {113}} (\bibinfo {year} {2018})}\BibitemShut {NoStop}%
\bibitem [{\citenamefont {Xu}\ \emph {et~al.}(2025)\citenamefont {Xu}, \citenamefont {Haller}, \citenamefont {Hegde}, \citenamefont {Meng},\ and\ \citenamefont {Schmidt}}]{xu2025quantum}%
  \BibitemOpen
  \bibfield  {author} {\bibinfo {author} {\bibfnamefont {C.}~\bibnamefont {Xu}}, \bibinfo {author} {\bibfnamefont {A.}~\bibnamefont {Haller}}, \bibinfo {author} {\bibfnamefont {S.}~\bibnamefont {Hegde}}, \bibinfo {author} {\bibfnamefont {T.}~\bibnamefont {Meng}},\ and\ \bibinfo {author} {\bibfnamefont {T.~L.}\ \bibnamefont {Schmidt}},\ }\bibfield  {title} {\bibinfo {title} {Quantum geometry in the dynamics of band-projected operators},\ }\href {https://arxiv.org/abs/2503.11425} {\bibfield  {journal} {\bibinfo  {journal} {arXiv preprint arXiv:2503.11425}\ } (\bibinfo {year} {2025})}\BibitemShut {NoStop}%
\bibitem [{\citenamefont {Lin}\ \emph {et~al.}(2018)\citenamefont {Lin}, \citenamefont {Liu}, \citenamefont {Ying}, \citenamefont {Chen},\ and\ \citenamefont {Wu}}]{lin2018explicit}%
  \BibitemOpen
  \bibfield  {author} {\bibinfo {author} {\bibfnamefont {X.}~\bibnamefont {Lin}}, \bibinfo {author} {\bibfnamefont {X.}~\bibnamefont {Liu}}, \bibinfo {author} {\bibfnamefont {F.}~\bibnamefont {Ying}}, \bibinfo {author} {\bibfnamefont {Z.}~\bibnamefont {Chen}},\ and\ \bibinfo {author} {\bibfnamefont {W.}~\bibnamefont {Wu}},\ }\bibfield  {title} {\bibinfo {title} {Explicit construction of diabatic state and its application to the direct evaluation of electronic coupling},\ }\href {https://doi.org/10.1063/1.5035114} {\bibfield  {journal} {\bibinfo  {journal} {The Journal of Chemical Physics}\ }\textbf {\bibinfo {volume} {149}} (\bibinfo {year} {2018})}\BibitemShut {NoStop}%
\bibitem [{\citenamefont {Fatehi}\ \emph {et~al.}(2013)\citenamefont {Fatehi}, \citenamefont {Alguire},\ and\ \citenamefont {Subotnik}}]{fatehi2013derivative}%
  \BibitemOpen
  \bibfield  {author} {\bibinfo {author} {\bibfnamefont {S.}~\bibnamefont {Fatehi}}, \bibinfo {author} {\bibfnamefont {E.}~\bibnamefont {Alguire}},\ and\ \bibinfo {author} {\bibfnamefont {J.~E.}\ \bibnamefont {Subotnik}},\ }\bibfield  {title} {\bibinfo {title} {Derivative couplings and analytic gradients for diabatic states, with an implementation for {B}oys-localized configuration-interaction singles},\ }\href {https://doi.org/10.1063/1.4820485} {\bibfield  {journal} {\bibinfo  {journal} {The Journal of Chemical Physics}\ }\textbf {\bibinfo {volume} {139}} (\bibinfo {year} {2013})}\BibitemShut {NoStop}%
\bibitem [{\citenamefont {Subotnik}\ \emph {et~al.}(2008)\citenamefont {Subotnik}, \citenamefont {Yeganeh}, \citenamefont {Cave},\ and\ \citenamefont {Ratner}}]{subotnik2008constructing}%
  \BibitemOpen
  \bibfield  {author} {\bibinfo {author} {\bibfnamefont {J.~E.}\ \bibnamefont {Subotnik}}, \bibinfo {author} {\bibfnamefont {S.}~\bibnamefont {Yeganeh}}, \bibinfo {author} {\bibfnamefont {R.~J.}\ \bibnamefont {Cave}},\ and\ \bibinfo {author} {\bibfnamefont {M.~A.}\ \bibnamefont {Ratner}},\ }\bibfield  {title} {\bibinfo {title} {Constructing diabatic states from adiabatic states: Extending generalized {M}ulliken--{H}ush to multiple charge centers with boys localization},\ }\bibfield  {journal} {\bibinfo  {journal} {The Journal of Chemical Physics}\ }\textbf {\bibinfo {volume} {129}},\ \href {https://doi.org/10.1063/1.3042233} {10.1063/1.3042233} (\bibinfo {year} {2008})\BibitemShut {NoStop}%
\bibitem [{\citenamefont {Littlejohn}\ \emph {et~al.}(2022)\citenamefont {Littlejohn}, \citenamefont {Rawlinson},\ and\ \citenamefont {Subotnik}}]{littlejohn2022parallel}%
  \BibitemOpen
  \bibfield  {author} {\bibinfo {author} {\bibfnamefont {R.}~\bibnamefont {Littlejohn}}, \bibinfo {author} {\bibfnamefont {J.}~\bibnamefont {Rawlinson}},\ and\ \bibinfo {author} {\bibfnamefont {J.}~\bibnamefont {Subotnik}},\ }\bibfield  {title} {\bibinfo {title} {The parallel-transported (quasi)-diabatic basis},\ }\href {https://pubs.aip.org/aip/jcp/article/157/18/184303/2842095/The-parallel-transported-quasi-diabatic-basis} {\bibfield  {journal} {\bibinfo  {journal} {The Journal of Chemical Physics}\ }\textbf {\bibinfo {volume} {157}} (\bibinfo {year} {2022})}\BibitemShut {NoStop}%
\bibitem [{\citenamefont {Souza}\ \emph {et~al.}(2001)\citenamefont {Souza}, \citenamefont {Marzari},\ and\ \citenamefont {Vanderbilt}}]{souza2001maximally}%
  \BibitemOpen
  \bibfield  {author} {\bibinfo {author} {\bibfnamefont {I.}~\bibnamefont {Souza}}, \bibinfo {author} {\bibfnamefont {N.}~\bibnamefont {Marzari}},\ and\ \bibinfo {author} {\bibfnamefont {D.}~\bibnamefont {Vanderbilt}},\ }\bibfield  {title} {\bibinfo {title} {Maximally localized {W}annier functions for entangled energy bands},\ }\href {https://link.aps.org/doi/10.1103/PhysRevB.65.035109} {\bibfield  {journal} {\bibinfo  {journal} {Physical Review B}\ }\textbf {\bibinfo {volume} {65}},\ \bibinfo {pages} {035109} (\bibinfo {year} {2001})}\BibitemShut {NoStop}%
\bibitem [{\citenamefont {Mostofi}\ \emph {et~al.}(2008)\citenamefont {Mostofi}, \citenamefont {Yates}, \citenamefont {Lee}, \citenamefont {Souza}, \citenamefont {Vanderbilt},\ and\ \citenamefont {Marzari}}]{mostofi2008wannier90}%
  \BibitemOpen
  \bibfield  {author} {\bibinfo {author} {\bibfnamefont {A.~A.}\ \bibnamefont {Mostofi}}, \bibinfo {author} {\bibfnamefont {J.~R.}\ \bibnamefont {Yates}}, \bibinfo {author} {\bibfnamefont {Y.-S.}\ \bibnamefont {Lee}}, \bibinfo {author} {\bibfnamefont {I.}~\bibnamefont {Souza}}, \bibinfo {author} {\bibfnamefont {D.}~\bibnamefont {Vanderbilt}},\ and\ \bibinfo {author} {\bibfnamefont {N.}~\bibnamefont {Marzari}},\ }\bibfield  {title} {\bibinfo {title} {Wannier90: A tool for obtaining maximally-localised {W}annier functions},\ }\href {https://www.sciencedirect.com/science/article/pii/S0010465507004936} {\bibfield  {journal} {\bibinfo  {journal} {Computer Physics Communications}\ }\textbf {\bibinfo {volume} {178}},\ \bibinfo {pages} {685} (\bibinfo {year} {2008})}\BibitemShut {NoStop}%
\bibitem [{\citenamefont {Yates}\ \emph {et~al.}(2007)\citenamefont {Yates}, \citenamefont {Wang}, \citenamefont {Vanderbilt},\ and\ \citenamefont {Souza}}]{yates2007spectral}%
  \BibitemOpen
  \bibfield  {author} {\bibinfo {author} {\bibfnamefont {J.~R.}\ \bibnamefont {Yates}}, \bibinfo {author} {\bibfnamefont {X.}~\bibnamefont {Wang}}, \bibinfo {author} {\bibfnamefont {D.}~\bibnamefont {Vanderbilt}},\ and\ \bibinfo {author} {\bibfnamefont {I.}~\bibnamefont {Souza}},\ }\bibfield  {title} {\bibinfo {title} {Spectral and fermi surface properties from wannier interpolation},\ }\href {https://link.aps.org/doi/10.1103/PhysRevB.75.195121} {\bibfield  {journal} {\bibinfo  {journal} {Physical Review B}\ }\textbf {\bibinfo {volume} {75}},\ \bibinfo {pages} {195121} (\bibinfo {year} {2007})}\BibitemShut {NoStop}%
\bibitem [{\citenamefont {Ozaki}(2024)}]{ozaki2024closest}%
  \BibitemOpen
  \bibfield  {author} {\bibinfo {author} {\bibfnamefont {T.}~\bibnamefont {Ozaki}},\ }\bibfield  {title} {\bibinfo {title} {Closest {W}annier functions to a given set of localized orbitals},\ }\href {https://link.aps.org/doi/10.1103/PhysRevB.110.125115} {\bibfield  {journal} {\bibinfo  {journal} {Physical Review B}\ }\textbf {\bibinfo {volume} {110}},\ \bibinfo {pages} {125115} (\bibinfo {year} {2024})}\BibitemShut {NoStop}%
\bibitem [{\citenamefont {Miserev}\ and\ \citenamefont {Sushkov}(2017)}]{miserev2017dimensional}%
  \BibitemOpen
  \bibfield  {author} {\bibinfo {author} {\bibfnamefont {D.}~\bibnamefont {Miserev}}\ and\ \bibinfo {author} {\bibfnamefont {O.}~\bibnamefont {Sushkov}},\ }\bibfield  {title} {\bibinfo {title} {Dimensional reduction of the luttinger hamiltonian and g-factors of holes in symmetric two-dimensional semiconductor heterostructures},\ }\href {https://link.aps.org/doi/10.1103/PhysRevB.95.085431} {\bibfield  {journal} {\bibinfo  {journal} {Physical Review B}\ }\textbf {\bibinfo {volume} {95}},\ \bibinfo {pages} {085431} (\bibinfo {year} {2017})}\BibitemShut {NoStop}%
\bibitem [{\citenamefont {Del~Vecchio}\ \emph {et~al.}(2025)\citenamefont {Del~Vecchio}, \citenamefont {Bosco}, \citenamefont {Loss},\ and\ \citenamefont {Moutanabbir}}]{del2025fully}%
  \BibitemOpen
  \bibfield  {author} {\bibinfo {author} {\bibfnamefont {P.}~\bibnamefont {Del~Vecchio}}, \bibinfo {author} {\bibfnamefont {S.}~\bibnamefont {Bosco}}, \bibinfo {author} {\bibfnamefont {D.}~\bibnamefont {Loss}},\ and\ \bibinfo {author} {\bibfnamefont {O.}~\bibnamefont {Moutanabbir}},\ }\bibfield  {title} {\bibinfo {title} {Fully tunable strong spin-orbit interactions in light hole germanium quantum channels},\ }\href {https://arxiv.org/abs/2506.14759} {\bibfield  {journal} {\bibinfo  {journal} {arXiv preprint arXiv:2506.14759}\ } (\bibinfo {year} {2025})}\BibitemShut {NoStop}%
\bibitem [{\citenamefont {Kloeffel}\ \emph {et~al.}(2014)\citenamefont {Kloeffel}, \citenamefont {Trif},\ and\ \citenamefont {Loss}}]{kloeffel2014acoustic}%
  \BibitemOpen
  \bibfield  {author} {\bibinfo {author} {\bibfnamefont {C.}~\bibnamefont {Kloeffel}}, \bibinfo {author} {\bibfnamefont {M.}~\bibnamefont {Trif}},\ and\ \bibinfo {author} {\bibfnamefont {D.}~\bibnamefont {Loss}},\ }\bibfield  {title} {\bibinfo {title} {Acoustic phonons and strain in core/shell nanowires},\ }\href {https://link.aps.org/doi/10.1103/PhysRevB.90.115419} {\bibfield  {journal} {\bibinfo  {journal} {Physical Review B}\ }\textbf {\bibinfo {volume} {90}},\ \bibinfo {pages} {115419} (\bibinfo {year} {2014})}\BibitemShut {NoStop}%
\bibitem [{\citenamefont {Luttinger}(1956)}]{luttinger1956quantum}%
  \BibitemOpen
  \bibfield  {author} {\bibinfo {author} {\bibfnamefont {J.~M.}\ \bibnamefont {Luttinger}},\ }\bibfield  {title} {\bibinfo {title} {Quantum theory of cyclotron resonance in semiconductors: General theory},\ }\href {https://link.aps.org/doi/10.1103/PhysRev.102.1030} {\bibfield  {journal} {\bibinfo  {journal} {Physical Review}\ }\textbf {\bibinfo {volume} {102}},\ \bibinfo {pages} {1030} (\bibinfo {year} {1956})}\BibitemShut {NoStop}%
\bibitem [{\citenamefont {Bir}(1974)}]{bir1974symmetry}%
  \BibitemOpen
  \bibfield  {author} {\bibinfo {author} {\bibfnamefont {G.~E.}\ \bibnamefont {Bir}},\ }\bibfield  {title} {\bibinfo {title} {Symmetry and strain-induced effects in semiconductors},\ }\href@noop {} {\bibfield  {journal} {\bibinfo  {journal} {Wiley}\ } (\bibinfo {year} {1974})}\BibitemShut {NoStop}%
\bibitem [{\citenamefont {Rotaru}\ \emph {et~al.}(2025)\citenamefont {Rotaru}, \citenamefont {Del~Vecchio},\ and\ \citenamefont {Moutanabbir}}]{rotaru2025hole}%
  \BibitemOpen
  \bibfield  {author} {\bibinfo {author} {\bibfnamefont {N.}~\bibnamefont {Rotaru}}, \bibinfo {author} {\bibfnamefont {P.}~\bibnamefont {Del~Vecchio}},\ and\ \bibinfo {author} {\bibfnamefont {O.}~\bibnamefont {Moutanabbir}},\ }\bibfield  {title} {\bibinfo {title} {Hole spin in direct bandgap germanium-tin quantum dot},\ }\href {https://link.aps.org/doi/10.1103/nr1v-sngw} {\bibfield  {journal} {\bibinfo  {journal} {Physical Review B}\ }\textbf {\bibinfo {volume} {112}},\ \bibinfo {pages} {125428} (\bibinfo {year} {2025})}\BibitemShut {NoStop}%
\bibitem [{\citenamefont {Ando}\ \emph {et~al.}(1982)\citenamefont {Ando}, \citenamefont {Fowler},\ and\ \citenamefont {Stern}}]{ando1982electronic}%
  \BibitemOpen
  \bibfield  {author} {\bibinfo {author} {\bibfnamefont {T.}~\bibnamefont {Ando}}, \bibinfo {author} {\bibfnamefont {A.~B.}\ \bibnamefont {Fowler}},\ and\ \bibinfo {author} {\bibfnamefont {F.}~\bibnamefont {Stern}},\ }\bibfield  {title} {\bibinfo {title} {Electronic properties of two-dimensional systems},\ }\href {https://link.aps.org/doi/10.1103/RevModPhys.54.437} {\bibfield  {journal} {\bibinfo  {journal} {Reviews of Modern Physics}\ }\textbf {\bibinfo {volume} {54}},\ \bibinfo {pages} {437} (\bibinfo {year} {1982})}\BibitemShut {NoStop}%
\bibitem [{\citenamefont {Fraj}\ \emph {et~al.}(2007)\citenamefont {Fraj}, \citenamefont {Sa{\"\i}di}, \citenamefont {Ben~Radhia},\ and\ \citenamefont {Boujdaria}}]{fraj2007band}%
  \BibitemOpen
  \bibfield  {author} {\bibinfo {author} {\bibfnamefont {N.}~\bibnamefont {Fraj}}, \bibinfo {author} {\bibfnamefont {I.}~\bibnamefont {Sa{\"\i}di}}, \bibinfo {author} {\bibfnamefont {S.}~\bibnamefont {Ben~Radhia}},\ and\ \bibinfo {author} {\bibfnamefont {K.}~\bibnamefont {Boujdaria}},\ }\bibfield  {title} {\bibinfo {title} {Band structures of {A}l{A}s, {G}a{P}, and {S}i{G}e alloys: A 30 k$\times$p model},\ }\href {https://doi.org/10.1063/1.2773532} {\bibfield  {journal} {\bibinfo  {journal} {Journal of Applied Physics}\ }\textbf {\bibinfo {volume} {102}} (\bibinfo {year} {2007})}\BibitemShut {NoStop}%
\bibitem [{\citenamefont {Reeber}\ and\ \citenamefont {Wang}(1996)}]{reeber1996thermal}%
  \BibitemOpen
  \bibfield  {author} {\bibinfo {author} {\bibfnamefont {R.~R.}\ \bibnamefont {Reeber}}\ and\ \bibinfo {author} {\bibfnamefont {K.}~\bibnamefont {Wang}},\ }\bibfield  {title} {\bibinfo {title} {Thermal expansion and lattice parameters of group {IV} semiconductors},\ }\href {https://www.sciencedirect.com/science/article/pii/S0254058496018081} {\bibfield  {journal} {\bibinfo  {journal} {Materials Chemistry and Physics}\ }\textbf {\bibinfo {volume} {46}},\ \bibinfo {pages} {259} (\bibinfo {year} {1996})}\BibitemShut {NoStop}%
\bibitem [{\citenamefont {Madelung}(2012)}]{madelung2012semiconductors}%
  \BibitemOpen
  \bibfield  {author} {\bibinfo {author} {\bibfnamefont {O.}~\bibnamefont {Madelung}},\ }\href {https://link.springer.com/10.1007/978-3-642-45681-7} {\emph {\bibinfo {title} {Semiconductors: group {I}V elements and {III}-{V} compounds}}}\ (\bibinfo  {publisher} {Springer Science \& Business Media},\ \bibinfo {year} {2012})\BibitemShut {NoStop}%
\bibitem [{\citenamefont {Van~de Walle}(1989)}]{van1989band}%
  \BibitemOpen
  \bibfield  {author} {\bibinfo {author} {\bibfnamefont {C.~G.}\ \bibnamefont {Van~de Walle}},\ }\bibfield  {title} {\bibinfo {title} {Band lineups and deformation potentials in the model-solid theory},\ }\href {https://link.aps.org/doi/10.1103/PhysRevB.39.1871} {\bibfield  {journal} {\bibinfo  {journal} {Physical Review B}\ }\textbf {\bibinfo {volume} {39}},\ \bibinfo {pages} {1871} (\bibinfo {year} {1989})}\BibitemShut {NoStop}%
\bibitem [{\citenamefont {Van~de Walle}\ and\ \citenamefont {Martin}(1986)}]{van1986theoretical}%
  \BibitemOpen
  \bibfield  {author} {\bibinfo {author} {\bibfnamefont {C.~G.}\ \bibnamefont {Van~de Walle}}\ and\ \bibinfo {author} {\bibfnamefont {R.~M.}\ \bibnamefont {Martin}},\ }\bibfield  {title} {\bibinfo {title} {Theoretical calculations of heterojunction discontinuities in the {S}i/{G}e system},\ }\href {https://link.aps.org/doi/10.1103/PhysRevB.34.5621} {\bibfield  {journal} {\bibinfo  {journal} {Physical Review B}\ }\textbf {\bibinfo {volume} {34}},\ \bibinfo {pages} {5621} (\bibinfo {year} {1986})}\BibitemShut {NoStop}%
\bibitem [{\citenamefont {Lawaetz}(1971)}]{lawaetz1971valence}%
  \BibitemOpen
  \bibfield  {author} {\bibinfo {author} {\bibfnamefont {P.}~\bibnamefont {Lawaetz}},\ }\bibfield  {title} {\bibinfo {title} {Valence-band parameters in cubic semiconductors},\ }\href {https://link.aps.org/doi/10.1103/PhysRevB.4.3460} {\bibfield  {journal} {\bibinfo  {journal} {Physical Review B}\ }\textbf {\bibinfo {volume} {4}},\ \bibinfo {pages} {3460} (\bibinfo {year} {1971})}\BibitemShut {NoStop}%
\bibitem [{\citenamefont {Polak}\ \emph {et~al.}(2017)\citenamefont {Polak}, \citenamefont {Scharoch},\ and\ \citenamefont {Kudrawiec}}]{polak2017electronic}%
  \BibitemOpen
  \bibfield  {author} {\bibinfo {author} {\bibfnamefont {M.}~\bibnamefont {Polak}}, \bibinfo {author} {\bibfnamefont {P.}~\bibnamefont {Scharoch}},\ and\ \bibinfo {author} {\bibfnamefont {R.}~\bibnamefont {Kudrawiec}},\ }\bibfield  {title} {\bibinfo {title} {The electronic band structure of $\mathrm{{G}e}_{1-x}\mathrm{{S}n}_x$ in the full composition range: indirect, direct, and inverted gaps regimes, band offsets, and the burstein--moss effect},\ }\href {https://iopscience.iop.org/article/10.1088/1361-6463/aa67bf/meta} {\bibfield  {journal} {\bibinfo  {journal} {Journal of Physics D: Applied Physics}\ }\textbf {\bibinfo {volume} {50}},\ \bibinfo {pages} {195103} (\bibinfo {year} {2017})}\BibitemShut {NoStop}%
\bibitem [{\citenamefont {Menendez}\ and\ \citenamefont {Kouvetakis}(2004)}]{menendez2004type}%
  \BibitemOpen
  \bibfield  {author} {\bibinfo {author} {\bibfnamefont {J.}~\bibnamefont {Menendez}}\ and\ \bibinfo {author} {\bibfnamefont {J.}~\bibnamefont {Kouvetakis}},\ }\bibfield  {title} {\bibinfo {title} {Type-{I} {G}e/$\mathrm{{G}e}_{1-x-y}$$\mathrm{{S}i}_x$$\mathrm{{S}n}_y$ strained-layer heterostructures with a direct {G}e bandgap},\ }\href {https://doi.org/10.1063/1.1784032} {\bibfield  {journal} {\bibinfo  {journal} {Applied Physics Letters}\ }\textbf {\bibinfo {volume} {85}},\ \bibinfo {pages} {1175} (\bibinfo {year} {2004})}\BibitemShut {NoStop}%
\bibitem [{\citenamefont {Fischetti}\ and\ \citenamefont {Laux}(1996)}]{fischetti1996band}%
  \BibitemOpen
  \bibfield  {author} {\bibinfo {author} {\bibfnamefont {M.~V.}\ \bibnamefont {Fischetti}}\ and\ \bibinfo {author} {\bibfnamefont {S.~E.}\ \bibnamefont {Laux}},\ }\bibfield  {title} {\bibinfo {title} {Band structure, deformation potentials, and carrier mobility in strained {S}i, {G}e, and {S}i{G}e alloys},\ }\href {https://doi.org/10.1063/1.363052} {\bibfield  {journal} {\bibinfo  {journal} {Journal of Applied Physics}\ }\textbf {\bibinfo {volume} {80}},\ \bibinfo {pages} {2234} (\bibinfo {year} {1996})}\BibitemShut {NoStop}%
\bibitem [{\citenamefont {Edward}\ \emph {et~al.}(1990)\citenamefont {Edward}, \citenamefont {Croke}, \citenamefont {McGill~Jr},\ and\ \citenamefont {Miles}}]{edward1990measurement}%
  \BibitemOpen
  \bibfield  {author} {\bibinfo {author} {\bibfnamefont {T.~Y.}\ \bibnamefont {Edward}}, \bibinfo {author} {\bibfnamefont {E.~T.}\ \bibnamefont {Croke}}, \bibinfo {author} {\bibfnamefont {T.~C.}\ \bibnamefont {McGill~Jr}},\ and\ \bibinfo {author} {\bibfnamefont {R.~H.}\ \bibnamefont {Miles}},\ }\bibfield  {title} {\bibinfo {title} {Measurement of the strain dependence of the {S}i/{G}e (100) valence band offset},\ }in\ \href {https://doi.org/10.1117/12.20822} {\emph {\bibinfo {booktitle} {Growth of Semiconductor Structures and High-{T}c Thin Films on Semiconductors}}},\ Vol.\ \bibinfo {volume} {1285}\ (\bibinfo {organization} {SPIE},\ \bibinfo {year} {1990})\ pp.\ \bibinfo {pages} {212--219}\BibitemShut {NoStop}%
\end{thebibliography}%

\end{document}